\documentclass[mnsc,nonblindrev]{informs3}

\OneAndAHalfSpacedXI

\usepackage{natbib}
 \bibpunct[, ]{(}{)}{,}{a}{}{,}%
 \def\bibfont{\small}%
\TheoremsNumberedThrough     
\ECRepeatTheorems

\EquationsNumberedThrough    

\usepackage[ruled]{algorithm2e} 

\usepackage[utf8]{inputenc}

\usepackage{hhline}
\usepackage{multirow}

\usepackage{mathtools}
\usepackage{algorithmic}

\usepackage{booktabs}
\usepackage{array}
\usepackage{tabularx}
\usepackage{multirow}
\usepackage{enumitem}
\usepackage{tikz}
\usepackage{etoolbox} 
\usepackage{siunitx} 
\usepackage{colortbl} 
\usepackage{svg}

\usetikzlibrary{matrix, shapes}

\newrobustcmd{\ccell}{{\cellcolor{gray!40}}}

\usepackage{subcaption}

\usepackage{comment}

\usepackage{hyperref}
\usepackage{cleveref}
\hypersetup{colorlinks=true, citecolor=blue, linkcolor=blue, urlcolor=blue}

\usepackage{wrapfig}

\usepackage{colortbl} 

\newcommand\studentset[1]{N_{#1}}

\newcommand\courseset[1]{M_{#1}}

\newcommand{\ep}{\varepsilon}

\newrobustcmd{\crow}{{\rowcolor{gray!40}}}
\newcommand{\mylabel}[2]{#2\def\@currentlabel{#2}\label{#1}}

\newcommand{\maximinshare}[2][]{
    \ensuremath{a^{\text{MaxiMin},#2,u_{#1}}} 
    }
\newcommand{\Epsmaximinshare}[3][]{
    \ensuremath{a^{\text{MaxiMin},(#2,#3),u_{#1}}} 
    }

\newcommand{\maximinsharehat}[2][]{
    \ensuremath{a^{\text{MaxiMin},#2,\hat{u}_{#1}}} 
    }

\renewcommand{\epsilon}{\varepsilon}

\newcommand{\x}{a}
\newcommand{\xii}{a_i}

\newcommand{\R}{\mathbb{R}}

\newboolean{releaseMode}
\newboolean{pageBreaksBetweenSections}
\newboolean{pageBreaksBetweenSubsections}

\setboolean{releaseMode}{false}
\setboolean{pageBreaksBetweenSections}{false}
\setboolean{pageBreaksBetweenSubsections}{false}

\ifthenelse{\boolean{releaseMode}}{
    \usepackage[colorinlistoftodos,prependcaption,disable]{todonotes}
    \newcommand{\para}[1]{}

}{ 
    \usepackage[colorinlistoftodos,prependcaption]{todonotes}
    \presetkeys{todonotes}{inline}{} 
    \newcommand{\para}[1]{\textbf{§ #1:}}

} 

\ifthenelse{\boolean{pageBreaksBetweenSubsections}}{
    \let\oldSubsection\subsection
    \renewcommand*{\subsection}{\newpage\oldSubsection}
    \setboolean{pageBreaksBetweenSections}{true}
}{} 

\ifthenelse{\boolean{pageBreaksBetweenSections}}{
    \let\oldSection\section
    \renewcommand*{\section}{\newpage\oldSection}
}{} 

\begin{document}




\RUNTITLE{Machine Learning-Powered Course Allocation}


\TITLE{Machine Learning-Powered Course Allocation}


\RUNAUTHOR{Soumalias et al.}
\ARTICLEAUTHORS{%
\AUTHOR{Ermis Soumalias}
\AFF{Department of Informatics, University of Zurich and ETH AI Center, \EMAIL{seuken@ifi.uzh.ch}}
\AUTHOR{Behnoosh Zamanlooy}
\AFF{McMaster University, \EMAIL{zamanlob@mcmaster.ca}}
\AUTHOR{Jakob Weissteiner}
\AFF{Department of Informatics, University of Zurich and ETH AI Center, \EMAIL{weissteiner@ifi.uzh.ch}}
\AUTHOR{Sven Seuken}
\AFF{Department of Informatics, University of Zurich and ETH AI Center, \EMAIL{seuken@ifi.uzh.ch}}
}




\ABSTRACT{%
We study the course allocation problem, where universities assign course schedules to students. The current state-of-the-art mechanism, \textit{Course Match}, has one major shortcoming: students make significant mistakes when reporting their preferences, which negatively affects welfare and fairness. To address this issue, we introduce a new mechanism, \textit{Machine Learning-powered Course Match (MLCM)}. At the core of MLCM is a machine learning-powered preference elicitation module that iteratively asks personalized pairwise comparison queries to alleviate students' reporting mistakes. Extensive computational experiments, grounded in real-world data, demonstrate that MLCM, with only ten comparison queries, significantly increases both average and minimum student utility by 7\%--11\% and 17\%--29\%, respectively. Finally, we highlight MLCM's robustness to changes in the environment and show how our design minimizes the risk of upgrading to MLCM while making the upgrade process simple for universities and seamless for their students.
}%




\maketitle






\section{Introduction}
The \emph{course allocation problem} arises when educational institutions assign schedules of courses to students \citep{budish2012multi}. 
Each course has a limited number of indivisible seats and monetary transfers are prohibited for fairness reasons. 
The key challenge is to accurately elicit the students' preferences. The combinatorial nature of the preferences makes this particularly difficult, as students may view certain courses as complements or substitutes \citep{budish2022can}.

\subsection{Course Match}
\label{sec:Course-Match-Introduction}
The state-of-the-art practical solution to the course allocation problem is the \emph{Course Match (CM)} mechanism by \citet{budish2017course}, which provides a good trade-off between efficiency, fairness, and incentives. In recent years, CM has been adopted at many institutions such as the Wharton School at the University of Pennsylvania\footnote{\href{https://mba-inside.wharton.upenn.edu/course-match/}{https://mba-inside.wharton.upenn.edu/course-match/}} and Columbia Business School\footnote{\href{https://students.business.columbia.edu/records-registration/course-match-registration}{https://students.business.columbia.edu/records-registration/course-match-registration}}.

CM uses a simple reporting language to elicit students' preferences over \textit{schedules} (i.e., course bundles). Concretely, CM offers students a graphical user interface (GUI) to enter a \emph{base value} between $0$ and $100$ for each course, and an \emph{adjustment value} between $-200$ and $200$ for each \textit{pair} of courses. Adjustments allow students to report complementarities and substitutabilities between courses, up to \emph{pairwise} interactions. The total value of a schedule is the sum of the base values reported for each course in that schedule plus any adjustments (if both courses are in the schedule).

Prior to the adoption of CM in practice, \citet{budish2022can} performed a lab experiment to evaluate CM. Regarding efficiency, they found that, on average, students were happier with CM compared to the \textit{Bidding Points Auction}, the previously used mechanism. Regarding fairness, students also found CM fairer. Regarding the reporting language, they found that students were able to report their preferences ``accurately enough to realize the efficiency and fairness benefits.'' Given these positive findings, Wharton was then the first university to switch to CM. 

\subsection{Preference Elicitation Shortcomings of Course Match}

However, \citet{budish2017course} were already concerned that the CM language may not be able to fully capture all students' preferences. Furthermore, they mentioned that some students might find it non-trivial to use the CM language and might therefore make \textit{mistakes} when reporting their preferences. Indeed, the lab experiment by \citet{budish2022can} revealed several shortcomings of CM in this regard.

First, students made very limited use of the CM language: on average, students only reported a base value for half of the 25 courses in the experiment. Additionally, the average number of pairwise adjustments was only $1.08$ (out of $300$ possible), and the median was $0$. This suggests that cognitive limitations negatively affect how well students can report their preferences using the CM language, unless students have essentially additive preferences. Second, in addition to not reporting part of their preferences, \citet{budish2022can} provided evidence that students are also \textit{inaccurate} when they do report their preferences.

\citet{budish2022can} found that both of these reporting mistakes negatively affected the welfare of CM. In their experiment, about $16\%$ of students would have preferred another course schedule, with a median utility difference for these schedules of $13\%$. Thus, preference elicitation in course allocation remains an important challenge.

\subsection{Machine Learning-powered Preference Elicitation}
\label{subsec:MLpoweredPreferenceElicitation}
To address this challenge, we turn to \emph{machine learning (ML)}. The high-level idea is to train a separate ML model for each student based on that student's reports after using the CM language (i.e., the GUI). These ML models can then be used in an \textit{ML-powered preference elicitation algorithm} that asks each student a sequence of carefully selected queries, thus enabling them to correct the reporting mistakes they made in the GUI. Based on those queries, the student's ML model is updated in real-time (in less than one second) and the next query is generated. At the end, the mechanism allocates schedules to students based on their trained ML models.

With our approach, we build on the ideas developed in a recent stream of papers on ML-powered preference elicitation.
\citet{lahaie2004applying} and \citet{blum2004preference} 
were the first to combine ML and mechanism design, studying the relation between learning theory and preference elicitation. \citet{brero2017, brero2018} 
were the first to integrate an ML-powered preference elicitation component into a practical combinatorial auction mechanism. They used support vector regression to learn bidders' value functions and to generate new queries in each auction round. In \citep{brero2021machine}, the authors proposed the MLCA mechanism and showed that it achieves higher efficiency than the widely-used combinatorial clock auction \citep{ausubel2006clock}. Recently, there has been a stream of papers improving the ML capability of the MLCA mechanism, which we discuss in 
\Cref{sec:RelatedWork}.

While these works are important pre-cursors to the present paper, there are several noteworthy differences. First, and most importantly, these papers used \emph{value queries} as the interaction paradigm (i.e., asking agents a query of the form ``What is your value for bundle \{XYZ\}''); 
however, in a setting without money, it is unnatural to report preferences in cardinal terms. 
Instead, we use \emph{pairwise comparison queries (CQs)} (i.e., ``Do you prefer course schedule A or B?'') because a pairwise CQ is a simpler type of query known to have low cognitive load \citep{conitzer2009eliciting,chajewska00}. Second, our goal is not to replace but to build on top of the CM language; thus, we must be able to handle the \textit{cardinal} input that students provide via the CM reporting language as well as the \textit{ordinal} feedback from answering CQs (e.g., $A \succ B$). Third, while an auctioneer can require bidders in an auction to participate in a \textit{synchronous} way (i.e., submitting a bid in every round), we must allow students to interact with the mechanism in an \textit{asynchronous} manner (i.e., allowing students to answer a sequence of CQs without having to wait on other students).

\subsection{Overview of Contributions}
In this paper, we introduce the \emph{machine learning-based Course Match (MLCM)} mechanism, which builds on top of CM by adding an ML-powered preference elicitation algorithm.

 First, students use the CM reporting language (i.e., the same GUI). As in CM, this input is required from all students. 
 Second, MLCM uses these initial reports to train a separate ML model for each student so that it can predict each student's utility for any possible course schedule. 
 Third, MLCM uses a carefully designed ML-powered preference elicitation algorithm to generate \emph{pairwise comparison queries} that are tailored to each student, and students simply answer which schedule they prefer. Based on this feedback, each student's ML model is retrained and the next query is generated. Importantly, this iterative phase is \emph{optional} -- each student can answer as many such queries as she wants (including none). However, the more queries she answers, the better the ML model will typically approximate her true preferences, which will benefit her in the last phase, where MLCM computes the final allocation based on all ML models (and in case a student has answered no queries, then only her GUI reports will be used for the final allocation calculation).\footnote{An alternative approach is to only use the trained ML models to update the student's GUI reports, and to then run the original CM mechanism on the updated reports. In \Cref{subsec:MLCM_practicability}, we show that this approach also works and leads to essentially the same efficiency, but in terms of run-time it is approximately 10 times slower.}


In \Cref{section: MLCM}, we describe all components of MLCM. 
First, we give a detailed description of MLCM (\Cref{subsec:mlcm_details}).
Then we formally introduce our query generation algorithm, which we call \textit{online binary insertion sort} (OBIS) (\Cref{subsec:query_generation}). We prove that the amount of information that can be inferred for a student grows superlinearly in the number of OBIS-generated CQs she answers. 
We provide a simple example, showcasing how MLCM using OBIS queries helps alleviate a student's reporting mistakes (\Cref{subsec:toy_example}). 
Finally, we extend the theoretical guarantees of CM to MLCM (\Cref{sec_theoretical_guarantees}). 
Importantly, we explain why MLCM is also \textit{strategyproof in the large}.\footnote{A mechanism is \emph{strategyproof in the large} if, for sufficiently many agents and any full-support i.i.d. distribution of the other agents' reports, reporting truthfully is approximately interim optimal \citep{azevedo2019strategy}.}

In \Cref{sec:MachineLearningInstantiation}, we instantiate the ML model used by MLCM with \emph{monotone-value neural networks (MVNNs)} \citep{weissteiner2022monotone}, an architecture specifically designed for combinatorial assignment problems. We introduce a training method that combines the cardinal input from the CM language and the ordinal feedback from the CQs in accordance with Bayesian theory.

In \Cref{sec_SPG}, we introduce a new course allocation simulation framework to evaluate MLCM. Its first component is a realistic student preference generator, designed so that each student's complete preferences can be encoded in a succinct Mixed Integer Program (MIP). This allows computing a benchmark allocation given the students' true preferences. Its second component models the students' reporting mistakes when interacting with the CM language. We calibrate the framework's parameters based on real-world data from the lab experiment of \citet{budish2022can}.\footnote{The source code for our course allocation simulator is already publicly available (we omit the hyperlink to preserve anonymity). 
We have submitted the source code for the simulator, for MLCM, and for all of our experiments as part of the supplementary material, and we will make the source code publicly available upon acceptance of this paper.}

In Section~\ref{Section_efficiency_results}, we empirically evaluate the performance of MLCM. We find that MLCM significantly outperforms CM in terms of \textit{average} as well as \textit{minimum student utility}, even with only one CQ (\Cref{subsec:experimental_results}).
We show that our results are robust to changes in the students' reporting mistakes (\Cref{subsec:Reporting Mistakes Robustness Study}),
that the expected benefit of a student unilaterally opting into MLCM is large (\Cref{subsec:Should Individual Students Opt Into MLCM?}),
and that MLCM also outperforms CM when students have simple additive preferences (\Cref{subsec:Results for Additive Preferences}). To explain the performance improvements of MLCM, we show that OBIS generates CQs that are surprising and informative for the ML models (\Cref{subsec:af_empirical_evaluation}). 
Finally, we highlight how gracefully MLCM's runtime scales with the number of courses (\Cref{sec:Scaling})
and the practicability of piloting MLCM for a university currently using CM (\Cref{subsec:MLCM_practicability}).

In Section~\ref{section_discussion}, we contextualize our results, explore the practicalities of implementing MLCM, and examine alternatives. In Section~\ref{sec:Conclusion}, we conclude, outline avenues for future work, and explain how our proposed mechanism can be applied to other combinatorial assignment problems as well.

\section{Related Work}
\label{sec:RelatedWork}
Our work is related to the research on course allocation and ML-based preference elicitation.

\subsection{Course Allocation}

The course allocation problem is an instance of the \emph{combinatorial assignment problem}, for which several impossibility results establish a tension between welfare, incentives, and fairness. For example, it is known that the only mechanisms for this problem that are ex-post Pareto efficient and strategyproof are dictatorships \citep{papai2001strategyproof,hatfield2009strategy}.

Multiple empirical studies have pointed out design flaws of course allocation mechanisms used in practice. \citet{budish2012multi} showed that the \emph{Harvard Business School (HBS) draft} mechanism \citep{draftbrams1979} creates significant incentives for students to misreport, leading to large welfare losses. 
The commonly used \emph{Bidding Points Auction} implicitly assumes that students have positive value for left-over virtual currency, which harms incentives and ultimately leads to allocations that are neither efficient nor fair \citep{sonmez2010course}.

Motivated by these design flaws, \citet{budish2011aceei} proposed a new mechanism for the combinatorial assignment problem called \emph{approximate competitive equilibrium from equal incomes (A-CEEI)}. \mbox{A-CEEI} circumvents the impossibility results previously mentioned by making slight compromises in all three of those dimensions of interest. Specifically, A-CEEI is approximately efficient, satisfies desirable fairness criteria (envy bounded by a single good and $(n+1)$-maximin share guarantee), and is strategyproof in the large. Later, \citet{budish2017course} introduced CM as the practical implementation of A-CEEI.  We provide a short review of A-CEEI and CM in Section~\ref{sec_prelim}.

While our research focuses on the combinatorial assignment approach to course allocation, there are also other modeling approaches.
\citet{diebold2014course} modeled course allocation as a two-sided matching problem, where, in contrast to our setting, instructors also have preferences over students. \cite{Bichler2021HowToAssign} studied the same setting where courses additionally have minimum quotas. \cite{kushnir2023} extended the A-CEEI mechanism to a similar two-sided matching setting where students have different priorities for courses.
\citet{bichler2021randomized} studied the assignment of tutorials for mandatory courses, which is more similar to a scheduling problem.

\subsection{Machine Learning-based Preference Elicitation}
Preference elicitation (PE) using comparison queries (CQs) has received a lot of attention in the ML community. Bayesian approaches are a natural candidate for PE due to their ability to explicitly model the uncertainty over agents’ utility functions.
\citet{preference_learning_with_gps} used GPs for the problem of learning preferences given a \textit{fixed} set of comparison queries. However, they did not answer the question of \textit{which} CQs to ask.  \citet{GP_preference_elicitation_2010} addressed this question by iteratively selecting the pairwise CQ that maximizes the \emph{expected value of information} (EVOI). But their approach is impractical in our setting, as the EVOI needs to be analytically calculated for each potential CQ. Concretely, the total number of potential CQs scales quadratically in the number of alternatives (course schedules in our case), which is already polynomial in the number of courses. \citet{multi_attribute_pe} showed how to speed up the evaluation of EVOI heuristics by introducing an approximate PE framework for performing efficient closed-form Bayesian belief updates and query selection. However, their approach does not work for course allocation because it is still too slow due to the very large number of possible CQs, and because it can only model additive utility functions. In general, GPs are not well suited to our setting due to the high dimensionality of the input space and the integrality constraints that make GP optimization intrinsically difficult.

\citet{Ailon_active_learning} took a different approach, proposing an active learning algorithm that, using pairwise comparison queries which may be non-transitive, can learn an almost optimal linear ordering of a set of alternatives with an almost optimal query complexity. This was further improved by \citet{Ailon2011ANA}. 
However, these approaches are impractical for course allocation because they would require more than one hundred thousand queries per student. This high query count stems from the fact that they do not exploit any notion of similarity between schedules. 

The PE approach we propose in this paper shares some similarities with prior work in \textit{multi-attribute utility theory (MAUT)} \citep{Keeney1993}. In MAUT, items (e.g., cars) are described by their values across various attributes (e.g., horsepower, color, price), and an agent’s utility for an item is the sum of the agent's values for the item’s individual attributes. Within this framework, multiple techniques have been proposed to learn agents' preferences. For example, \textit{adaptive conjoint analysis} (e.g., \citet{Green1991Adaptive}) first obtains an approximate ordering of attribute importance and then asks agents pairwise CQs. Another technique was proposed by \citet{boutilier2006}, who ask pairwise CQs that minimize the agent's regret. However, the MAUT framework is not a good fit for course allocation because courses are not items that are described by multiple attributes; rather, courses are bundles of items. While one could map course allocation into the MAUT framework by treating a schedule as an item with an attribute for each potential course, this would be ineffective for two reasons. First, the techniques developed for preference learning in MAUT work well with a smaller number of attributes (e.g., 5-10), but scale badly to a setting with 100s of attributes (i.e., potential courses contained in a schedule). Second, the MAUT framework can only handle additive preferences well, as modeling pairwise interactions between attributes would already lead to an explosion of the feature space.

Most related to our work is the research on ML-powered combinatorial auctions using support vector regression \citep{brero2017, brero2018, brero2021machine}. \citet{weissteiner2020deep} extended this work by using neural networks, further increasing efficiency. \citet{weissteiner2022fourier} introduced Fourier-sparse approximations for the
problem of learning combinatorial preferences. Finally, \citet{weissteiner2022monotone} introduced \emph{monotone-value neural networks (MVNNs)}, which are specifically designed to learn \textit{monotone} combinatorial preferences, resulting in a further efficiency increase.

\section{Preliminaries}
\label{sec_prelim}

In this section, we first present our formal model and then review A-CEEI and CM.

\subsection{Formal Model} \label{subsec_formal_model}
Let $\studentset{} = \{1, \dots, n \}$ denote the set of students indexed by $i$, and let $\courseset{} = \{ 1, \dots, m \}$  denote the set of courses indexed by $j$. Each course $j$ has a \textit{capacity} $q_j \in \mathbb{N}_{>0}$. Each student $i$ has a set $\Psi_i \subseteq 2^{M}$ of \textit{permissible course schedules}.
The set $\Psi_i$ encapsulates both scheduling constraints as well as any student-specific constraints (e.g., prerequisites). An indicator vector $x \in \mathcal{X} = \{0,1 \}^m$  denotes a course schedule where $x_j = 1$ iff course $j \in M$ is part of schedule $x$. We let $a= (a_i)_{i \in \studentset{}} \in \mathcal{X}^n$ denote an \textit{allocation} of course schedules to students, where $a_i$ is the course schedule of student $i$. A course $j$ is \textit{oversubscribed} in an allocation $a$ iff $\sum_{i=1}^n a_{ij} > q_j$ and \textit{undersubscribed}  iff $\sum_{i=1}^n a_{ij} < q_j$. We denote the set of feasible allocations by 
$\mathcal{F} = \{ a \in \mathcal{X}^n: \sum_{i = 1}^n a_{ij} \le q_j \;  \forall j \in M, a_i \in \Psi_i \; \forall i \in \studentset{} \}$.
Students' preferences over course schedules are represented by their (private) \textit{utility functions} $u_i: \mathcal{X} \xrightarrow[]{} \mathbb{R}_{+}$, $i \in N$, i.e., $u_i(x)$ represents student $i$'s utility for course schedule $x$.

\subsection{Approximate Competitive Equilibrium from Equal Incomes (A-CEEI)}
\citet{budish2011aceei} proposed the A-CEEI mechanism as an approximation to the competitive equilibrium from equal incomes (CEEI). A-CEEI simulates a virtual economy where students are assigned budgets that are approximately (but not exactly) equal.
To introduce A-CEEI formally, we represent each student $i$'s complete ordinal preferences by $\preceq_i$. Each student $i$ is allocated a budget $b_i \in [1, 1+\beta],\, \beta>0$. Next, \textit{approximate market-clearing prices} $p^* \in \mathbb{R}^m_{\ge 0}$ are calculated (where $p^*_j$ is the price for course $j$) such that, when each student $i$ ``purchases'' her favorite permissible schedule $a_i^*$ within her budget, the market approximately clears. Formally, given an allocation $a^*$, the \textit{clearing error} $z_j$ for course $j$ and price $p^*_j$ is
\begin{align}
    z_j \coloneqq \begin{cases}
\sum_i a_{ij}^* - q_j & p^*_j > 0,   \\ 
\max \left \{ \sum_i a_{ij}^* - q_j, 0  \right \} & p^*_j = 0.
\end{cases}
\end{align}

The \textit{clearing error} of the allocation $a^*$ is defined as $\alpha \coloneqq \sqrt{ \sum_j z_j^2 }$. 
Following \citet{budish2011aceei}, we say that a price vector $p^*$ \emph{approximately clears} the market if $\alpha \le \sqrt{\sigma m} / 2$, with $\sigma = \min \{2k, m \}$, where $k$ is the maximum number of courses in a permissible schedule.\footnote{\citet{budish2011aceei} proved that for any $\beta > 0$ such a price vector always exists.} Each student $i$ is allocated her utility-maximizing schedule $a_i^*$ that is permissible and within her budget. Formally,
\begin{equation}
    a_i^* \in \argmax_{\preceq_i} \left [a_i \in \Psi_i: \sum_j a_{ij} p^*_j \le b_i   \right ]. 
\end{equation}

Then $[a^*, b, p^*]$ constitutes an $(\alpha, \beta)$-A-CEEI.

\subsection{Course Match (CM)} \label{subsection: CM_prelims}

\begin{wrapfigure}[13]{r}[0pt]{0.5\textwidth}
    \centering
    \vskip -0.5cm
    \resizebox{0.5\textwidth}{!}{
    \includegraphics[trim = 0 0 60 0, clip]{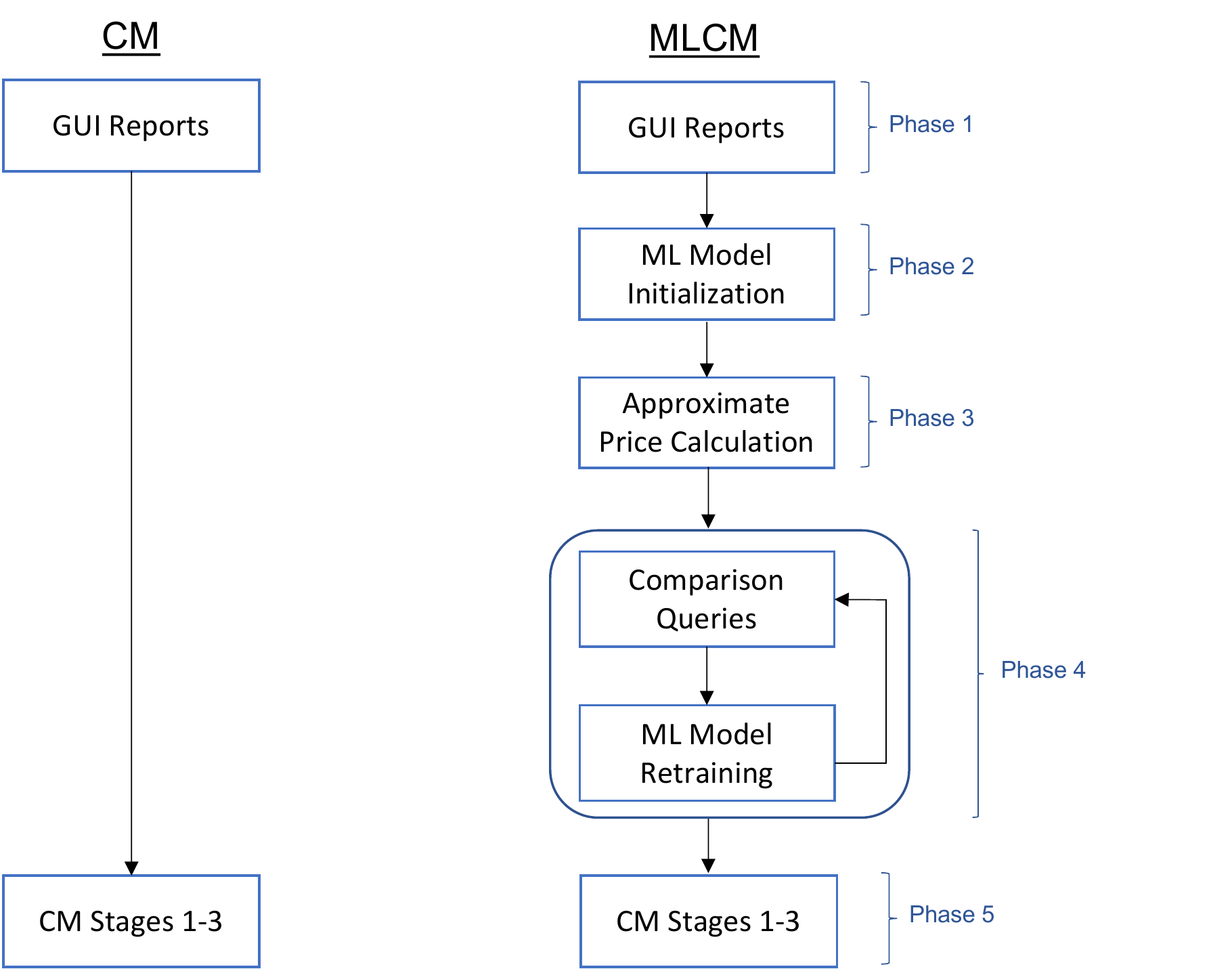}
    }
    \caption{Schematic overview of CM and MLCM}
    \label{fig:cm_mlcm}
\end{wrapfigure}

A-CEEI has many attractive properties (see Section~\ref{sec:RelatedWork}), but it cannot be directly implemented in practice for multiple reasons. First, A-CEEI assumes access to the students' \textit{full} ordinal preferences. Second, it only \emph{approximately} clears the market, which implies that  some courses could be oversubscribed (which would violate a hard capacity constraint in many business schools, where seats cannot easily be added to a classroom). Third, the combinatorial allocation problem is PPAD-complete \citep{othman2016complexity}; thus, we do not have a polynomial-time algorithm to solve it. To address these challenges, \citet{budish2017course} introduced \textit{Course Match (CM)} as a practical implementation of A-CEEI. In CM, students first report their preferences using the GUI (see Section~\ref{sec:Course-Match-Introduction}); the final allocation is then computed in three stages (see Figure~\ref{fig:cm_mlcm}). 

\paragraph{Stage 1:} 
In Stage 1, CM finds a price vector that constitutes an A-CEEI based on a heuristic algorithm that iteratively examines price vectors based on their clearing error. In the original CM implementation, the heuristic algorithm employed was tabu search \citep{tabusearch}. Recently,
\citet{rubinstein2023practical} proposed a novel algorithm specifically designed for this task that outperforms tabu search in terms of computation speed by several orders of magnitude. An A-CEEI for a large business school can now be computed on a conventional laptop in under ten minutes, while the same task previously required about a day on a large compute cluster. 
For both heuristic algorithms, for every price vector examined, a MIP has to be solved for every student to determine her utility-maximizing schedule within her budget. 
Even though the allocation determined by Stage 1 is an A-CEEI,
it could, in theory,
be infeasible due to some courses being slightly oversubscribed.\footnote{
For all real-world instances it has been applied to, the heuristic algorithm of \citet{rubinstein2023practical} was able to perfectly clear the market in Stage~1, implying no difference between the Stage~1, Stage~2, or Stage~3 allocations.}

\paragraph{Stage 2:} 
In Stage 2, CM removes oversubscription (if any), by iteratively increasing the price of the most oversubscribed course until no oversubscribed courses are left. 
\paragraph{Stage 3:} 
In Stage 3, CM reduces the undersubscription caused by Stage 2 by first increasing all students' budgets by a fixed percentage and then allowing students, one after the other, to ``purchase'' courses that still have seats available.

\section{Machine Learning-powered Course Match (MLCM)}
\label{section: MLCM}

In this section, we introduce our \emph{ML-powered Course Match} (MLCM) mechanism and discuss its theoretical properties. Here, we describe MLCM for a generic ML model $\mathcal{M}$. In Section~\ref{sec:MachineLearningInstantiation}, we instantiate the ML model using monotone-value neural networks.

\subsection{Details of MLCM} \label{subsec:mlcm_details}
MLCM proceeds in five phases (see \Cref{fig:cm_mlcm}).

\paragraph{Phase 1: Preference Reporting via GUI.}
In Phase 1, students initially report their preferences using the same reporting language (and same GUI) as in CM (see Section~\ref{sec:Course-Match-Introduction}). After this phase, each student can decide whether she also wants to use MLCM's ML-based preference elicitation feature (which we will simply call the ``ML feature'' going forward). If a student decides to ``opt out'' of the ML feature, then MLCM treats that student's preference reports in the same way as CM would, without employing any ML.

\paragraph{Phase 2: ML Model Initialization.}
In Phase $2$, for each student $i$ who has not opted out of the ML feature, MLCM creates an initial ML model of her utility function based on her GUI reports from Phase 1. To do so, MLCM creates a \emph{cardinal} training data set $D_{i, {\text{card}}}$ consisting of $\ell$ schedule-value pairs implied by student $i$'s reports from Phase 1, i.e., $D_{i, {\text{card}}} = \{(x_{ik}, \hat{u}_i(x_{ik})\}_{k=1...\ell}$ (see electronic companion \Cref{A_bootstrap} for details) and then uses $D_{i, {\text{card}}}$ to train an initial ML model $\mathcal{M}_i^0: \mathcal{X} \rightarrow \mathbb{R_+}$, where $\mathcal{M}_i^0 (x)$ denotes the ML model's prediction of student $i$'s utility for schedule $x$. Our choice of ML model architecture is detailed in \Cref{sec:MachineLearningInstantiation}.

\paragraph{Phase 3: Approximate Price Calculation.}
Next, MLCM runs Stage 1 of CM to calculate approximately market-clearing prices for all courses. 
These prices are not final, but only used to steer the preference elicitation in Phase~4.\footnote{
\label{footonote_apx_prices}
If it was desired that students could immediately start answering comparison queries after reporting their initial preferences via the GUI, one could use last year's prices for Phase 4 and set the price of new courses to some average price. Of course, if students' preferences for courses had changed significantly compared to last year, or if the new courses were particularly popular or unpopular, then using last year's prices could lower the effectiveness of the ML-powered elicitation phase. Nonetheless, in  electronic companion~\ref{subsec:app_price_sensitivity}, we experimentally show that MLCM is robust to miss-specification of the approximate prices of Phase 3.} 
For students who opted out of the ML feature, MLCM uses their GUI reports; for every other student $i$, it uses the values predicted by the ML model $\mathcal{M}_i^0$.

\paragraph{Phase 4: ML-based Preference Elicitation.}
In this phase, MLCM uses OBIS, a novel ML-powered algorithm to generate a sequence of CQs tailored to each individual student. 
OBIS is described in detail in \Cref{subsec:query_generation}.

\paragraph{Phase 5: Computing the Final Allocation.}
MLCM runs Stages 1--3 of CM to determine the final allocation. For those students who opted out of the ML feature, MLCM uses their GUI reports; for every other student $i$, it uses the values predicted by the ML model $\mathcal{M}_i$.

\subsection{Online Binary Insertion Sort (OBIS)}
\label{subsec:query_generation}
In this section, we introduce our query generation algorithm, \textit{Online Binary Insertion Sort (OBIS)}.

OBIS is designed to generate a sequence of CQs that maximizes the worst case number of schedule pairs for which we can infer their relation with respect to the student's preferences. To illustrate why this is challenging, we first present two straw men approaches.

The first straw man is the \textit{random algorithm}: at each iteration, ask the student to compare two random  schedules within her budget. 
This algorithm has two shortcomings. First, it does not focus on the most important part of the allocation space -- schedules that are \textit{valuable} yet \textit{attainable} for the student.
Second, it does not leverage the student's past responses to ask the next CQ.

The second straw man, the \textit{na\"{i}ve algorithm}, alleviates both of these limitations. 
At each iteration, it asks the student to compare between her most preferred schedule found so far and the schedule within her budget with the highest predicted utility that she has not already been queried about.
This algorithm, unlike the random algorithm, focuses on the most important part of the allocation space and leverages past information when generating the next CQ. 
It has one additional advantage: in case the student prefers the 
newly-generated schedule,
we can infer that she prefers that schedule over \textit{all} schedules that she was previously queried about, and we can incorporate that information into our training algorithm (see \Cref{subsec:Integrating comparison queries into MVNNs}). 
However, in the opposite case, we can only infer that she prefers her previously most preferred schedule to the newly-generated one. 


This observation motivates OBIS's design. During the entire preference elicitation phase, OBIS maintains a list of all schedules a student has already been queried about, sorted with respect to the student's preferences. Just like the na\"{i}ve algorithm, it identifies the schedule within the student's budget with the highest predicted utility that she has not already been queried about. In contrast to the na\"ive algorithm, OBIS does not just compare that schedule with the student's most preferred schedule found thus far. Instead, OBIS asks the same queries that binary insertion sort would ask when inserting the newly-generated schedule into the already sorted list, with the student giving the answers to those queries online. With each CQ that the student answers, binary insertion sort reduces the search space of the sorted list in half. For the half that was excluded, we can infer the ordinal relation between the newly-generated schedule and any schedule in the excluded half. 
Consequently, in contrast to the na\"{i}ve algorithm, even in the worst case, the number of schedule pairs for which we can infer their relation  grows superlinearly in the number of OBIS-generated CQs. We show this result formally in \Cref{proposition:amortized_impact}.
OBIS is detailed in \Cref{alg_query_generation}.


\begin{algorithm}[t]
\caption{Online Binary Insertion Sort (OBIS)}
\label{alg_query_generation}
\begingroup 
\renewcommand{\baselinestretch}{1} 
\selectfont 
\textbf{Input}: Current ML model $\mathcal{M}_i$, cardinal data set $D_{i, \text{card}}$, price vector $p$, permissible schedules $\Psi_i$, budget $b_i $, student's answered comparison queries $Q_i$, list of sorted schedules based on student's past responses $S_i$, newly-generated schedule to be inserted $x$ 
\\
\textbf{Output}: Next CQ  
\begin{algorithmic}[1] 
\IF {$x$ can be inserted into $S_i$ based on the already answered CQs} \label{alg_cq_gen:bS_done}
\STATE $S_i \gets$  Sort($S_i,x, Q_i)$ \label{alg_cq_gen:sort_list}
\STATE $D_{i,\text{ord}} \gets$  All pairwise orderings inferred from $S_i$ \label{alg_cq_gen:get_all_orderings}
\STATE $\mathcal{M}_i \gets$  Train($D_{i, \text{card}},D_{i,\text{ord}}$) \label{alg_cq_gen:train_model}
\STATE $x \gets \argmax_{x' \in \Psi_i, x' \notin S_i: x' \cdot p \le b_i} \mathcal{M}_{i}(x')$ \label{alg_cq_gen:get_best_bundle}
\ENDIF
\STATE $CQ \gets$  NextBinaryInsertionSortQuery($S_i, x, Q_i$) \label{alg_cq_gen:call_bsquery}
\STATE \textbf{return} $CQ$ \label{alg_cq_gen:return}
\end{algorithmic}
\endgroup 
\end{algorithm}

First, OBIS checks if, based on the student's past responses, it already has enough information to sort the newly-generated schedule $x$ into the list of schedules $S_i$ (Line~\ref{alg_cq_gen:bS_done}). 
If that is the case, the sorted list $S_i$ is extended to include $x$ (Line~\ref{alg_cq_gen:sort_list}),  all pairwise orderings that can now be inferred from $S_i$ are added to the ordinal data set $D_{i,\text{ord}}$ (Line~\ref{alg_cq_gen:get_all_orderings}), 
the ML model is retrained with all available data (Line~\ref{alg_cq_gen:train_model}), and the next schedule $x$ with the highest predicted utility is determined (Line~\ref{alg_cq_gen:get_best_bundle}). Then, the \textit{NextBinaryInsertionSortQuery} sub-routine (see \Cref{alg_nextbinarysearchqueries} in electronic companion \ref{app:nextbinarysearchquery}) identifies the next CQ (Line~\ref{alg_cq_gen:call_bsquery}). If a new schedule $x$ was just generated in Line~\ref{alg_cq_gen:get_best_bundle}, then this sub-routine effectively starts a new binary insertion sort aiming to insert $x$ into $S_i$. Otherwise, it continues binary insertion sort where it left off with the previous query. For our choice of ML model (see \Cref{sec:MachineLearningInstantiation}), both training the ML model and determining the schedule with the highest predicted utility can be done in less than half a second, enabling the real-time generation of CQs.
\medskip

Recall that, for the two straw men algorithms, in the worst case, the average number of additional schedule pairs for which the student's preference relation can be inferred is only \emph{one} per CQ. In contrast, for OBIS, this number is quickly much larger than one. \Cref{proposition:amortized_impact} formalizes this.



\begin{proposition}    
\label{proposition:amortized_impact}
Let $l$ be the current length of the student's sorted list of schedules $S$ at the start of an iteration of binary insertion sort. Then, in the worst case, the average number of additional schedule pairs for which the student's preference relation can be inferred is $\frac{l}{ \lceil \log_2 (l + 1) \rceil }$ across all CQs generated until the next schedule is inserted into $S$.
\end{proposition}

The proof is provided in electronic companion \ref{sec_appendix_query_gen_proofs}. 

To illustrate \Cref{proposition:amortized_impact}, consider a student who has answered $25$ OBIS CQs. With $25$ CQs, the minimum length $l$ of the student's sorted list of schedules is $10$.\footnote{To see this, note that increasing a list of length $l$ by $1$ element using binary insertion sort requires at most $ \lceil \log_2(l+1) \rceil $ queries. Thus, creating a list of length $10$ requires at most $\sum_{n = 0}^{9} \lceil \log_2(n+1) \rceil = 25$ queries.} Then, for the last $4$ CQs, the average number of additional schedule pairs for which the preference relation can be inferred is $\frac{9}{ \lceil \log_2 (9 + 1) \rceil } = 2.25$. In total, for any of the  ${10 \choose 2} = 45$ possible schedule pairs, we can determine which of the two schedules the student prefers. 

In \Cref{subsec:af_empirical_evaluation}, we experimentally compare OBIS against the two straw men. We show that the ordinal dataset that can be inferred indeed grows significantly faster for OBIS, while OBIS also produces CQs that are more surprising for the ML model, resulting in larger changes to its weights. In \Cref{subsec:experimental_results}, we show that these advantages of OBIS result in significant welfare improvements.

\subsection{A Worked Example}\label{subsec:toy_example}
In this section, we present an  example to illustrate MLCM and OBIS. The example provides intuition for how the CQs help correct students' initial reporting mistakes. 
See electronic companion~\ref{app_sec:example_forgetting_base_values} for an example where the student also forgets to report some of her base values. 

\begin{example}[Correcting reporting mistakes]\label{example:noisy_report_II}

We consider a setting with four courses $M\coloneqq\{1,2,3,4\}$ with capacity 1 each. There is a single student $N\coloneqq\{1\}$ who has a budget of $b=1$ and who wants a schedule consisting of at most two courses (and there are no other student-specific constraints). 
We assume the student answers four CQs, making no mistakes when answering them. 
The prices of the courses are $p=(0.6,0.6,0.3,0.3)$; thus, the student can afford all schedules of size two, except for $\{1,2\}$.  For ease of exposition, we assume that the student has additive preferences and that the GUI only allows for additive reports. The example is presented in \Cref{tab:toy_example_2}.\footnote{We omit the subscript $i$ for all variables in this example, as there is only a single student.} The student's true utility function $u$ is presented in row 1. The student's GUI reports $u^{GUI}$ (which include reporting mistakes) are shown in row 2. The student has reported a base value for every course but has made small mistakes in all her reports. 
We use linear regression as the ML architecture so that each linear coefficient can be interpreted as an estimate of the student's base value for a course. 
Let $\mathcal{M}^{t}(x)$ denote the student's ML model trained on the data from $t\in \mathbb{N}_0$ CQs.

\begin{table}[t!]
\centering
\begin{sc}
	\begin{tabular}{
                cccccccc
			}
		\toprule
            \multicolumn{1}{c}{Preference}&\multicolumn{4}{c}{Courses}& {Utility Maximizing}&  \multicolumn{1}{c}{Utility}& \multicolumn{1}{c}{Answer to}\\
		\multicolumn{1}{c}{Model} & \multicolumn{1}{c}{1} & \multicolumn{1}{c}{2} & \multicolumn{1}{c}{3} & \multicolumn{1}{c}{4}&  {Schedule $a^*$}&{$u(a^*)$}& {CQ}  \\
            \cmidrule(lr){1-1}
		\cmidrule(lr){2-5}
		\cmidrule(lr){6-6}
            \cmidrule(lr){7-7}
            \cmidrule(lr){8-8}
		{$u$}   & 85& 70 & 50  & 40  & {\{1,3\}} & 135& \\
		{$u^{GUI}$}  & 75& 77 & 42  & 45  & {\{2,4\}} & 110&\\
            \midrule
		{$\mathcal{M}^{0}$} & 75& 77 & 42  & 45  & {\{2,4\}} & 110& 
  {$\{1,4\} \succ \{2,4\}$}\\
		{$\mathcal{M}^{1}$} & 80& 72 & 42  & 45  & {\{1,4\}} & 125& {$\{1,3\} \succ \{1,4\}$} \\
        {$\mathcal{M}^{2}$} & 80& 72 & 47  & 40  & {\{1,3\}} & 135& {$\{1,4\} \succ \{2,3\}$} \\
        {$\mathcal{M}^{3}$} & 80& 72 & 47  & 40  & {\{1,3\}} & 135& {$\{2,3\} \succ \{2,4\}$} \\
        {$\mathcal{M}^{4}$} & 80& 72 & 47  & 40  & {\{1,3\}} & 135&  \\
		\bottomrule
	\end{tabular}
\vskip 0.1cm
\caption{Worked example illustrating the OBIS preference elicitation algorithm. Each row represents the linear coefficients (corresponding to the base values) that uniquely define the corresponding function, the current utility maximizing schedule $a^*$, its corresponding utility $u(a^*)$ and the answer to the CQ.}
\label{tab:toy_example_2}
\end{sc}
\end{table}

First, given $u^{GUI}$, MLCM constructs a cardinal training data set $D_{card}$.\footnote{For this example, $D_{card}$ consists of all schedules of size $1$ and $2$ and their respective values, as implied by the student's reports. For more details on the creation of the cardinal dataset see electronic companion~\ref{A_bootstrap}.}
Next, MLCM fits the linear regression model $\mathcal{M}^{0}$ on $D_{card}$.\footnote{We train the linear regression models using gradient descent as this allows us to train them on both cardinal and ordinal data (see Section~\ref{subsec:Integrating comparison queries into MVNNs}).} In Table~\ref{tab:toy_example_2}, we see that $\mathcal{M}^{0}$ perfectly fits $u^{GUI}$. 
Next, MLCM generates the first CQ consisting of the two admissible schedules within budget that have the highest and second highest predicted utility with respect to $\mathcal{M}^{0}$: schedule $\{2,4\}$ with $\mathcal{M}^{0}(\{2,4\})=122$, and schedule $\{1,4\}$ with $\mathcal{M}^{0}(\{1,4\})=120$. 
The student answers that she prefers $\{1,4\}$ with a true utility of $u(\{1,4\})=125$ over $\{2,4\}$ with true utility $u(\{2,4\})=110$. \
Next, OBIS creates the first sorted list of schedules for the student 
$S^{1} =  \left [ \{1,4\} \succ \{2,4\} \right ]$ (Line~\ref{alg_cq_gen:sort_list} of \Cref{alg_query_generation}).
Based on $S^{1}$, OBIS updates the ordinal training data set as $D_{\text{ord}}^{1}=\{\{1,4\} \succ \{2,4\}\}$ (Line~\ref{alg_cq_gen:get_all_orderings} of \Cref{alg_query_generation}) and the ML model $\mathcal{M}^{1}$ is trained
(Line~\ref{alg_cq_gen:train_model} of \Cref{alg_query_generation}). 
We see that the regression coefficient for course 1 increased from 75 to 80 and the coefficient for course 2 decreased from 77 to 72. 
Now the model $\mathcal{M}^{1}$ correctly predicts the ordinal ranking of the schedules involved in the first CQ (i.e., $\mathcal{M}^{1}(\{1,4\})=125>117=\mathcal{M}^{1}(\{2,4\})$). Thus, MLCM has alleviated the student's initial reporting mistake on her two most preferred courses, which led to an increase of her utility for her (predicted) utility-maximizing schedule $u(a_1^*)$ from $110$ to $125$. This completes the first iteration of the binary insertion sort of OBIS, here only leading to one CQ.

Next, OBIS uses the ML model $\mathcal{M}^{1}$ to identify the not already-queried schedule with the highest predicted utility within the student's budget, namely $\{ 1, 3 \}$ (Line~\ref{alg_cq_gen:get_best_bundle} of \Cref{alg_query_generation}). 
Then, OBIS checks whether  $\{1, 3 \}$ can already be sorted into $S^{1}$ (Line~\ref{alg_cq_gen:bS_done} of \Cref{alg_query_generation}). Given that OBIS does not yet have enough information to do this, it starts another iteration of binary insertion sort (Line~\ref{alg_cq_gen:call_bsquery} of \Cref{alg_query_generation}). This  results in the CQ consisting of  $\{ 1, 3\}$ and $\{ 1, 4\}$.\footnote{When performing binary insertion sort, we break ties in favor of the schedule with the higher predicted utility.} The student answers that she prefers the schedule $\{1, 3\}$. Thus, the list of schedules 
$\{1,3\}, \{1, 4\}$ and $\{2, 4\}$ can now be fully sorted, completing the second iteration of binary insertion sort.

After training the ML model $\mathcal{M}^{2}$ based on the updated ordinal training data set $D_{\text{ord}}^{2}=\{\{1,4\} \succ \{2,4\}, \{1,3\} \succ \{1,4\}, \{1,3\} \succ \{2,4\}\}$, the model now predicts the correct ordinal ranking of all admissible schedules. Thus, MLCM has already identified the true utility-maximizing schedule $\{1,3\}$. The next schedule identified by OBIS is $\{2,3\}$, and another iteration of binary insertion sort is started. Since the student's answers to all further CQs do not contradict the model's predicted ordinal ranking, the linear coefficients of $\mathcal{M}^{3}$ and $\mathcal{M}^{4}$ remain unchanged.
\end{example}

\subsection{Theoretical Properties of MLCM} \label{sec_theoretical_guarantees}
\citet{budish2011aceei} showed that A-CEEI satisfies \textit{envy-bounded by a single good}, \textit{$(n+1)$-maximin share guarantee}, and \textit{Pareto efficiency}. If CM had access to the full true ordinal preferences, then the Stage~1 allocation of CM would also satisfy the same properties. In electronic companion~\ref{A_guarantees}, we prove that the same properties also hold in an \emph{approximate} sense for the Stage 1 allocation of MLCM if the students' preferences are captured \textit{approximately} via the ML models $\mathcal{M}_i$.

Regarding incentives, \citet{budish2011aceei} showed that A-CEEI is \emph{strategyproof in the large (SP-L)}. \citet{budish2017course} argued that CM is also SP-L. Their argument proceeds in two steps. First, as the number of students increases, $p^*$ calculated at the end of Stage 2 will be exogenous to the students' reports, and hence the students become price-takers. Second, since the market-clearing error target $\alpha \le \sqrt{\sigma m} / 2$ is independent of the number of students, the probability that Stage 3 affects a student's allocation goes to zero as the market grows. In MLCM, the main difference is that we use the trained ML models instead of the GUI reports when calculating prices. However, students are still price-takers, and the probability that Stage 3 affects an individual student still goes to zero. 
Thus, the same arguments as in \citet{budish2017course} apply, and we therefore argue that MLCM is also SP-L.\footnote{Recently, \citet{rubinstein2023practical} uncovered a possible manipulation of the original CM mechanism. To alleviate this, they added a new constraint to the price calculation algorithm called \textit{Contested Envy-Free-but-for-Tie-Breaking}. We added the same constraint to MLCM to maintain the same incentive property as CM.}

\section{Machine Learning Instantiation} \label{sec:MachineLearningInstantiation}
In this section, we instantiate the ML model architecture used by MLCM.

\subsection{Desiderata}
\label{subsec:Machine Learning Model Desiderata}
There are four important desiderata for the choice of the ML model architecture. First, it must be able to learn the students' preferences with a small number of queries. Second, it must be expressive enough to capture possibly complex preferences (including substitutes and complements). Third, we must be able to train the ML model using both the \textit{cardinal} input that students provide via the CM reporting language as well as the \textit{ordinal} input from answering CQs. Fourth, in Phases 3--5 of MLCM, we must be able to quickly compute each student's utility-maximizing schedule, given the ML model predictions, budget, and prices. 
Concretely, we must be able to quickly solve the following optimization problem:
\begin{align}\label{eq:util_max_problem}
x^* \in \argmax_{x \in \Psi_i, x \cdot p \le b_i} \mathcal{M}_i(x)
\end{align}
Thus, a key requirement on the model architecture is that \Cref{eq:util_max_problem} can be solved fast enough. For this, we adopt as a requirement that \Cref{eq:util_max_problem} can be translated into a succinct MIP.

\subsection{Monotone-Value Neural Networks (MVNNs)}

\textit{Monotone-value neural networks (MVNNs)} \citep{weissteiner2022monotone} are a recently introduced network architecture designed to capture monotone combinatorial valuations. Specifically, MVNNs incorporate at an architectural level the prior information that the value of the empty bundle is zero and that the value function of an agent is \textit{monotone} (which is the case when the utility function satisfies \emph{free disposal}). The same properties are true in our domain: a student's utility for a schedule with no courses is zero, and the utility function satisfies free disposal (as a few days before classes begin, there is a drop/add period during which students can drop courses from their schedules \citep{budish2017course}). 
Because MVNNs incorporate this prior knowledge, they outperform standard neural networks in low-sample regimes \citep{weissteiner2022monotone,weissteiner2023bayesian,soumalias2023machine}. This makes them particularly well suited for our domain, where students can only be expected to answer a few queries. Thus, MVVNs satisfy the first desideratum described above.

Furthermore, \citet{weissteiner2022monotone} proved that MVNNs are universal in the class of monotone value functions. For our domain, this means that MVNNs can, in principle, capture any monotone utility function a student may have, satisfying the second desideratum. In Section~\ref{subsec:Integrating comparison queries into MVNNs}, we show how MVNNs can be trained using both cardinal and ordinal data, thus satisfying the third desideratum. Finally, \citet{weissteiner2022monotone} provided a succinct MIP formulation for MVNNs, which can be readily adopted to solve Equation~\ref{eq:util_max_problem}, thus satisfying the fourth desideratum. Because MVNNs fulfill all four desiderata, we choose MVNNs as the model architecture for MLCM.

\subsection{Training MVNNs using cardinal and ordinal data}
\label{subsec:Integrating comparison queries into MVNNs}
Our  method for training MVNNs using both GUI reports (i.e., regression/cardinal data) and CQs (i.e., ranking/ordinal data) is inspired by the \textit{Combined Ranking and Regression (CRR)} method by \citet{sculley2010combined}. CRR utilizes both regression and ranking data to train regression models. For regression data, CRR employs a standard regression loss.  For ranking data, CRR first calculates the probability that an agent prefers one alternative $x$ over another $x'$ as the sigmoid of the difference of their predicted values, i.e., $\mathbb{P}[x \succ x'] = \sigma (\mathcal{M}(x) - \mathcal{M}(x'))$.\footnote{Note that this is equivalent to the Bradley-Terry model \citep{bradleyterry1952}, which has found extensive application in aligning large language models with human preferences (see, e.g., \cite{azar2023}).} This probability is then interpreted as the model's prediction for x $\succ x'$, so that a standard binary classification loss can be used, with the true ordinal ranking between $x$ and $x'$ being the true label. Then, CRR employs a gradient-based method that randomly chooses one of the two losses in each gradient step.

Our approach differs from CRR in that we do not randomly switch between regression and ranking data during training; instead, we adopt a two-step process. First, we train our model exclusively on the GUI reports (i.e., regression data), applying high regularization to prevent overfitting to the mistakes in those reports. Second, we fine-tune the model using the answered CQs (i.e., ranking data), employing much lower regularization to allow the model the flexibility to better fit the students' answers to the CQs. This approach is in line with Bayesian theory, given that the answers to the CQs are less noisy due to their lower cognitive load \citep{conitzer2009eliciting, chajewska00}. See electronic companion~\ref{A_scully} for more details on our training procedure.

\section{Course Allocation Simulation Framework} \label{sec_SPG}
In this section, we describe our student preference generator (\Cref{sec:StudentPreferenceGenerator}), how we model students' reporting mistakes (\Cref{subsec:noise}), and how we calibrate the parameters of our simulation to real-world data (\Cref{subsection:calibrating}). 
For additional details, please see electronic companion~\ref{A_SPG}--\ref{A_mistake_calibration}.

\subsection{Student Preference Generator}
\label{sec:StudentPreferenceGenerator}
We have two goals for our student preference generator. First, students' preferences (and their reports) should be realistic, i.e., they should closely match real-world data on the usage of CM. Second, we must be able to encode each student's complete preferences as a MIP so that we can compute an (optimal) benchmark allocation given the students’ true preferences.

\paragraph{Correlation.} One of the key features the simulator must capture is some notion of \emph{correlation} between students' preferences. To this end, we divide courses into \textit{popular} and \textit{unpopular}. Popular courses are those that many (but not necessarily all) students have a high value for. Concretely, for every student $i$, we randomly select a set of \emph{favorite} courses from the set of popular courses. Then, for each course, student $i$'s base value is drawn from some distribution, where the mean of that distribution is high for her favorite courses and low for all others. Note that a smaller number of popular courses implies higher correlation, as more students will have the same favorite courses. Thus, we can use the number of popular courses to control the degree of correlation.

\paragraph{Complementarities and Substitutabilities.}
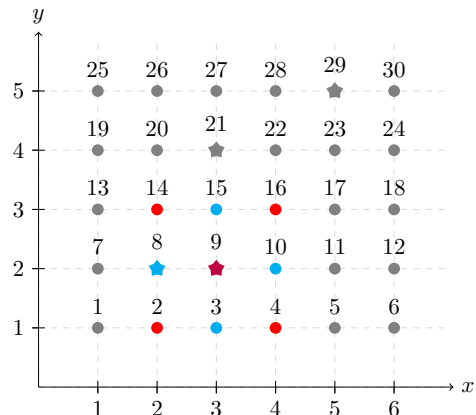
\begin{wrapfigure}[10]{r}[0pt]{0.5\textwidth}
    \centering
    \vskip -0.05cm
    \resizebox{!}{0.34\textwidth}{
    \begin{tikzpicture}
\draw[help lines, color=gray!30, dashed] (0,0) grid (6.9,5.9);

\draw[->] (0,0)--(7,0) node[right]{$x$};
\draw[->] (0,0)--(0,6) node[above]{$y$};

\foreach \x in  {1,2,3,4,5,6} 
\draw[shift={(\x,0)},color=black] (0pt,3pt) -- (0pt,-3pt);

\foreach \x in {1,2,3,4,5,6} 
\draw[shift={(\x,0)},color=black] (0pt,0pt) -- (0pt,-3pt) node[below] 
{$\x$};

\foreach \x in  {1,2,3,4,5} 
\draw[shift={(0,\x)},color=black] (-3pt,0pt) -- (3pt,0pt);
\foreach \x in {1,2,3,4,5} 
\draw[shift={(0,\x)},color=black] (0pt,0pt) -- (-3pt,0pt) node[left] 
{$\x$};

\node[circle,inner sep=2pt,fill=gray,label=:{$1$}] at (1,1) {};
\node[circle,inner sep=2pt,fill=red,label=:{$2$}] at (2,1) {};
\node[circle,inner sep=2pt,fill=cyan,label=:{$3$}] at (3,1) {};
\node[circle,inner sep=2pt,fill=red,label=:{$4$}] at (4,1) {};
\node[circle,inner sep=2pt,fill=gray,label=:{$5$}] at (5,1) {};
\node[circle,inner sep=2pt,fill=gray,label=:{$6$}] at (6,1) {};

\node[circle,inner sep=2pt,fill=gray,label=:{$7$}] at (1,2) {};
\node[star,star points=5, ultra thick,inner sep=2pt,fill=cyan,label=:{$8$}] at (2,2) {};
\node[star,star points=5, ultra thick, inner sep=2pt,fill=purple,label=:{${9}$}] at (3,2) {};
\node[circle,inner sep=2pt,fill=cyan,label=:{${10}$}] at (4,2) {};
\node[circle,inner sep=2pt,fill=gray,label=:{${11}$}] at (5,2) {};
\node[circle,inner sep=2pt,fill=gray,label=:{${12}$}] at (6,2) {};

\node[circle,inner sep=2pt,fill=gray,label=:{${13}$}] at (1,3) {};
\node[circle,inner sep=2pt,fill=red,label=:{${14}$}] at (2,3) {};
\node[circle,inner sep=2pt,fill=cyan,label=:{${15}$}] at (3,3) {};
\node[circle,inner sep=2pt,fill=red,label=:{${16}$}] at (4,3) {};
\node[circle,inner sep=2pt,fill=gray,label=:{${17}$}] at (5,3) {};
\node[circle,inner sep=2pt,fill=gray,label=:{${18}$}] at (6,3) {};

\node[circle,inner sep=2pt,fill=gray,label=:{${19}$}] at (1,4) {};
\node[circle,inner sep=2pt,fill=gray,label=:{${20}$}] at (2,4) {};
\node[star,star points=5, ultra thick,inner sep=2pt,fill=gray,label=:{${21}$}] at (3,4) {};
\node[circle,inner sep=2pt,fill=gray,label=:{${22}$}] at (4,4) {};
\node[circle,inner sep=2pt,fill=gray,label=:{${23}$}] at (5,4) {};
\node[circle,inner sep=2pt,fill=gray,label=:{${24}$}] at (6,4) {};

\node[circle,inner sep=2pt,fill=gray,label=:{${25}$}] at (1,5) {};
\node[circle,inner sep=2pt,fill=gray,label=:{${26}$}] at (2,5) {};
\node[circle,inner sep=2pt,fill=gray,label=:{${27}$}] at (3,5) {};
\node[circle,inner sep=2pt,fill=gray,label=:{${28}$}] at (4,5) {};
\node[star,star points=5, ultra thick, inner sep=2pt,fill=gray,label=:{${29}$}] at (5,5) {};
\node[circle,inner sep=2pt,fill=gray,label=:{${30}$}] at (6,5) {};

\end{tikzpicture}
    }
    \vskip -0.15 in
    \caption{A latent space with 30 courses.}
    \label{fig_latent_space}
\end{wrapfigure}

To model complementarities and substitutabilities between courses, we build on prior experimental work modeling bidders in combinatorial auctions \citep{lsvm, gsvm}.
Concretely, following \citet{lsvm}, we assume that courses lie on a latent space and that the distance between courses in that space defines the students' view on the complementarity/substitutability of those courses. \Cref{fig_latent_space} depicts an example latent space with $\courseset{} = \{1, \dots, {30}\}$. The popular courses are  $\{8, {9}, {21}, {29}\}$ and are marked with a star. Assume that student $i$'s favorite courses are $9$ and $21$. From those, we randomly draw a set of
\emph{centers} -- which is course $\color{purple} 9$ in \Cref{fig_latent_space}. All courses with $L_1$ distance smaller or equal to $1$ from that center form the set of substitutes $\color{cyan} \{9, 3, 8, {10}, {15}\}$. All courses with $L_{\infty}$ distance smaller or equal to $1$ from that center (that are not in the set of substitutes) and the center $\color{purple} 9$ form the set of complements $\color{red} \{9, 2, 4, {14}, {16}\}$. The number of centers and the distances control the 
\textit{degree} of complementarity and substitutability.
\smallskip

\begin{remark}
For the simulator, only the \textit{induced} preferences matter. Of course, one could use other topologies that result in the same or very similar induced preferences. 
However, using our proposed common latent space enables a concise, mathematically tractable formalization of substitutes and complements. With this, our framework provides us with granular, interpretable control over 
the degree of complementarities/substitutabilities. In \Cref{subsection:calibrating}, we show that this design allows us to produce instances very similar to those observed in \citet{budish2022can}.
\end{remark}
\smallskip

\paragraph{Total utility calculation.} We calculate a student's utility for a schedule based on her values for single courses and the number of courses from each set of complements and substitutes in the schedule. In electronic companion~\ref{A_SPG}, we provide mathematical details for the preference generator. In electronic companion~\ref{A_MIP}, we provide the corresponding MIP formulation.

\subsection{Students' Reporting Mistakes}
\label{subsec:noise}
Students can make mistakes when reporting their base values for courses or when reporting pairwise adjustments. To capture these reporting mistakes, we use the following four parameters: 
\begin{enumerate}[leftmargin=*,topsep=0pt,partopsep=0pt, parsep=0pt]
\item $f_{b} \in [0,1]$ denotes the probability that a student forgets one of her \textit{base values}. We assume that a student forgets to report a base value for her lower-valued courses first.
\item $f_{a} \in [0,1]$ denotes the probability a student forgets to report one of her adjustments (for adjustments for which she has not forgotten to report a base value for either of the involved courses).
%
\item $\sigma_b \in \mathbb{R}_{\geq 0}$ denotes the standard deviation of the additive Gaussian noise $\mathcal{N}(0, \sigma_b^2)$ with which a student reports her \emph{base values}. Thus, for a course $c$ for which a student does not forget to report her base value, her base value report is calculated as $\widehat{v} (c) = v(c) + \epsilon$, where $v(c)$ is the student's true base value for $c$, and where $\epsilon \sim \mathcal{N}(0, \sigma_b^2) $. 
\item $\sigma_a \in \mathbb{R}_{\geq 0}$ controls the support of the multiplicative uniform noise $\mathcal{U}[-\sigma_a, + \sigma_a]$ with which a student reports her \textit{adjustments values}. Thus, for a pair of courses $c,c'$ for which a student does not forget to report an adjustment value, her adjustment value report is calculated as $\widehat{\alpha}(c, c') = \alpha (c, c') \cdot (1 + \epsilon)$, where $\alpha(c,c')$ is the student's true adjustment value and $\epsilon \sim \mathcal{U}[- \sigma_a, \sigma_a]$.
\end{enumerate}

\subsection{Calibration of the Simulation Framework} \label{subsection:calibrating}
We calibrate the parameters of our framework (i.e., the preference generator and the reporting mistakes) to match the experimental results from \citet{budish2022can}. The key metrics are:
\begin{enumerate}[leftmargin=*,topsep=0pt,partopsep=0pt, parsep=0pt]
    \item Students only report a base value for $49.9$\% of the courses.\label{finding_1}
    \item The number of courses with a reported value in $[50,100]$ is approximately equal to the number of courses with a reported value in $[0,50]$.\label{finding_2}
    \item Students report between 0 and 10 pairwise adjustments.\label{finding_3}
    \item Students report an average of $1.08$ pairwise adjustments; the median is equal to $0$.  \label{finding_4}
    \item Students were asked to compare the schedule they received with several other course schedules. Their answers were consistent with their reported preferences in $84.41$\% of the cases, and in case of disagreements, the median utility difference, based on their reports, was $13.35$\%.\label{finding_5}
\end{enumerate}
\begin{table}[b!]
	\robustify\bfseries
	\centering
	\begin{sc}
	\resizebox{1\textwidth}{!}{
	\setlength\tabcolsep{4pt}

        \begin{tabular}{lrrrrrrrrr}
        \toprule
         \multicolumn{1}{c}{\textbf{Setting}}&  \multicolumn{3}{c}{\textbf{\#Courses with value}} & \multicolumn{4}{c}{\textbf{\#Adjustments}} & \multicolumn{1}{c}{\textbf{Accuracy}} & \multicolumn{1}{c}{\textbf{If disagreement}}\\
        \cmidrule(l{2pt}r{2pt}){2-4}
        \cmidrule(l{2pt}r{2pt}){5-8} 
         &                          \multicolumn{1}{c}{$>0$} &                           \multicolumn{1}{c}{$(0,50) $} &              \multicolumn{1}{c}{$[50,100]$} &                      \multicolumn{1}{c}{Mean} & \multicolumn{1}{c}{Median} &    \multicolumn{1}{c}{Min} & \multicolumn{1}{c}{Max} & \multicolumn{1}{c}{\textbf{in \%}}  &                    \multicolumn{1}{c}{\textbf{utility difference in \%}} \\
        \cmidrule(l{2pt}r{2pt}){1-1}
        \cmidrule(l{2pt}r{2pt}){2-4}
        \cmidrule(l{2pt}r{2pt}){5-8}
        \cmidrule(l{2pt}r{2pt}){9-9}
        \cmidrule(l{2pt}r{2pt}){10-10}
              $9$ Popular Courses &  12.49 $\pm$ \scriptsize 0.02 &   6.32 $\pm$ \scriptsize 0.09 &  6.17 $\pm$ \scriptsize 0.09 &   0.98 $\pm$ \scriptsize 0.05 &         1 &   0 &   10 &         81 $\pm$ \scriptsize 1 &         -15.52 $\pm$ \scriptsize 0.76 \\    
        \midrule
             $6$ Popular Courses &  12.49 $\pm$ \scriptsize 0.02 &   6.33 $\pm$ \scriptsize 0.08 &  6.16 $\pm$ \scriptsize 0.08 &  1.08 $\pm$ \scriptsize 0.06 &    1 &      0 &     10 &  83 $\pm$ \scriptsize 1 &   -13.34 $\pm$ \scriptsize 0.71 \\  
             \midrule
             {\cite{budish2022can}}  &   12.45\hphantom{ $\pm$ \scriptsize 0.02}  &  6.17\hphantom{ $\pm$ \scriptsize 0.02}  &  6.27\hphantom{ $\pm$ \scriptsize 0.02} & 1.08\hphantom{ $\pm$ \scriptsize 0.02}  & 0 &  0 &         10 &         84\hphantom{ $\pm$ \scriptsize 1}  &          -13.35 $\pm$ \scriptsize 0.41 \\
        \bottomrule
        \end{tabular}
        }
        \vskip 0.1cm
\caption{Results of calibrating our simulator to the experimental findings of \cite{budish2022can} for settings with 6 and 9 popular courses. 2,000 students in total. Shown are average results and 95\% CIs.}
\label{tab:MistakeCalibration_shortened}
\vskip -0.5cm
\end{sc}
\end{table}
\Cref{tab:MistakeCalibration_shortened} shows the results of our calibration procedure. For the setting with 6 popular courses, we set $(f_{b},f_{a},\sigma_b,\sigma_a)=(0.5, 0.4825, 17, 0.2)$, and for the setting with 9 popular courses, we set $(f_{b},f_{a},\sigma_b,\sigma_a)=(0.5, 0.48, 23, 0.2)$. With these parameters, we match metrics (\ref{finding_1})-(\ref{finding_3}) within at most 3\%. For metric (\ref{finding_4}), our mean is $10$\% lower (in one of two settings), while our median is slightly higher (1 instead of 0). Regarding metric (\ref{finding_5}), both accuracy and scaled median utility difference are within $3$ percentage points of the reported one. Thus, our framework produces instances very similar to those described in \cite{budish2022can}.

\section{Experimental Evaluation} \label{Section_efficiency_results}

In this section, we experimentally evaluate the performance of MLCM using our simulation framework from \Cref{sec_SPG}. We provide the source code for all experiments in the supplementary material.

\subsection{Experiment Setup}  \label{subsec:experiment_setup}
We consider a setting with 25 courses and 100 students.\footnote{We selected 25 courses to match the number of courses in the lab experiments of \citet{budish2022can}, allowing us to properly calibrate the mistake profile of the students (see \Cref{subsection:calibrating}). However, since this calibration does not depend on the number of students, we increased the number of students from about $20$, as reported in \citet{budish2022can}, to $100$. Consequently, the course capacity increased from approximately 5 to between 22 and 30, depending on the supply ratio, which means that the reports from any single  student have a smaller impact on the (over-)demand for individual courses. This makes this setting more realistic and reduces variance during experimentation, while still keeping the runtime for our experiments manageable. 
} We consider settings with 6 and 9 popular courses (where 6 popular courses corresponds to extreme correlation). As in the lab experiment of \citet{budish2022can}, every student wants a schedule of (at most) five courses. Thus, for $100$ students, the total demand is $500$ seats. The \emph{supply ratio} (SR) denotes the fraction of the total number of seats to the total demand. Thus, an SR of $1.5$ means that there are a total of $750$ course seats (split equally among all courses). We consider SRs of $1.1$, $1.25$ and $1.5$.

To compare CM and MLCM, we assume that both receive the same reports from the CM reporting language, with the students' reporting mistakes calibrated as described in \Cref{subsection:calibrating}. For MLCM, each student additionally answers $1$, $5$, $10$, $15$, or $20$ CQs generated by the OBIS algorithm. 
We denote these mechanisms as \textsc{CM} and \textsc{MLCM (1/5/10/15/20 OBIS CQs)}, respectively. We assume that students make no mistakes when answering CQs.\footnote{Recall that pairwise CQs are known to have low cognitive load \citep{conitzer2009eliciting,chajewska00}. Furthermore, \citet{budish2022can} used the students' answers to pairwise CQs as \textit{ground truth} for the same problem.
Nevertheless, in electronic companion~\ref{subsec:app:CQ_robustness}, we show that the performance gains of MLCM remain high, even when the students answer a large fraction of their CQs incorrectly.}

We further consider five benchmarks. The first is \mbox{\textsc{CM\textsuperscript{*} (Full Preferences)}}, which is a modified version of CM that takes as input the students' \emph{full}, \textit{true} cardinal preferences. Importantly, \textsc{CM\textsuperscript{*}} includes interactions between three or more courses, which cannot be expressed using CM's reporting language. The second benchmark is \mbox{\textsc{CM (No Mistakes)}}, which is standard CM, but assuming no reporting mistakes. 
The next two benchmarks are  \mbox{\textsc{MLCM ($10$/$20$ RANDOM CQs)}} and \mbox{\textsc{MLCM ($10$/$20$ na\"{i}ve CQs)}}, where each student answers $10$/$20$ CQs generated by the random and na\"{i}ve algorithms (see \Cref{subsec:query_generation}), respectively. 
Finally, we also use \textit{random serial dictatorship (RSD)} as another benchmark, with the same GUI reports as input as CM.

We use the same $500$ instances to test each mechanism. 
For both CM and MLCM, we use the algorithm introduced in \citet{rubinstein2023practical} to calculate an A-CEEI. 
For details on the hyperparameter optimization procedure, the selected hyperparameters, and the computing infrastructure, see  electronic companion~\ref{sec:app_Experiment Details}.
Unless otherwise noted, CQs denote OBIS-generated CQs.

\begin{table}[t!] 
    \robustify\bfseries
    \centering
    \begin{sc}
    {
    \setlength\tabcolsep{4pt}
    \begin{tabular}{llrrr}
    \toprule
    & & \multicolumn{2}{c}{\textbf{Student Utility (in \%)}}  & \multicolumn{1}{c}{\textbf{Time}}\\
    \cmidrule(l{2pt}r{2pt}){3-4}
    \textbf{Mechanism} & \textbf{Parametrization} & \multicolumn{1}{c}{\textbf{Average}} & \multicolumn{1}{c}{\textbf{Minimum}}  & \multicolumn{1}{c}{\textbf{(in min)}} \\
    \cmidrule(l{2pt}r{2pt}){1-2}
    \cmidrule(l{2pt}r{2pt}){3-4}
    \cmidrule(l{2pt}r{2pt}){5-5}
    CM\textsuperscript{*} & Full Preferences & 100.0 $\pm$ \scriptsize 0.0 & 73.1 $\pm$ \scriptsize 0.4 & 68.1 \\
    CM & No Mistakes & 98.8 $\pm$ \scriptsize 0.1 & 72.7 $\pm$ \scriptsize 0.4 & 52.6 \\
    \cmidrule(l{2pt}r{2pt}){1-2}
    \cmidrule(l{2pt}r{2pt}){3-4}
    \cmidrule(l{2pt}r{2pt}){5-5}
    RSD & - & 76.5 $\pm$ \scriptsize 0.2 & 30.3 $\pm$ \scriptsize 0.6 & 0.0 \\
    \ccell CM & \ccell - &  \ccell 79.3 $\pm$ \scriptsize 0.2 & \ccell 41.5 $\pm$ \scriptsize 0.5 & \ccell 24.1 \\
    \cmidrule(l{2pt}r{2pt}){1-2}
    \cmidrule(l{2pt}r{2pt}){3-4}
    \cmidrule(l{2pt}r{2pt}){5-5}
    MLCM & \hphantom{0}1 OBIS CQ & 80.1 $\pm$ \scriptsize 0.2 & 42.6 $\pm$ \scriptsize 0.5 & 8.0 \\
    & \hphantom{0}5 OBIS CQs & 85.2 $\pm$ \scriptsize 0.2 & 49.0 $\pm$ \scriptsize 0.5 & 7.1 \\
    & 10 OBIS CQs & 87.6 $\pm$ \scriptsize 0.2 & 52.2 $\pm$ \scriptsize 0.5 & 7.0 \\
    & 15 OBIS CQs & 89.2 $\pm$ \scriptsize 0.2 & 53.9 $\pm$ \scriptsize 0.5 & 8.4 \\
    & 20 OBIS CQs & 90.2 $\pm$ \scriptsize 0.2 & 55.1 $\pm$ \scriptsize 0.5 & 8.3 \\
    \cmidrule(l{2pt}r{2pt}){2-2}
    \cmidrule(l{2pt}r{2pt}){3-4}
    \cmidrule(l{2pt}r{2pt}){5-5}
    & 10 random CQs & 79.9 $\pm$ \scriptsize 0.2 & 42.2 $\pm$ \scriptsize 0.5 & 8.4 \\
    & 20 random CQs & 80.9 $\pm$ \scriptsize 0.2 & 44.1 $\pm$ \scriptsize 0.5 & 8.3 \\
    \cmidrule(l{2pt}r{2pt}){2-2}
    \cmidrule(l{2pt}r{2pt}){3-4}
    \cmidrule(l{2pt}r{2pt}){5-5}
    & 10 na\"{i}ve CQs & 86.7 $\pm$ \scriptsize 0.2 & 46.3 $\pm$ \scriptsize 0.6 & 7.0 \\
    & 20 na\"{i}ve CQs & 88.8 $\pm$ \scriptsize 0.2 & 47.7 $\pm$ \scriptsize 0.6 & 7.9 \\
    \cmidrule(l{2pt}r{2pt}){1-2}
    \cmidrule(l{2pt}r{2pt}){3-4}
    \cmidrule(l{2pt}r{2pt}){5-5}
    MLCM-Projected & 10 OBIS CQs & 87.6 $\pm$ \scriptsize 0.2 & 51.8 $\pm$ \scriptsize 0.5 & 79.6 \\
    & 20 OBIS CQs & 90.2 $\pm$ \scriptsize 0.2 & 55.2 $\pm$ \scriptsize 0.5 & 69.3 \\
    \bottomrule
    \end{tabular}
    }
    \end{sc}
    \vskip 0.1cm
    \caption{Comparison of different mechanisms for a supply ratio of 1.25, 9 popular courses, and default parameterization for reporting mistakes.  Standard CM is highlighted in grey. Shown are averages over 500 runs. We normalize average and minimum utility by the average utility of \textsc{CM\textsuperscript{*}} and also show their 95\% CIs.}
    \label{tab:welfare_results_sr_1.25_pop_9}
\end{table}

\subsection{Welfare Results}  \label{subsec:experimental_results}
In Table \ref{tab:welfare_results_sr_1.25_pop_9}, we present results for an SR of $1.25$ (which is very close to Wharton's SR; see \cite{budish2022can}) and $9$ popular courses (corresponding to medium correlation). 
In electronic companion~\ref{A_welfare_results}, we also provide results for SRs of $1.1$ and $1.5$ and for $6$ popular courses; the results are qualitatively the same for all SRs and slightly worse for $6$ popular courses. 
We report average and minimum utility of the final allocation; we normalize the utility results by the average utility of \textsc{CM\textsuperscript{*} (Full Preferences)} so that all utility metrics can be reported in percent. 
Additionally, we report the runtime of each mechanism for computing an allocation.\footnote{In contrast to CM, by default, MLCM requires two A-CEEI computations (one in Phase 3 and one in Phase 5). Thus, MLCM's total runtime is twice that reported in Table \ref{tab:welfare_results_sr_1.25_pop_9}. However, the A-CEEI computation in Phase 3 could be avoided if last year's prices would be used instead (see \Cref{footonote_apx_prices} for a discussion on this alternative approach).}


In Table \ref{tab:welfare_results_sr_1.25_pop_9}, we see that, even with a single OBIS-generated CQ, MLCM already significantly outperforms CM in terms of both average and minimum student utility, and its performance improves further as the number of CQs increases.\footnote{See electronic companion~\ref{A_significance_tests} for statistical tests for all such statements.} 
For example, with $10$ CQs, compared to CM, \textsc{MLCM} increases average utility from $79.3\%$ to $87.6\%$, an $8.3$ percentage points (pp) increase (which is a $10.5$\% relative increase) and minimum utility from $41.5\%$ to $52.2\%$ (a $25.8$\% relative increase). 
With $20$ CQs, \textsc{MLCM} increases average utility from $79.3\%$ to $90.2\%$ (a $13.7$\% relative increase) and minimum utility from $41.5\%$ to $55.1\%$ (a $32.8$\% relative increase).

Next, we compare MLCM using OBIS against the two straw men algorithms. We see that asking $10$ \emph{random} CQs only improves average student utility by $0.6$ pp and minimum student utility by $0.7$ pp, in contrast to $8.3$ pp and $10.7$ pp for MLCM using OBIS. Thus, asking random CQs is ineffective. Next, we see that asking $10$ \emph{na\"ive} CQs improves average student utility by $7.4$ pp and minimum student utility by $4.8$ pp. Thus, the na\"ive algorithm performs reasonably well on average student utility but a lot worse (compared to OBIS) on minimum student utility. These results underscore the importance of asking ``smart'' CQs and align with our theoretical results: in \Cref{subsec:query_generation}, we showed that OBIS has the best worst-case guarantee regarding the amount of information it can infer from the generated CQs, which explains why it performs so much better than the na\"ive algorithm on minimum student utility. In \Cref{subsec:af_empirical_evaluation}, we experimentally investigate what is driving the advantage of the OBIS algorithm compared to the random and na\"{i}ve algorithms.

Next, we turn to RSD, another benchmark mechanism. Compared to RSD, CM increases average student utility by $2.8$ pp. By comparison, MLCM with 10 CQs increases average student utility over CM by a much larger amount, namely $8.3$ pp (i.e., an $11.1$ pp increase over RSD). 
Second, we observe that CM increases minimum student utility by $10.8$ pp over RSD. By comparison, MLCM increases minimum student utility by a similar amount over CM, namely $10.7$ pp (i.e., a $21.5$ pp increase over RSD).
Overall, these results show that MLCM not only improves upon CM but does so to an even greater extent than the improvement CM offers over RSD.


Finally, we observe that the performance difference between \mbox{\textsc{CM\textsuperscript{*} (Full Preferences)}} and \mbox{\textsc{CM (No Mistakes)}} is very small. This implies that the hyperparameters of the simulation lead to preferences that can be captured very well by the CM language, even though CM only allows for pairwise adjustments. Thus, MLCM’s advantage over CM primarily stems from its ability to correct students' reporting mistakes rather than its potential to handle more complex preferences.

\subsection{Reporting Mistakes Robustness Study}\label{subsec:Reporting Mistakes Robustness Study}
We now vary how many reporting mistakes students make in the CM language. We keep the setting fixed to a SR of $1.25$ and $9$ popular courses (see electronic companion~\ref{A_mistake_robustness_results} for a $\text{SR}$ of $1.1$ or $1.5$ and 6 popular courses). We multiply all parameters of the students' mistake profile (i.e., $f_{b}$, $f_{a}$, $\sigma_b$ and $\sigma_a$, see \Cref{sec_SPG}) by a common constant $\gamma$. For $\gamma < 1$, students make fewer mistakes than in the default mistake profile; for $\gamma > 1$, the opposite is true.
For each $\gamma$, we run the same $100$ instances for CM and MLCM, with each student answering $10$ CQs in MLCM. 

\begin{figure}[t!]
    \centering
    \includegraphics[width=1.\columnwidth]{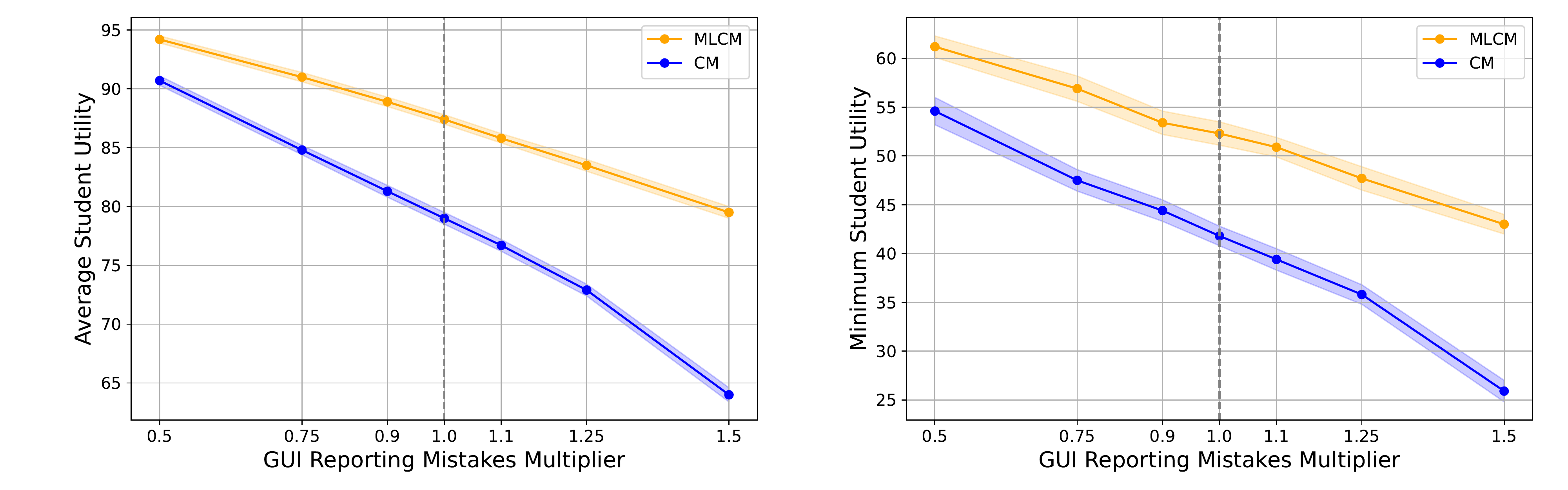}
    \caption{CM language reporting mistakes robustness experiment for a supply ratio of 1.25 and 9 popular courses. Shown are average results in \% over 100 runs including 95\% CI.}
    \label{fig:noise_robustness}
\end{figure}

\Cref{fig:noise_robustness} shows the results of this robustness study. 
We observe that MLCM significantly outperforms CM for all $\gamma \in [0.5, 1.5]$. Note that a $\gamma$ of $0.5$ corresponds to students being much more precise in reporting their preferences, because $\gamma$ does not linearly affect the students' mistakes.\footnote{ 
Concretely, for $\gamma = 0.5$, students make about $50$\% \textit{fewer} mistakes compared to $\gamma=1$ (as measured by pairwise comparisons) but additionally, the \textit{severity} of these mistakes (i.e., utility difference in case of disagreements) is only $20$\% as large (see electronic companion~\ref{A_mistake_calibration}).
Thus, the mean absolute error (MAE) of the students' reported versus true utilities over the schedule space (in case of disagreements) is only approximately $9.1\%$ of the MAE for $\gamma = 1$.}
As $\gamma$ increases, the performance of both CM and MLCM monotonically decreases 
and the relative performance gap between the two mechanisms gets significantly larger. For example, for $\gamma=1.5$, the relative performance increase of MLCM for average and minimum student utility is $24.2$\% and $66.0$\%, respectively. These results were achieved without retuning MLCM's hyperparameters for each $\gamma$, demonstrating MLCM's robustness to changes in the students' reporting mistake profile.

\subsection{Should an Individual Student Opt Into the ML Feature?}\label{subsec:Should Individual Students Opt Into MLCM?}

Suppose MLCM is implemented in practice, and an individual student faces the decision of whether to opt into MLCM’s ML feature. How much would the student benefit if (a) no other student opts into the ML feature, or (b) everyone else also opts in?

\paragraph{No Other Student Opts into the ML Feature.}

We first study the scenario in which only the focal student contemplates opting in. We conduct two simulations: one where no students have opted in and another where just one student has.\footnote{To make the experiment computationally feasible, for each setting, we use the price vector that would result if no student would opt in (assuming that the student who is considering to opt in is a price taker).} Our analysis encompasses 100 instances, with 100 students in each instance, totaling 10,000 students. We report averages over those 10,000 students.

\Cref{table_first_student_to_switch} shows the results for an SR of $1.25$.
The focal student prefers the ``MLCM schedule'' in at least $71.9$\% of the cases (across all settings), while she prefers the ``CM schedule'' in at most $17.1$\% of the cases.\footnote{In the lab experiment of \citet{budish2022can}, they compared CM to the previous mechanism used at Wharton (i.e., the BPA). In those experiments, only $42.4$\% of the students preferred CM's allocation, $31.8$\% preferred BPA's allocation, while the remaining students were indifferent.} Furthermore, the \emph{expected relative gain} from opting in is at least $9.6\%$. As the number of CQs increases, the benefit from opting in becomes even larger. 
Finally, the improvement is larger in settings with more popular courses and almost identical for all SRs (see electronic companion~\ref{A_unilateral_switch}). In electronic companion~\ref{A_unilateral_switch}, we perform an analogous  experiment for the scenario where all other students opt into MLCM's ML feature. These results are almost identical.

\begin{table}[t!]
	\robustify\bfseries
	\centering
	\begin{sc}
	\resizebox{0.85\columnwidth}{!}{
	\setlength\tabcolsep{4pt}

\begin{tabular}{ccc r  r  r r r r}
\toprule
\multicolumn{3}{c}{\textbf{Setting}} & \multicolumn{3}{c}{\textbf{Preferred Mechanism}} & \multicolumn{3}{c}{\textbf{Gain from Opting Into MLCM}}  \\
\cmidrule(l{2pt}r{2pt}){1-3}
\cmidrule(l{2pt}r{2pt}){4-6}
\cmidrule(l{2pt}r{2pt}){7-9}
                    \multicolumn{1}{c}{\textbf{SR}} & \multicolumn{1}{c}{\textbf{\#PoP}} & \multicolumn{1}{c}{\textbf{\#CQs}}& \multicolumn{1}{c}{\textbf{MLCM}} & \multicolumn{1}{c}{\textbf{CM}} & \multicolumn{1}{c}{\textbf{Indiff.}} & \multicolumn{1}{c}{\textbf{Expected}} & \multicolumn{1}{c}{\textbf{if pref MLCM}} &  \multicolumn{1}{c}{\textbf{if pref CM}} \\
\cmidrule(l{2pt}r{2pt}){1-3}
\cmidrule(l{2pt}r{2pt}){4-6}
\cmidrule(l{2pt}r{2pt}){7-9}
1.25 & 9 & 10  & 78.5\% & 10.4\% & 11.1\% & 14.5\% & 19.7\% & -8.9\% \\
1.25 & 9 & 15  & 83.0\% & 8.7\%  & 8.3\%  & 17.2\% & 21.7\% & -8.5\% \\
1.25 & 9 & 20  & 85.2\% & 8.0\%  & 6.8\%  & 18.7\% & 22.7\% & -8.1\% \\
\cmidrule(l{2pt}r{2pt}){1-3}
\cmidrule(l{2pt}r{2pt}){4-6}
\cmidrule(l{2pt}r{2pt}){7-9}
1.25 & 6 & 10 &    71.9\% &   17.1\% &     11.0\% &       9.6\% &  15.6\% & -9.2\% \\
1.25 & 6 & 15 &    76.0\% &   15.0\% &      9.0\% &      11.3\% &  16.7\% & -9.2\% \\
1.25 & 6 & 20 &    78.3\% &   14.0\% &      7.7\% &      12.4\% &  17.5\% & -9.3\% \\
\bottomrule
\end{tabular}
}
\end{sc}
\vskip 0.1cm
\caption{Expected relative gain from opting into MLCM's ML feature when no other student opts in. Shown are average results across 10,000 students per setting (\textsc{SR},\textsc{\#PoP},\textsc{\#CQs}). CIs for all metrics are $\approx$0.}
\label{table_first_student_to_switch}
\vskip -0.2 cm
\end{table}

\subsection{Results for Additive Preferences}\label{subsec:Results for Additive Preferences}
In this subsection, we test the robustness of MLCM to changes in the \textit{true} students' preferences. Specifically, we  repeat all experiments described in the previous subsections for the simple case of students having \emph{additive} true preferences.

We use our student preference generator (see \Cref{sec:StudentPreferenceGenerator}) to generate additive student utility functions and calibrate the mistake profile of the students so that both their accuracy and the reported utility difference in case of disagreements match the results of \citet{budish2022can} (see \Cref{tab:app:noise_calibration_additive_preferences} in electronic companion~\ref{sec:app:Results_for_Additive Preferences}). We then repeat the welfare experiment described in \Cref{subsec:experimental_results}, the reporting mistakes robustness experiment described in \Cref{subsec:Reporting Mistakes Robustness Study}, as well as  the experiment investigating whether an individual student should opt into MLCM's ML feature described in \Cref{subsec:Should Individual Students Opt Into MLCM?}. The results of those experiments are qualitatively very similar to those for the original (non-additive) preferences and can be found in electronic companion~\ref{sec:app:Results_for_Additive Preferences}. 
For example, for the most realistic SR of 1.25, MLCM with 10 CQs increases average and minimum student utility over CM by 9.0\% and 24.5\%, respectively, compared to increases of 10.5\% and 25.8\% in the non-additive setting. Overall, these results underscore MLCM's robustness and its applicability to a large range of settings.\footnote{We did not retune the hyperparameters for the additive setting, which further illustrates MLCM's robustness.}

\subsection{What is driving the performance advantage of the OBIS algorithm?}
\label{subsec:af_empirical_evaluation}
In this subsection, we identify the two factors driving the performance advantage of the OBIS algorithm by closely examining the queries it generates compared to those generated by the random and na\"{i}ve algorithms described in \Cref{subsec:query_generation}. 
To this end, we first train an MVNN on the GUI reports of each student and calculate an A-CEEI (i.e., we perform the first three phases of MLCM). 
Then we simulate Phase $4$ of MLCM using all three algorithms to generate new CQs. 
After each CQ, we report the size of the ordinal dataset (i.e., the number of schedule pairs) that we can infer for that student based on her responses thus far, as well as whether the student's response to her last CQ agrees with her MVNN's prediction. 
We use an SR of $1.25$ and $9$ popular courses.


\begin{figure}[t!]
    \vskip -0.45cm
    \centering
    \begin{subfigure}{0.45\textwidth}
        \centering
        \captionsetup{font=small}
        \includegraphics[width=\textwidth]{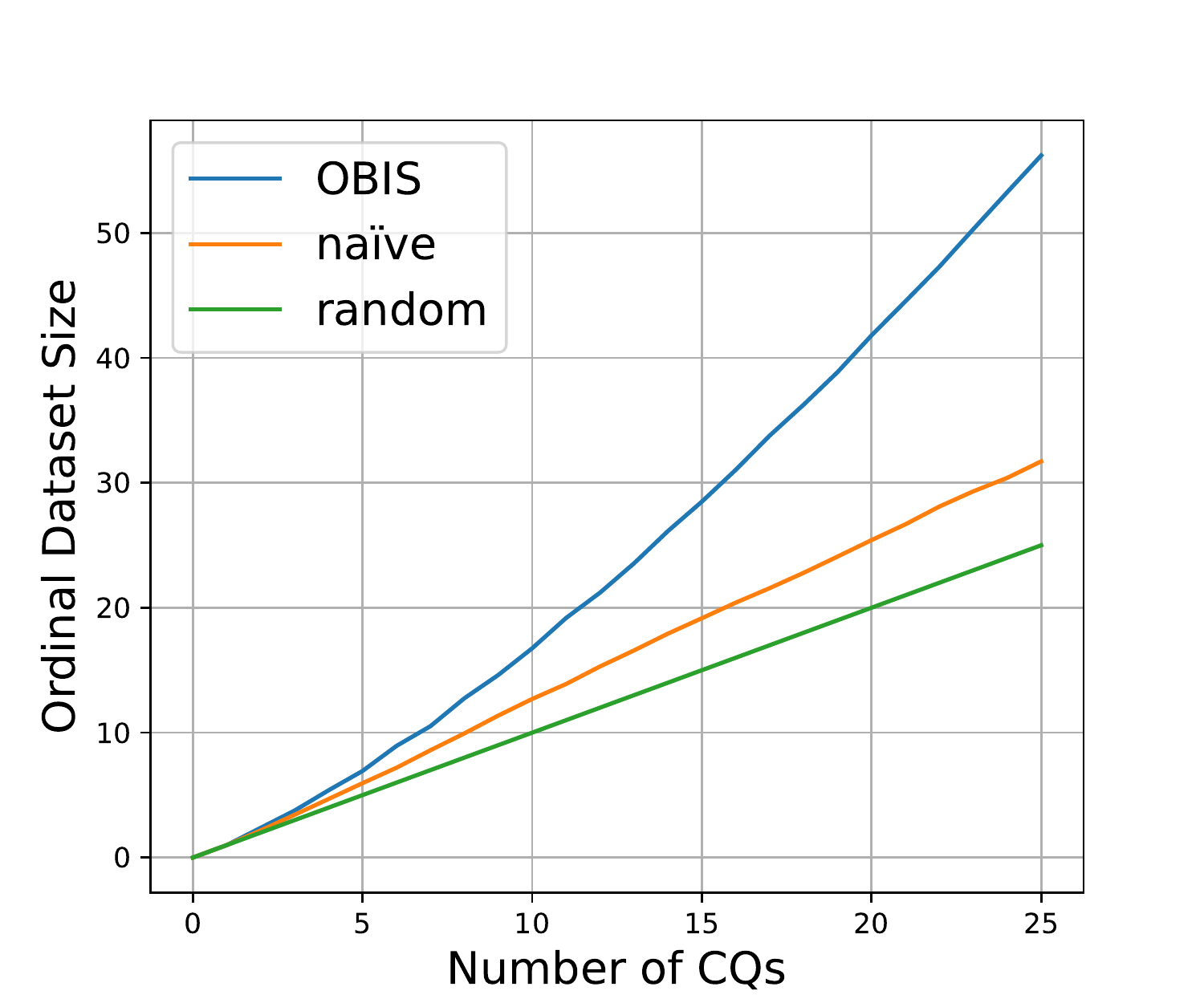}
        \caption{}
        \label{fig:ordinal_dataset_size}
    \end{subfigure}
    \hfill
    \begin{subfigure}{0.45\textwidth}
        \centering
        \captionsetup{font=small}
        \includegraphics[width=\textwidth]{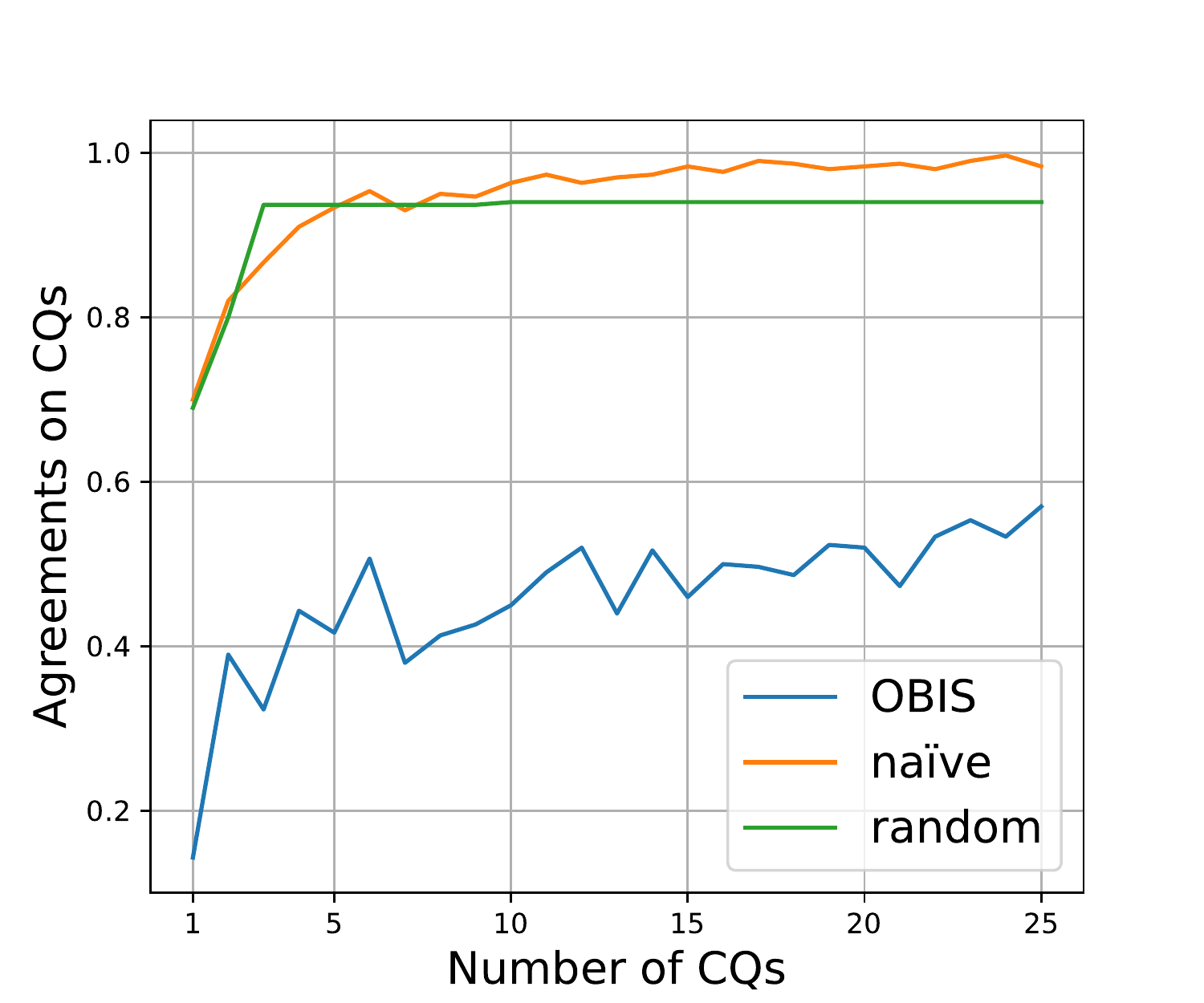}
        \caption{}
        \label{fig:agreements_on_cqs}
    \end{subfigure}
    \caption{Comparison of different query generation algorithms with respect to the ordinal dataset size they induce and how often the students' answers agree with the ML predictions for the generated CQs. Shown are averages over 300 students.}
    \label{fig:af_comparison}
\end{figure}

As we can see in \Cref{fig:ordinal_dataset_size},
the ordinal dataset than can be inferred grows much faster for OBIS compared to the other two algorithms.
For example, for $25$ OBIS CQs, the ordinal dataset has an average size of $56.2$, in contrast to only $31.7$ for the na\"{i}ve algorithm and $25$ for the random algorithm. 
This result is in line with our theoretical results from \Cref{subsec:query_generation}: in \Cref{proposition:amortized_impact}, we proved that,
in the worst case, the average increase of the ordinal dataset per CQ is $\frac{l}{ \lceil \log_2 (l + 1) \rceil }$ (where $l$ is the current size of the student's sorted list),
as opposed to only $1$ for the other two algorithms. Note that, for 25 CQs, the worst case guarantee for OBIS is 45 (and 20 for the other two algorithms). This suggests that, in practice, for all three algorithms, their average performance with respect to this metric is very close to their worst-case guarantee. 

Next, we investigate the extent to which a student's response to her last CQ corresponds with her MVNN's prediction. \Cref{fig:agreements_on_cqs} shows that for OBIS-generated queries, the student’s response matches the MVNN’s prediction in about 50-60\% of the cases. For the other two algorithms, this match rate jumps to over 90\% after just five CQs. Specifically, with the random algorithm, this high rate is observed because, after only five CQs, the MVNN has adequately learned the student’s preferences, enabling it to correctly predict the student's more preferred schedule  between two random schedules most of the time. Regarding the na\"ive algorithm, the explanation is similarly straightforward: after just five CQs, the MVNN has identified an almost optimal schedule for the student within her budget. The subsequent schedules that are compared against this almost optimal one are generally slightly less preferred by the student, which the MVNN reliably predicts.

Asking CQs that the ML model is uncertain about (and thus cannot reliably predict) is highly effective from an ML perspective. Intuitively, training on these data points leads to significant model improvements, as this is where the model performs poorly.
In electronic companion \ref{sec_appendix_query_gen_proofs}, we prove that if our \Cref{alg_mixed_training} is used for training with cross entropy loss for the ordinal dataset, then the queries maximizing the ML model's uncertainty also maximize its expected loss (with the expectation taken over the probability distribution induced by the model). In active learning, it is a common strategy to select data points that are expected to result in the largest model change, as estimated by some proxy (for which we use training loss).
This is motivated by the observation that the model's performance can only improve significantly if the model itself is changed meaningfully. This principle has been successful in various active learning problems (e.g., \citet{cai2017emcm_app1,cai2017emcm_app2}).



To conclude, in \Cref{subsec:query_generation}, we proved that the OBIS algorithm has a better worst-case guarantee regarding the size of the inferred ordinal dataset compared to the other two algorithms.
In this section, we have shown empirically that OBIS also induces a larger ordinal dataset in practice. Moreover, OBIS generates CQs that yield a larger expected model change. 
These advantages ultimately translate into the concrete welfare improvements for MLCM observed in \Cref{subsec:experimental_results}.

\subsection{Scaling up to a Larger Number of Courses}
\label{sec:Scaling}
All experiments described so far were performed in settings with $25$ courses to match the lab experiment of \citet{budish2022can}. 
As the number of courses increases, the time required to solve the MIP that determines a student's most preferred schedule at a given price vector increases. 
This affects the A-CEEI calculation and the generation of CQs by OBIS. In this section, we show that MLCM easily scales to settings with a large number of courses.

\begin{wrapfigure}[10]{r}[0pt]{0.5\textwidth}
    \vskip -0.85cm
    \includegraphics[width=0.5 \columnwidth, trim={0 0 0 2.5cm}, clip]{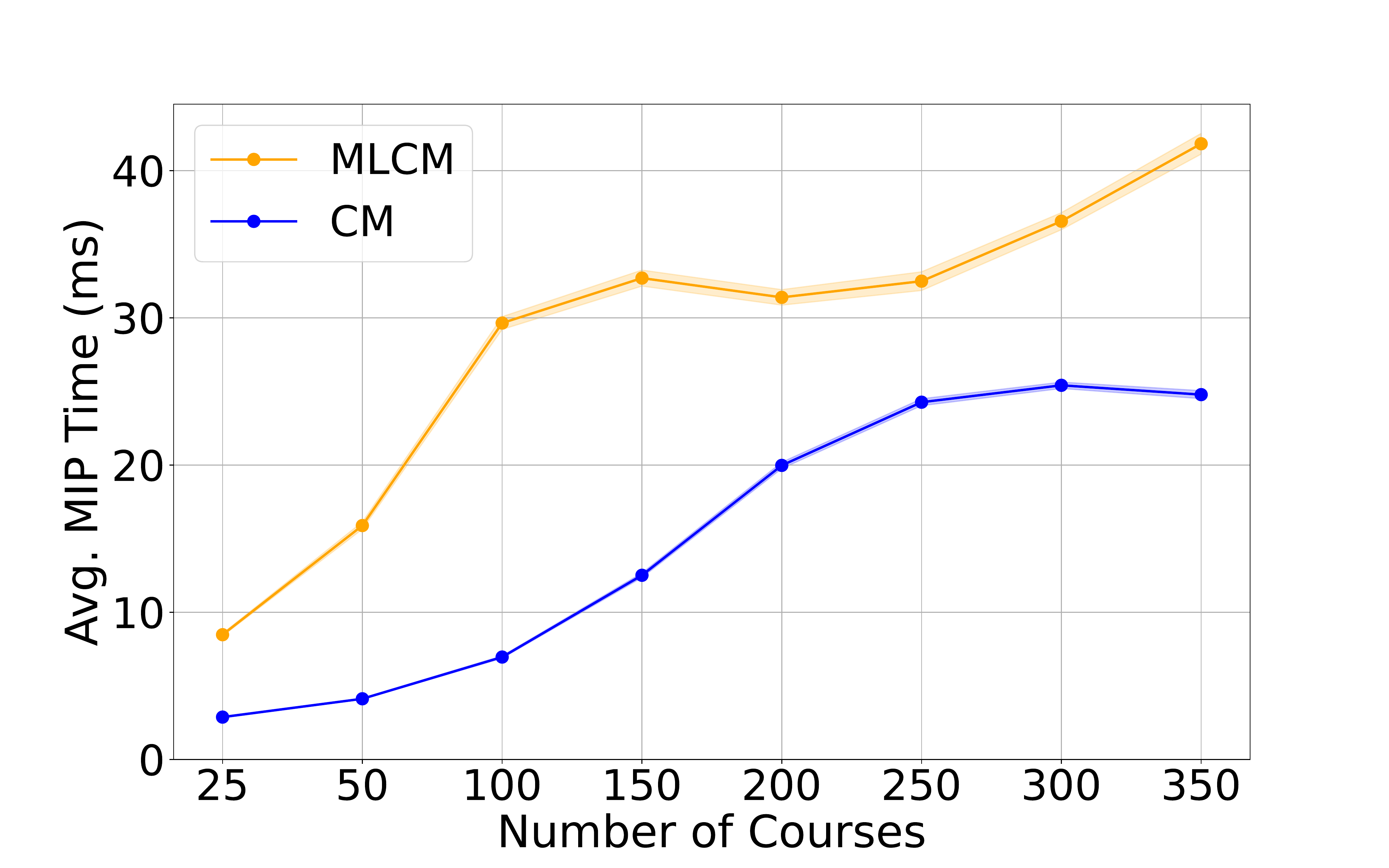}
    \vskip -0.1cm
    \caption{MIP solution time as a function of the number of courses. Shown are averages over 100 students and 20 price vectors and 95\% CIs.}
    \label{fig:mip_solution_times}
\end{wrapfigure}

To show this, we conduct the following experiment: we generate  instances varying the number of courses from $25$ to $350$.
For each instance, we train an MVNN on the GUI reports of each student and measure the time required 
to determine the student's optimal schedule within her budget.
For each number of courses, we test $100$ different students on the same $20$ random price vectors.

\Cref{fig:mip_solution_times} shows how gracefully the MIP solution time scales with the number of courses.\footnote{For this experiment, we used the same MVNN architecture as for all experiments in \Cref{Section_efficiency_results}, only changing the input layer to support the additional number of courses. Note that the structure of the students' preferences does not become significantly more complicated when increasing the number of courses (as suggested by Wharton's CM user manual, encouraging students willing to take up to 5 courses to report a base value for at least 10 courses, even though the university offers over 300 courses per semester \citep{CM_Guide}). Thus, our original network architecture can still  capture the students' preferences.} 
We see that a $14$-fold increase in the number of courses (from $25$ to $350$) only leads to a $5$-fold increase in the MIP solution time. Furthermore, across all settings, the MIP solution time for MLCM is at most 4.2 times the solution time for CM. Note that, for 25 courses, despite a three times longer MIP solution time, the A-CEEI calculation time for MLCM is actually three times faster than for CM  (see \Cref{tab:welfare_results_sr_1.25_pop_9}).\footnote{This is due to an optimization only possible for MLCM (see electronic companion~\ref{subsec:app_speedup}).} Given that for 350 courses, the MIP solution time for MLCM is only 1.5 times longer than for CM, the A-CEEI calculation time for MLCM is also faster than for CM in large settings. Regarding the \textit{real-time} component of MLCM (i.e., generating CQs), even with 350 courses, 
generating the next CQ requires less than $0.5$ seconds (where most of the time is required for retraining the ML model, not for solving the MIP). Thus, runtime is not a concern for MLCM, even for institutions with a large number of courses.


\subsection{On the Practicability of Piloting MLCM} \label{subsec:MLCM_practicability}

Regarding the practicability of piloting MLCM for institutions currently using CM, there are two noteworthy findings. First, as just discussed in \Cref{sec:Scaling}, MLCM's runtime is comparable to that of CM, even for institutions with a large number of courses. Second, we also consider a variant of MLCM, which we call \textsc{MLCM-Projected}  (see \Cref{tab:welfare_results_sr_1.25_pop_9}). While this variant  also uses MLCM's preference elicitation module, it differs in the A-CEEI calculation: it projects the students' ML models into CM's reporting language such that the A-CEEI can be computed via CM's standard algorithm (for details, see electronic companion~\ref{subsec:app_projection}).
As we  see in \Cref{tab:welfare_results_sr_1.25_pop_9}, \textsc{MLCM-Projected} matches MLCM in average and minimum student utility. While the time required to compute an A-CEEI is approximately three times higher compared to CM ($70$-$80$ minutes instead of $24$ minutes),\footnote{The runtime increase is due to the richness of the learned preferences (i.e., the projected preferences consist of more base values and adjustments than the original GUI reports). Furthermore, we cannot apply the same technical optimization (detailed in \Cref{subsec:app_speedup}) as for the standard MLCM.} the runtime is still sufficiently low not to be a concern in practice. Overall, these two findings highlight the ease of piloting MLCM in institutions already using CM.



\section{Discussion}
\label{section_discussion}
In our most realistic setting, MLCM with 10 OBIS CQs achieves a performance increase over CM of $10.5$\% and $25.5$\% for average and minimum student utility, respectively.
Furthermore, the expected gain of an individual student answering 10 CQs is over $10$\%, independent of whether any other student answers any CQs. 
To put this in perspective, these improvements are larger than what CM achieved over RSD. This underscores the significance of MLCM's large performance increase.

To realize these improvements, the students incur an additional cognitive cost because they have to answer some CQs on top of reporting their preferences using the GUI. 
However, given that the students are already encouraged to input their cardinal values to the GUI for at least $10$ courses \citep{CM_Guide}, given that pairwise CQs have a comparatively lower cognitive cost \citep{conitzer2009eliciting,chajewska00}, and given  the large utility improvements that students can expect, we consider this additional cognitive cost justified. In practice, to motivate students to answer more CQs, one could show each student her expected benefit for answering the next CQ.\footnote{Concretely, for the generated CQ, we can determine the probability with which a student will provide each answer based on the Bradley-Terry model and her current ML model (see \Cref{subsec:Integrating comparison queries into MVNNs}). 
We can train two versions of the student's model, one for each possible answer to the CQ, and measure the utility improvements according to the newly-trained models. The weighted average of these two improvements is the student's estimated expected improvement.}

Recall that MLCM uses a separate ML model for each student.
An alternative approach would have been to instead use a \textit{single} ML model for all students, leveraging possible similarities between students in the learning task. 
However, this approach might open up new possibilities for strategic manipulations since a student could then influence other students' utility representations via her reports. Additionally, this design would incentivize students to answer their CQs as late as possible so that they benefit from the ML model having been improved by the other students' answers.

MLCM uses the ML models not only to generate CQs in the iterative preference elicitation phase, but also to determine the final allocation. 
An alternative approach would have been to only use the ML models to \textit{suggest corrections} to the students' GUI reports, which the students could then either accept or reject, 
and then run CM with the revised GUI reports.
While this approach might seem attractive at first glance, it has multiple drawbacks. The main one is that it would require significantly more interaction between the students and the mechanism: the students would have to first report their preferences to the GUI, then answer some CQs, and finally go back to correct their original GUI reports. 
This last step in particular might be cognitively too demanding for most students, as it involves reviewing ``suggested corrections'' for possibly tens of courses, which are all in conflict with what the students originally reported via the GUI. 
Additionally, this alternative approach might introduce new reporting mistakes, since it requires that students input their cardinal base values and adjustments a second time. Finally, it does not address the fact that the CM language may not be able to fully capture every student's preferences.


Several institutions employing CM in practice (e.g., University of Toronto, Columbia Business School) use a different reporting language from the one described in \Cref{sec:Course-Match-Introduction}.
In those institutions, each student places the courses she is interested in into four different priority groups and additionally ranks the courses within each group. These preferences are then translated into cardinal values so that a course has higher utility than all courses the student placed into lower priority groups and all courses in the same group that she ranked lower \citep{Columbia_Guide}.  
The only interactions between courses that students can report are pairs of courses they do not want to be assigned simultaneously.
Since the students' preferences are translated into cardinal values, MLCM is also compatible with this simpler reporting language without any changes. However, we did not consider this language in our experiments as there is no data available on the students' reporting mistakes using this language, in contrast to the original CM reporting language. Recall that \citet{budish2022can} were already concerned that the CM language may not be able to fully capture all students' preferences. The alternative reporting language is even more restrictive, not allowing students to report base values for individual courses nor any complementarities. Given this, it is even more likely that the students' preferences are not captured correctly, suggesting that MLCM would likely be able to improve the students' utilities to an even larger degree.

\section{Conclusion}
\label{sec:Conclusion}
In this paper, we have introduced the \emph{Machine Learning-powered Course Match (MLCM)} mechanism. In contrast to prior work on course allocation, we have focused on alleviating the students' reporting mistakes, which we have found to have a larger impact on welfare than the choice of the mechanism itself. Our experimental results have demonstrated that MLCM significantly increases both average and minimum student utility compared to CM across a wide range of settings.

When designing MLCM, we pursued the following desiderata: the mechanism should handle both cardinal and ordinal inputs; participation in the ML-powered preference elicitation phase should be optional for students; the mechanism should perform well with few queries; it should accommodate asynchronous student interactions; and it should be computationally efficient enough to enable real-time interactions.

We have shown that MLCM is very robust to changes in the environment. These include changes to the degree of correlation in students’ preferences and the ratio of supply to demand (\Cref{A_welfare_results}), changes in the number of reporting mistakes students make in the GUI (\Cref{subsec:Reporting Mistakes Robustness Study}), changes in the complexity of the students’ preferences (\Cref{subsec:Results for Additive Preferences}), changes in the quality of the approximate prices used by the query-generation algorithm (\Cref{subsec:app_price_sensitivity}), and changes in students’ accuracy when answering CQs (\Cref{subsec:app:CQ_robustness}).

Throughout the design of MLCM, we have followed the \textit{minimalist market design} framework \citep{sonmez2023minimalist}. First, we identified efficiency and fairness as the two design objectives for course allocation that are not achieved satisfactorily under the current mechanism. Second, we identified students' reporting mistakes as the main obstacle preventing the current mechanism from achieving these objectives. Third, we proposed minimal changes to CM's current design by introducing an ML-powered preference elicitation algorithm that alleviates CM's shortcomings. We expect that this market design approach will make it more likely for institutions to adopt MLCM.

In a similar vein, we have also taken great care to reduce friction and potential risks for both institutions and their students when upgrading from CM to MLCM. First, we have designed MLCM such that participating in the ML-powered preference elicitation phase is \textit{optional} for each student. Second, MLCM supports \textit{informed consent}: even after trying out the ML feature, each student can decide to abort the process and revert to her GUI reports. Finally, we have introduced \textsc{MLCM-Projected}, a variant of MLCM that is compatible with the standard CM implementation to compute the final allocation. Thus, by designing MLCM to include optional participation, informed consent, and a variant compatible with the current CM infrastructure, we have significantly minimized risks and streamlined the transition process both for institutions and their students.

There are several promising directions for future work. One such direction is fielding MLCM in an institution currently using CM to assess its real-world efficacy. As discussed above, we have designed MLCM to make such a transition as easy as possible while mitigating potential risks. To evaluate the welfare improvements of MLCM, each student could be sent a survey after the final allocation has been determined, asking her to compare two schedules: the one she was allocated under MLCM and the one she would have been allocated under CM. Additionally, it would be interesting to see how many students opt into the ML feature and how many CQs they answer.

A second direction involves investigating the design of the query generation procedure in more depth. Integrating ML models that can handle \textit{model uncertainty} (e.g., \citep{heiss2022nomu}) would make it feasible to accurately gauge the ML model's uncertainty regarding students' preferences for schedules that have not yet been queried. This would enable the adoption of query generation strategies that leverage the principle of \textit{expected value of information}, ensuring that each query maximally contributes to the model's understanding of student preferences.

A third direction involves exploring the most effective ways to communicate to students what the ML algorithm has inferred about their preferences, and the reasons behind these inferences (e.g., ``Based on your response to query X, we infer that your utility for course A is lower than what you initially reported in the GUI"). Incorporating these explanatory features into the interface could bolster students' trust in the ML feature, potentially leading to higher adoption rates and, consequently, enhancing overall welfare.

Finally, we emphasize that our ML-powered preference elicitation approach has the potential to be applied across a broad spectrum of combinatorial assignment and matching problems. A particularly impactful application is refugee resettlement, a domain where recent research has framed resettlement as a matching problem that considers the preferences of both local communities and refugees \citep{teytelboym_refugees, refugees3}. An intriguing direction for future work is exploring how our ML-powered approach can assist refugees in more accurately reporting their preferences -- such as employment opportunities and proximity to family members previously resettled. For example, our preference elicitation algorithm could be integrated into mechanisms such as the one proposed by \cite{Ahani2021Placement}, who use machine learning to suggest optimal placements for refugees. This could lead to a matching process that is both more efficient and more equitable.





\bibliographystyle{informs2014} 
\bibliography{CApapers}



%
%
%

\ECSwitch


\renewcommand{\theHchapter}{A\arabic{chapter}}  
\ECHead{Electronic Companion}




\section{Cardinal Dataset from Students' GUI Reports} \label{A_bootstrap}

In this section, we describe the process of building the cardinal dataset $D_{i, \text{card}}$ for each student $i\in N$, given their GUI reports.

From a conceptual standpoint, our idea for incorporating the students' reports in the learning framework is straightforward: For every student $i$, we want to use their reports to the GUI and create a bootstrap dataset $D_{i, \text{card}}$ where every datapoint is a tuple of the form (schedule, utility). $D_{i, \text{card}}$ is then used to initialize $\mathcal{M}_i$.

As explained in \Cref{subsection: CM_prelims}, in the current implementation of Course Match, the students have two means at their disposal to express their preferences: base values and pairwise adjustments.
Based on the student $i$'s report, there are three distinct sets of schedules we can generate for our dataset $D_{i, \text{card}}$: 

\begin{itemize}
    \item \textbf{Bundles of type 1 (T1):} 
    This set specifically includes two kind of schedules:
    \begin{enumerate}
        \item Bundles containing a single course, for which the corresponding student has declared a base value. 
        \item  Bundles containing two courses for which the corresponding student has declared both base values plus the corresponding pairwise adjustment.
    \end{enumerate} 
    The value of each such schedule is calculated exactly as in the current Course Match implementation, i.e., by summing the reported base values plus the reported adjustments whose courses are contained in that schedule. 
    
    These schedules constitute the most solid pieces of information in our dataset, as for any schedule of this type, we can be sure that there cannot be any terms that the student did not report, either because they forgot to, or because it was impossible to do so in the CM reporting language. 
    \item \textbf{Bundles of type 2 (T2)}: Each schedule of T2 contains five courses, for all of which the student reported a base value 
    but not necessarily all corresponding pairwise adjustments.
    The value of each such schedule is calculated in the same way as in the current CM implementation, i.e., it is simply the sum of base values plus any adjustments reported by the student for the courses it contains. 
    
    These schedules are important for two reasons: First, they push the distribution of the training set closer to  the part of the space that matters in practice, a student participating in the mechanism is  allocated a schedule of about five courses; not one or two. 
    Additionally, these schedules implicitly ``show'' the learning model 
    the combinatorial structure of the students' utility functions. For most ML algorithms tested, training them without schedules of this type lead to significantly worse generalization performance. 
    \item \textbf{Bundles of type 3 (T3)}: Each schedule of this type contains five courses, but for at least one of them, the student did not even submit a base value when reporting their preferences in the CM language. For every such course, we impute a base value for it. 
    Specifically, the imputed value is the expected value of the prior value distribution of a course, but now conditioned on the fact that a student did not report a base value for it in the CM language. No additional pairwise adjustments are imputed that a student did not report. Having imputed those base values, the value of each schedule of T3 can be calculated in the same way as for T1 and T2. 
    
    Concretely, suppose that a student didn't report a base value for a course. In the current implementation, this course will be treated as if the student had declared a zero base value for it. Thus, there are two possible explanations for a missing base value: Either the student actually had a zero value for the specific course, or they had a non-zero base value, but they actually forgot to report it. Given the assumption that students tend to forget to report a base value for courses towards the lower-end of their valuation, we model the base value $v$ of a course that wasn't reported by the student as:  
\begin{align}
    v = 
     \begin{cases}
  0 & \text{w.p.  } p_z \\
  \sim U[l, h] & \text{w.p.  } 1 - p_z
  \end{cases}
\end{align}
And thus its expected value is: 
\begin{align}
    \mathbb{E}[v] = \left ( 1 - p_z \right ) \left ( \frac{h - l}{2} + l \right )  = \left ( 1 - p_z \right )\frac{h + l}{2}.
\label{eq:expected_value}
\end{align}

In order to generate the schedules of T3 in our experiments, we set $p_z=0$, $l=0$, and $h$ equal to the lowest non-zero reported base value. 
    
\end{itemize}

\section{Next Binary Search Query} \label{app:nextbinarysearchquery}
In this section, we provide a detailed description of the NextBinaryInsertionSortQuery sub-process mentioned in \Cref{subsec:query_generation}. 
Just like (offline) binary insertion sort, in each iteration of the main loop (Line~\ref{alg:nextquery:while_loop}), as long as the next comparison point is included in the list $Q_i$ of CQs already answered by the student (Line~\ref{alg:nextquery:cq_included}), the search space is reduced in half, based on the student's answer to the CQ (Lines~\ref{alg:nextquery:left} and~\ref{alg:nextquery:right}), until a CQ is found whose answer is not included in $Q_i$, and that query is returned (Line~\ref{alg:nextquery:found_cq}). 
Note that this will always happen, as by definition $x$ is a query that cannot be already sorted into the list $S_i$, based on the student's already answered CQs $Q_i$. 

\begin{algorithm}
\caption{NextBinaryInsertionSortQuery}
\label{alg_nextbinarysearchqueries}
\begingroup 
\renewcommand{\baselinestretch}{1} 
\selectfont 
\textbf{Input}: List of sorted schedules based on student's past responses $S_i$, schedule to be inserted $x$, student's answered comparison queries $Q_i$
\\
\textbf{Output}: Next CQ to ask the student $x_1 \prec x_2 $
\begin{algorithmic}[1] 
\STATE $L \gets 0$
\STATE $R \gets n - 1$
\WHILE{$L \le R$}  \label{alg:nextquery:while_loop}
    \STATE $M \gets \lfloor (L + R) / 2 \rfloor$
    \IF{$(x, S_i[M]) \in Q_i$ }  \label{alg:nextquery:cq_included}
    \IF{$x  \succ_i S_i[M]$}
        \STATE $L \gets M + 1$ \label{alg:nextquery:left}
    \ELSE
        \STATE $R \gets M - 1$ \label{alg:nextquery:right}
    \ENDIF
    \ELSE
        \STATE \textbf{return} $CQ = (x, S_i[M])$ \label{alg:nextquery:found_cq}
    \ENDIF
\ENDWHILE
\end{algorithmic}
\endgroup 
\end{algorithm}

\section{Theoretical Properties of \Cref{alg_query_generation}} 
\label{sec_appendix_query_gen_proofs}
In this subsection, we provide all proofs relating to the OBIS algorithm from \Cref{subsec:query_generation,subsec:af_empirical_evaluation}.


\proof{Proof of \Cref{proposition:amortized_impact}}
To insert an element into the sorted list $S$ of length $l$, binary insertion sort starts with the full list as its search space, and it iteratively reduces that search space in half, by asking the student the query comparing the element to be inserted with the median of the remaining search space, until only a single viable position remains. 
Given that the search space is reduced in half with each CQ, the precise position to insert the new element will be determined in at most 
$\lceil \log_2 (l + 1) \rceil$ such CQs. 
At this point, the new element can be inserted into the correct place in the list, and the student's preference between that and any of the other $l$ elements of the list can be determined. 
Thus, this set of at most $\lceil \log_2 (l + 1) \rceil$ CQs resulted in the student's preference relation to be revealed for any of the $l$ pairs involving the new element $x$ and any element of the list $S$. 
And so, for those CQs, the number of such pairs grew by at least an amortized 
$\frac{l}{ \lceil \log_2 (l + 1) \rceil }$ per CQ. 
\Halmos 
\endproof

\begin{remark}
Note that for $l \ge 2$ (i.e., after the first CQ that a student answers), this number is greater than one, and 
$\lim_{l \xrightarrow[]{} \infty} \frac{l}{ \lceil \log_2 (l + 1) \rceil } =  \infty$. 
Thus, this rate is indeed superlinear (i.e., it exceeds any function that is linear in $l$, as $l \xrightarrow[]{} \infty$). 
Furthermore, the more CQs that a student answers, the larger the length $l$ of her list $S$ is. 
Thus, the more CQs that a student answers, the largest the amortized growth rate of the number of inferred ordinal pairs per CQ becomes. 
\end{remark}

Finally, we provide \Cref{lemma:BCE_maximization} referenced in \Cref{subsec:af_empirical_evaluation}.

\begin{lemma}
\label{lemma:BCE_maximization}
Let $\mathcal{M}(\cdot | \mathcal{D})$ be a regression ML model trained on dataset $\mathcal{D}$, and let $p_{\mathcal{M}(\cdot | \mathcal{D})} (x_1 \succ x_2)  = \frac{1}{1 + e^{-(\mathcal{M}(x_1 | \mathcal{D}) - \mathcal{M}(x_2 | \mathcal{D})) }}$ be interpreted as the probability with which $x_1 \succ x_2$, based on $\mathcal{M}(\cdot | \mathcal{D})$.
Finally, let $\text{BCE}_c$ be Binary Cross Entropy loss, with the log function output clamped to be at least some very large negative number $c$.\footnote{This is the case in most ML libraries to ensure numerical stability, see e.g.,  
\href{https://pytorch.org/docs/stable/generated/torch.nn.BCELoss.html}{PyTorch documentation}.}
Then, $\mathbb{E}_{p_{\mathcal{M}(\cdot | \mathcal{D})}}  
\left [\text{BCE}_c ((p_{\mathcal{M}(\cdot | \mathcal{D})}(x_1 \succ x_2), p_{\mathcal{M}(\cdot | \mathcal{D})}(x_1 \prec x_2)), (\boldsymbol{1}_{x_1 \succ x_2 }, \boldsymbol{1}_{x_1 \prec x_2 }) )\right ]$
is maximized by any comparison query involving bundles $x_1, x_2$ that minimize 
$\left | p_{\mathcal{M}(\cdot | \mathcal{D})}(x_1 \succ x_2) -  p_{\mathcal{M}(\cdot | \mathcal{D})}(x_1 \prec x_2) \right|$.
\end{lemma}

\proof{Proof of \Cref{lemma:BCE_maximization}}
Let $p_{\mathcal{M}(\cdot | \mathcal{D})}(x_1 \succ x_2) = p$. 
Assuming that ties in the student's preferences are broken lexicographically,
for the expected binary cross entropy loss of a comparison query including schedules $x_1$ and $x_2$ we have: 
\begin{align}
& \mathbb{E}_{p_{\mathcal{M}(\cdot | \mathcal{D})}}  
\left [\text{BCE}_c ((p_{\mathcal{M}(\cdot | \mathcal{D})}(x_1 \succ x_2), p_{\mathcal{M}(\cdot | \mathcal{D})}(x_1 \prec x_2)), (\boldsymbol{1}_{x_1 \succ x_2 }, \boldsymbol{1}_{x_1 \prec x_2 }) )\right ]   \nonumber \\
& =  \mathbb{E}_{p_{\mathcal{M}(\cdot | \mathcal{D})}}  \left [\text{BCE}_c ((p_{\mathcal{M}(\cdot | \mathcal{D})}(x_1 \succ x_2), 1- p_{\mathcal{M}(\cdot | \mathcal{D})(x_1 \succ x_2)}(x_1 \succ x_2)), (\boldsymbol{1}_{x_1 \succ x_2 }, 1 - \boldsymbol{1}_{x_1 \succ x_2 }) )
\right ]  \nonumber \\ 
& = p \cdot \text{BCE}_c ((p_, 1- p), (1, 0)) 
+ (1 - p) \cdot \text{BCE}_c ((p_, 1- p), (0, 1))  \nonumber \\ 
& =p \left [ -p  \max \{ \log 1, c \} - (1-p) \max \{\log 0, c\} \right ] \nonumber \\ 
   & \hphantom{5} + (1 - p) \left [ - p  \max \{\log 0, c \} - (1-p) \max \{ \log 1, c\} \right ] \nonumber \\ 
& = -p^2 \log1 - p(1-p) c - p(1-p) c - (1-p)^2 \log 1 \nonumber \\ 
& = -2p (1 - p) c \label{eq:expected_cross_entropy_calculations}
\end{align}
which, given that $c < 0$ and $0 \le p \le 1$, is a concave function that is maximized at $p = \frac{1}{2}$.
\Halmos
\endproof

\section{Inferring missing base values - A worked example} \label{app_sec:example_forgetting_base_values}
In \Cref{example:Inferring_missing_base_values_I}, we show how MLCM can infer missing base values, a common error observed in the lab experiment by \citet{budish2022can}.

\begin{table}[h!]
\centering
\begin{sc}
	\begin{tabular}{
                S[table-format=2]
			S[table-format=2]
			S[table-format=2]
			S[table-format=2]
			S[table-format=2]
			S[table-format=2]
			S[table-format=3]
                S[table-format=2]
			}
		\toprule
            \multicolumn{1}{c}{Preference}&\multicolumn{4}{c}{Courses}& {Utility Maximizing}&  \multicolumn{1}{c}{Utility}& \multicolumn{1}{c}{Elicited}\\
		\multicolumn{1}{c}{Model} & \multicolumn{1}{c}{1} & \multicolumn{1}{c}{2} & \multicolumn{1}{c}{3} & \multicolumn{1}{c}{4}&  {Schedule $a^*$}&{$u(a^*)$}& {CQ}  \\
            \cmidrule(lr){1-1}
		\cmidrule(lr){2-5}
		\cmidrule(lr){6-6}
            \cmidrule(lr){7-7}
            \cmidrule(lr){8-8}
		{$u$}   & 85& 70 & 40  & 0  & {\{1,3\}} & 125& \\
		{$u^{GUI}$}  & 75& 76 & 0  & 0  & {\{2\}} & 70&\\
            \midrule
		{$\mathcal{M}^{0}$} & 75& 76 & 38  & 38  & {\{2,3\}} & 110& {$\{2,3\} \succ \{2,4\}$}\\
		{$\mathcal{M}^{1}$} & 75& 76 & 41  & 33  & {\{2,3\}} & 110& {$\{1,3\} \succ \{2,3\}$} \\
        {$\mathcal{M}^{2}$} & 78& 71 & 44  & 36  & {\{1,3\}} & 125& {$\{2,3\} \succ \{1,4\}$} \\
        {$\mathcal{M}^{3}$} & 78& 72 & 45  & 32  & {\{1,3\}} & 125& {$\{1,4\} \succ \{2,4\}$} \\
        {$\mathcal{M}^{4}$} & 78& 72 & 45  & 32  & {\{1,3\}} & 125&  \\
		\bottomrule
	\end{tabular}
\vskip 0.1cm
\caption{Worked example illustrating the OBIS preference elicitation algorithm. Each row represents the linear coefficients (corresponding to the base values) that uniquely define the corresponding function, the current utility maximizing schedule $a^*$, its corresponding utility $u(a^*)$ and the answer to the CQ.}
\label{tab:toy_example_3}
\end{sc}
\end{table}

\begin{example}[Inferring Missing Base Values]\label{example:Inferring_missing_base_values_I} In this example, we assume that the student forgot to report a base value for courses 3 and 4. This is treated in the original CM reporting language as inserting a base value equal to zero in the GUI, i.e., $u^{GUI}(\{3\})=u^{GUI}(\{4\})=0$ (as can be seen in the second row of \Cref{tab:toy_example_3}). First, note that our cardinal dataset generation procedure (see \Cref{A_bootstrap} for details) implies that any schedule for which a student forgot to report a base value has the same initial predicted base value (as defined in \Cref{eq:expected_value}), which is not necessarily zero but lower than the base value of any non-forgotten course. This results in  $\mathcal{M}^{0}(\{3\}) = \mathcal{M}^{0}(\{4\}) = 38$ instead of zero.  

After this point, MLCM proceeds again in the same way as in \Cref{example:noisy_report_II} until the student stops answering CQs.

\end{example}

\section{Theoretical Properties of MLCM} \label{A_guarantees}

In this section, we show that, if the preferences are captured \textit{approximately} via the ML models $\mathcal{M}_i$, then the same theoretical properties as for the CM Stage 1 allocation (i.e, 
 \textit{envy-bounded by a single good}, \textit{$(n+1)$-maxmin share guarantee}, and \textit{Pareto efficiency}) also hold in an \emph{approximate} sense for the Stage 1 allocation of MLCM. First, we recall our notation.

  \paragraph{Notation} $\studentset{}$ is the set of students, $n$ is the number of students, $M$ is the set of courses, $m$ is the number of courses, $k$ is the maximum number of courses allowed in a schedule, $u_i(\cdot)$ is the true utility function for student $i$, $x$ is a schedule of courses (i.e. course schedule), $\x$ an allocation, 
  and $\xii$ denotes the schedule student $i$ receives in allocation $\x$, i.e., student $i$'s allocation. The set of feasible allocations is represented by $\mathcal{F}$. Furthermore, we slightly overload the notation and denote by $a_i$ \emph{both} the indicator vector representing the allocation for student $i\in N$ 
 (i.e., $a_i\in \{0,1\}^m$) as well as the corresponding set (i.e., $a_i\in 2^M$). With this, we denote by $j\in a_i$ that $j\in M$ is contained in $a_i$ and by $a_i\setminus \{j\}$ the allocation $a_i$ without the course $j\in M$ (we also use this for other generic schedules $x$ or $x^{\prime}$ ).
  
  First, we define our notion of utility function approximation, approximate fairness and welfare.

\begin{definition}[$\ep$-approximation of a utility function]
For $\ep \geq 0$,  a function $\hat{u}(\cdot)$ is an $\ep$-approximation of the true utility function $u: \{0,1\}^{m} \xrightarrow{} \R$ if
\begin{equation}
 \sup_{x \in \{0,1\}^{m}} |\hat{u}(x) -  u(x)| < \ep.
\end{equation}
\end{definition}

\begin{definition}[Envy $\ep$-bounded by a single good] An allocation $a$ satisfies envy $\ep$-bounded by a single good if for all $i,i^{\prime} \in N$ 
\begin{enumerate}
    \item $u_i(a_i) \ge u_i (a_{i^{\prime}}) - \ep$ or \\ 
    \item There exists some good $j \in a_{i^{\prime}}$ such that $u_i(a_{i}) \ge u_i(a_{i^{\prime}} \setminus \{j \} ) - \ep$.
\end{enumerate}
\end{definition}
That is, if student $i$ envies student $i^{\prime}$, by removing some single good from student $i^{\prime}$'s schedule we can make $i$'s envy at most $\ep$.

\begin{definition}[$l$-maximin share, \citet{budish2011aceei}] Let $Z^*$ denote a $l$-maximin split, i.e., 
\begin{equation}
Z^*\coloneqq\argmax\limits_{\substack{\{z_1,\ldots,z_l\}:z_k \in \mathcal{X}\\ \sum_{k=1}^lz_{kj}\le q_j}} \left ( \min\limits_{k \in \{1,\ldots,l\}}u_i(z_k)\right).
\end{equation}
Then, student $i$'s $l$-maximin share $\maximinshare[i]{l}$ for her given utility function $u_i$ is defined as
\begin{equation}
    \maximinshare[i]{l}\coloneqq\argmin\limits_{z \in Z^*} u_i(z).
\end{equation}
In words, $\maximinshare[i]{l}\in \{0,1\}^m$ is the course schedule an student obtains when she selects the utility maximizing feasible partition consisting of $l$ schedules of the set of all courses given that an adversary assigns her the worst schedule from that proposed partition.
\end{definition}

\begin{definition}[$(l,\ep)$-maximin share] For any $\ep\ge 0$ student $i$'s $(l,\ep)$-maximin share is any course schedule $\Epsmaximinshare[i]{l}{\ep}$ for which student $i$ has utility at most $\ep$ less than her maximin share, i.e,
\begin{equation}
    u_i(\Epsmaximinshare[i]{l}{\ep}) \ge u_i \left(\maximinshare[i]{l}\right) - \ep.
\end{equation}
\end{definition}

\begin{definition}[$(l,\ep)$-maximin share guarantee] Any feasible allocation $a=(a_i)_{i=1}^n\in \mathcal{F}\subset \mathcal{X}^n$ where \emph{all} students $i\in N$ get a schedule $a_i$ they weakly prefer to their $(l,\ep)$-maximin share $\Epsmaximinshare[i]{l}{\ep}$ (w.r.t. their true utility functions, i.e., $\left\{u_i\right\}_{i\in N}$) is said to satisfy the \textit{$(l,\ep)$-maximin share guarantee}.
\end{definition}

In \Cref{prop:app:theoretical_results}, we now prove our main theoretical result.
\begin{proposition}\label{prop:app:theoretical_results} 
Let $[a^*, b, p^* ]$ be an $(\alpha, \beta)$-A-CEEI calculated using the $\ep$-approximation of the true utility functions  $\{{\hat{u}}_i\}_{i\in \studentset{}}$, then:
\begin{enumerate}
    \item\label{itm:1propA1} If $\beta \le \frac{1}{k-1}$ with $k$ being the maximum number of courses per student, then $a^*$ satisfies envy $2 \ep$-bounded by a single good w.r.t. the true utility functions $\{{u}_i\}_{i\in \studentset{}}$. Moreover, this bound is tight.
    \item\label{itm:2propA1} If there exists some $\delta \ge 0$ such that
    $p^* \in \mathcal{P}(\delta, b)$, \footnote{$\mathcal{P}(\delta, b)=\{p \in [0, \max_{i} b_i]^{m}: \sum_{j} p_j q_j 
\leq \sum_i b_i (1+\delta)\}.$} and $\beta < (1 - \delta n)/ n (1 + \delta)$, then $a^*$ satisfies the $(n+1, 2 \ep)$-maximin share guarantee w.r.t. the true utility functions $\{{u}_i\}_{i\in \studentset{}}$.  Moreover, this bound is tight.
    \item\label{itm:3propA1} Given $p^*$, the true utility of every student $i\in N$ for the schedule she receives in the allocation $a^*$ (i.e., ${u}_i(a^*_i)$) is within $2 \ep$ of her true utility for her most preferred schedule she could afford. 
    \item \label{itm:4propA1} Given $p^*$,  the allocation $a^*$ is $2 \ep$-Pareto efficient w.r.t. the true utility functions $\{{u}_i\}_{i\in \studentset{}}$, i.e., $\nexists\,\, a^{\prime}\in \mathcal{F}$ such that $\forall i\in N$ it holds that:
    \begin{equation}
        u_i(a_i^{\prime}) 
        \geq 
        u_i(a_i^{*}) - 2\ep.
    \end{equation}
\end{enumerate}
\end{proposition}

\proof{Proof.}
\begin{enumerate}
    \item Let $ \hat{u} = (\hat{u}_1, \hat{u}_2, \dots \hat{u}_n)$ be the profile of learned utility functions. Using Theorem 3 of \citet{budish2011aceei}\footnote{To apply Theorem 3 of \citet{budish2011aceei}, we need $\beta \le \frac{1}{k-1}$.}, we have that for those learned utility functions the allocation $a^*$ satisfies envy bounded by a single good i.e., for any $i, i^{\prime} \in S$ either:
\begin{enumerate}
    \item $\hat{u}_i(a^*_i) \ge \hat{u}_i (a^*_{i^{\prime}}) $ or 
    \item There exists some good $j \in a_{i^{\prime}}$ such that $\hat{u}_i(a^*_{i}) \ge \hat{u}_i(a^*_{i^{\prime}} \setminus \ \{ j \} )$.
\end{enumerate}
Since $\hat{u}_i (\cdot)$ is an $\varepsilon$-approximation of the true utility function $u_i(\cdot)$ we have that in the first case: 
\begin{gather}
    u_i(a^*_i) + \varepsilon \ge \hat{u}_i(a^*_i) \ge \hat{u}_i (a^*_{i^{\prime}}) \ge u_i (a^*_{i^{\prime}}) - \varepsilon \implies
    u_i(a^*_i) \ge u_i (a^*_{i^{\prime}}) - 2  \varepsilon \label{singlegood:1}
\end{gather}
Similarly in the second case: 
\begin{gather}
    u_i(a^*_i) + \varepsilon \ge \hat{u}_i(a^*_i) \ge \hat{u}_i (a^*_{i^{\prime}} \setminus \{ j \}) \ge u_i (a^*_{i^{\prime}}) - \varepsilon \implies
    u_i(a^*_i) \ge u_i(a^*_{i^{\prime}} \setminus \{j \} ) - 2 \varepsilon \label{singlegood:2} 
\end{gather}
From \Cref{singlegood:1,singlegood:2} it follows immediately that $a^*$ satisfies envy $2 \varepsilon$-bounded by a single good.

Next we provide an example that shows that the bound is tight. Assume that there are 3 courses, $a,b$ and $c$ with capacities $q_a = 2$ and $  q_b = q_c = 1$. Moreover, assume that there are 2 students with the following utility functions:
\begin{itemize}
        \item $u_1( \{ a,b  \}) = u_1 (\{ a, b, c \} )  = 1$, and $0$ for any other schedule.
        \item $u_2( \{ a  \}) = u_2( \{ b  \}) =  0.5 - \varepsilon , u_2(\{ a,c \} ) = u_2(\{ b , c\}) = 0.5 + \varepsilon , u_2( \{ a , b \}) = u_2 ( \{ a,b,c \} ) = 1$, and $0$ for any other schedule.
\end{itemize}
    Take $\ep, \ep' > 0$ and the learned utility functions of the $2$ students to be: 
    \begin{itemize}
        \item $\hat{u}_1( \{ a,b  \}) = 1 - \varepsilon , \hat{u}_1 \{ a, b, c \} = 1$, and $0$ for any other schedule.
        \item $\hat{u}_2( \{ a  \}) = \hat{u}_2( \{ b  \}) = \hat{u}_2( \{ a,c  \}) = \hat{u}_2 ( \{  b, c \})  = 0.5 $ \\ $ \hat{u}_2(\{ a,b \}) = \hat{u}_2(\{ a,b,c \}) = 1 $, and and $0$ for any other schedule. 
    \end{itemize}
    Then,  a $(0,\beta)$-A-CEEI ( $ \beta \le \frac{1}{k - 1 } = \frac{1}{2}$) can be formed with the following elements:
    \begin{itemize}
        \item $   {a}^*_1  = \{ a , b,  c \},   {a}^*_2 = \{ a  \}  $
        \item $  {b}_1 = 1 + \beta,   {b}_2 = 1$
        \item $  {p}^*_a =   \beta,    {p}^*_b = 1  ,   {p}^*_c = 0$.
    \end{itemize}
    
    In this  $(0,\beta)$-A-CEEI, the envy of student $2$ with respect to her true utility function is exactly $2 \varepsilon$-bounded by a single good: 
    \begin{align}
        u_2 ( {a}^*_2) = 0.5 -   \varepsilon = u_2 ( {a}^*_1 \setminus \{ b \}) - 2 \varepsilon.
    \end{align}
    Therefore, the bound is tight.
 
\item Let $ \hat{u} = (\hat{u}_1, \hat{u}_2, \dots, \hat{u}_n)$ be the profile of learned utility functions. Using Theorem 2 of \cite{budish2011aceei},\footnote{To apply Theorem 2 of \citet{budish2011aceei}, we need the existence of $\delta \ge 0$ such that
    $p^* \in \mathcal{P}(\delta, b )$
    and $\beta < (1 - \delta n)/ n (1 + \delta)$.} we have that for those learned utility functions the allocation $a^*$ satisfies the $(n+1)$-maximin share guarantee. Thus, for any $i \in N$:
\begin{align}
    \hat{u}_i(a^*_i) & \ge \hat{u}_i \left ( \maximinsharehat[i]{n+1} \right ) \label{p2_eq1}\\ 
    & \ge \hat{u}_i \left ( \maximinshare[i]{n+1} \right ) \\
    & \ge  {u}_i \left (\maximinshare[i]{n+1}\right )  -  \ep \label{p2_eq2}
\end{align}
where \eqref{p2_eq1} follows from the definition of the $(n+1)$-maximum share guarantee for the A-CEEI w.r.t. the learned utilities $\{\hat{u}_i\}_{i\in N}$ and  \eqref{p2_eq2} is true because $\hat{u}_i$ is per assumption an $\ep$-approximation of $u_i$.
Then,  we have that
\begin{equation}
    u_i(a^*_i) + \ep \ge \hat{u}_i(a^*_i) \ge {u}_i \left (\maximinshare[i]{n+1}\right )  -  \ep, 
\end{equation}
and therefore
\begin{equation}
    u_i(a^*_i) \ge {u}_i \left (\maximinshare[i]{n+1} \right )  - 2 \ep.
\end{equation}
Hence the allocation $a^*$ satisfies the $(n+1, 2\ep)$-maximin share guarantee with respect to the true utilities $\{u_i\}_{i\in N}$.

Next, we provide an example that shows this bound is tight. Assume that there are 4 courses, $a,b,c$ and $d$ with capacities $q_a =  q_b = q_c = q_d = 1$. Moreover, assume that there are 2 students with utility functions:
    \begin{itemize}
        \item $u_1 ( \{ a \} ) = 1,  u_1 ( \{ b \} ) =  0.5 + \ep , u_1 ( \{ c \} ) = 0.5 - \ep, u_1 ( \{ d \} ) = 0   $ and $u_1 ( \{ b,c \} ) = u_1 ( \{ b,d \} ) = u_1 ( \{ c,d \} ) = 0.5 + \ep$.
        \item $u_2( \{ a  \}) = 0.8 , u_2 ( \{ a, b\}) = u_2 ( \{ a, d \}  ) = 0.9 , u_2 ( \{ a, b, d \}  ) = 1 $ and $0$ for any other schedule.
    \end{itemize}
Furthermore, assume that the learned utility functions of the $2$ students are given as follows: 
    \begin{itemize}
        \item $\hat{u}_1( \{ a  \}) = 1, \hat{u}_1( \{ b \})  = \hat{u}_1( \{ c \}) = \hat{u}_1( \{ b,c   \}) = \hat{u}_1( \{ b,d  \}) = \hat{u}_1( \{ c,d \}) = 0.5$ and 0 for any other schedule.
        \item $\hat{u}_2(  \cdot ) = u_2 (\cdot) $. 
    \end{itemize}
Then,  a $(0,0.25)$-A-CEEI ($ \beta = 0.25 \le \frac{1 - \delta n}{ n (1 + \delta ) } \overset{ \delta = 0 }{=}  \frac{1}{2}$) is given by: 
    \begin{itemize}
        \item $ {a}^*_1  = \{  c \}, {a}^*_2 = \{ a, b, d  \}  $
        \item ${b}_1 = 1 , {b}_2 = 1 + \beta = 1.25$
        \item ${p}^*_a = 1.1,  {p}^*_b = {p}^*_c = 0.15  , {p}^*_d = 0$.
    \end{itemize}
    
    A maximin $(n+1)$-split for student $1$ w.r.t. her true utility function is  $ \left \{ \{a \},  \{ b \} , \{ c , d \}   \right \} $ and her true utility for her $(n+1)$-maximin share is $0.5 + \ep$. Hence, the utility of student $1$ with respect to her true utility function is exactly $2 \ep$ less than her $(n+1)$-maximin share of the endowment: 
    \begin{align}
        u_1 ({a}^*_1) &= 0.5 -  \ep  \nonumber  \\ 
        &= u_1  \left ( \maximinshare[1]{n+1} \right ) - 2 \ep .
    \end{align}
    \item 
    For any $i \in \studentset{} $,  from the definition of the $(\alpha, \beta)$-A-CEEI, it holds that
\begin{equation}
    a^*_i = \argmax_{x\in \left \{ x' \in \mathcal{X} :  p^* \cdot  x' \le b_i   \right \}} \hat{u}_i(x).
\end{equation}

Therefore,
\begin{align*}
    u_i(a_i^* ) + \ep & \ge \hat{u}_i(a^*_i) \\ 
    & = \hat{u}_i \left(\argmax_{x\in \left \{ x' \in \mathcal{X} :  p^* \cdot  x' \le b_i   \right \}} \hat{u}_i(x)\right)  \\
    & \ge \hat{u}_i \left(\argmax_{x\in \left \{ x' \in \mathcal{X} :  p^* \cdot  x' \le b_i   \right \}} u_i(x)\right)  \\
    & \ge u_i \left(\argmax_{x\in \left \{ x' \in \mathcal{X} :  p^* \cdot  x' \le b_i   \right \}} u_i(x)\right)- \ep
\end{align*}
Hence, we have that
\begin{equation}
    u_i(a_i^* )  \ge u_i \left(\argmax_{x\in \left \{ x' \in \mathcal{X} :  p^* \cdot  x' \le b_i   \right \}} u_i(x)\right) - 2 \ep.
\end{equation}

\item Assume that $a^*$ is not $2\ep$- Pareto efficient. This implies that there exists an $a^{\prime}\in \mathcal{F}$, which is a $2 \ep$-Pareto improvement of $a^*$ in economy $ \left (N, M, (q^*_j)_{j = 1}^m, ({u}_i)_{i=1}^n  \right )$, i.e., the utility of every student $i\in N$ in allocation $a^{\prime}$ is more than $2 \ep$ higher than her utility in $a^*$. By \cref{itm:3propA1}, for every $i \in N$ it holds that $p^* \cdot a^{\prime}_i > p^* \cdot a_i^*$. This implies that 
$ \sum_i p^* \cdot a^{\prime}_i > \sum_i  p^* \cdot a_i^*$. This is a  contradiction since prices are non-negative and $a^*$ allocates all units of positive-priced goods.
\end{enumerate}
\Halmos
\endproof

Intuitively, \cref{itm:4propA1} of \Cref{prop:app:theoretical_results} says that if we fix the prices of  A-CEEI, the students cannot aggregate their budgets and then spend them in such a way that they are all benefited by more than $2 \ep$, w.r.t. their true utility function. 

Using the wording of \citet{budish2011aceei}, an implication of \cref{itm:4propA1} of \Cref{prop:app:theoretical_results} is that the allocation induced by an A-CEEI on the $\ep$-approximate utility functions will not admit any $2 \ep$-Pareto-improving trades among the students, but may admit Pareto-improving trades among sets of students and the administrator.

\section{Integrating Comparison Queries into MVNNs} \label{A_scully}
In this section, we detail the integration of GUI reports (regression data) and CQs (classification data) into the training of MVNNs.

\begin{algorithm}[t!]
\caption{Mixed training for regression model $\mathcal{M}$}
\label{alg_mixed_training}
\textbf{Input}: $\{X_{reg}, y_{reg}\}$, $\{X_{class}, y_{class}\}$ \\
\textbf{Parameters}: epochs $t_{reg}$, learning rate $\eta_{reg}$, regularization parameter $\lambda_{reg}$, epochs $t_{class}$, learning rate $\eta_{class}$, regularization parameter $\lambda_{class}$ \\
\textbf{Output}: Parameters of trained ML model $\mathcal{M}$
\begin{algorithmic}[1] 
\STATE $\theta_0 \leftarrow{}$ initialize parameters of the ML model $\mathcal{M}$ \label{alg_init}
\FOR{$i = 1$ to $t_{reg}$} 
    \STATE $loss_{reg} \leftarrow 0$
    \FOR{each $(x,y)$ in $\{X_{reg}, y_{reg}\}$}
        \STATE $\hat{y} \leftarrow \mathcal{M}^{\theta_{i-1}}(x)$
        \STATE $loss_{reg} \leftarrow loss_{reg} + l_{reg}(y, \hat{y}) + \lambda_{reg} \sum{\theta_{i-1}^2}$ \label{mixed_training:reg_loss}
    \ENDFOR
    \STATE $\theta_i \leftarrow \operatorname{ADAM}(\theta_{i-1}, loss_{reg}, \eta_{reg})$ \label{alg_mixed_training_adam1}
\ENDFOR \label{alg_reg_end}
\FOR{$i = t_{reg} + 1$ to $t_{reg} + t_{class}$} \label{alg_class_start}
    \STATE $loss_{class} \leftarrow 0$
    \FOR{each $((x_1, x_2),y)$ in $\{X_{class}, y_{class}\}$}
        \STATE $\hat{y}_1 \leftarrow \mathcal{M}^{\theta_{i-1}}(x_1)$
        \STATE $\hat{y}_2 \leftarrow \mathcal{M}^{\theta_{i-1}}(x_2)$
        \STATE $\hat{y} \leftarrow \frac{1}{1 + e^{-(\hat{y}_1 - \hat{y}_2)}}$ \label{mixed_training:probability}
        \STATE $loss_{class} \leftarrow loss_{class} + l_{class}(y, \hat{y}) + \lambda_{class} \sum{\theta_{i-1}^2}$ \label{mixed_training:classification_loss}
    \ENDFOR
    \STATE $\theta_i \leftarrow \operatorname{ADAM}(\theta_{i-1}, loss_{class}, \eta_{class})$ \label{alg_mixed_training_adam2}
\ENDFOR \label{alg_class_end}
\STATE \textbf{return} $\theta_{t_{reg} + t_{class}}$ \label{alg_return}
\end{algorithmic}
\end{algorithm}

The high-level idea of \Cref{alg_mixed_training} is to first train the regression model on the GUI reports of the student with high regularization, and then train on her CQ answers with a lower one. 
When training on a data point from the regression dataset, a standard regression loss is used, comparing the real-valued output of $\mathcal{M}(\cdot)$ against the inferred value of the student for that schedule, based on her GUI reports (Line~\ref{mixed_training:reg_loss}). 
When training on a CQ, we use the sigmoid function $f(x) = \frac{1}{1+e^{-x}}$ to convert the real-valued outputs of $\mathcal{M}(\cdot)$ for the two schedules $x_1, x_2$ involved in the CQ to the $[0,1]$-interval, where the value of $\frac{1}{1 + e^{-(\hat{y}_1 - \hat{y}_2)}}$ is interpreted as the predicted probability that schedule $x_1$ is more valuable than schedule $x_2$ for the student (Line \ref{mixed_training:probability}). 
This predicted probability is compared against the true probability of the student preferring schedule $x_1$ which is $1$, if that was the answer she gave to the corresponding CQ, and $0$ otherwise (Line \ref{mixed_training:classification_loss}). 
For the ML model we use $\operatorname{MVNNs}$ \citep{weissteiner2022monotone} for the reasons outlined in \Cref{sec:MachineLearningInstantiation}.
For the regression and classification losses, we use mean absolute error (MAE) and binary cross entropy loss (BCE) respectively, both with $l_2$ regularization.
In both cases, the ADAM algorithm \citep{kingma2017adam} is then used to update the parameters of the MVNN (Lines \ref{alg_mixed_training_adam1} and \ref{alg_mixed_training_adam2}). 
Our algorithm terminates after $t_{reg} + t_{class}$ epochs and returns the selected parameters (Line~\ref{alg_return}).

\section{Student Preference Generator} \label{A_SPG}

In this section, we describe in detail our proposed student preference generator.

Our design goal for the student preference generator is two-fold: First, it should be realistic. That is, the preferences combined with the modeling of students' mistakes should closely approximate the metrics on reported preferences given in 
\cite{budish2022can}. Second, we should be able to formulate the generated preferences in a succinct MIP as this is required by all stages of CM. 

We build our preference generator based on the following two assumptions. 
First, a single student's value for a course schedule depends on their value for every single course and the complementarities/substitutabilities between the courses within a schedule.
Second, the high-valued courses amongst students are correlated and fall under the category of popular courses. This is because students have the same reasoning for considering a course to be high-valued. These reasons could be a popular topic, a good instructor, an interest in the same minors, etc. 

Thus, a student's utility for a schedule of courses depends on the following two things: 
\begin{itemize}[align=left, leftmargin=*]
    \item[\textbf{D1}] The \emph{individual value} of each course in that schedule. 
    \item[\textbf{D2}] The \emph{complementarities and substitutabilities} between courses in that schedule. 
\end{itemize} 
    
Recall, that $\courseset{} \neq \emptyset$ and and $\studentset{} \neq \emptyset$ denote the set of all courses and the set of students, respectively. To imitate the effect of students having correlated high-valued courses, we choose $\courseset{p} \subseteq \courseset{}$ to represent the \emph{popular courses} amongst students and  $\courseset{np} = \courseset{} \setminus  \courseset{p}$ to represent the \emph{non-popular courses}. We also assume that the set of high-valued courses for each student is a subset of the popular courses. We call these the student's \emph{favorite courses}. For each student, their \emph{base value} for each course in their favorite and non-favorite courses is drawn independently from $U(l_p, u_p)$ and $U(l_{np}, u_{np})$, respectively, where $U(a, b)$ denotes the uniform distribution on the interval $(a,b)$, $0 \leq l_{np} \leq l_p$, and $0 \leq u_{np} \leq u_p$.

To simulate the substitutabilities and complementarities between courses in students' preferences, we build on the prior work on preference generators for spectrum auctions (i.e., the Local Synergy Value Model (LSVM) \citep{lsvm} and the Global Synergy Value Model (GSVM) \citep{gsvm}).

Specifically, we assume that the courses are arranged in a latent space where the complementarity and substitutability relation between courses is a function of their distance to each other, i.e.,
\begin{itemize}
    \item If two courses are ``too close'', then the contents of the courses have too much overlap and hence they are substitutes for each other.
    \item If they are ``close'' but not ``too close'',  they are complements.
    \item If they are ``far away'', they are neither complements nor substitutes to each other.
\end{itemize}

\begin{figure}[t!]
    \centering
    \begin{tikzpicture}
\draw[help lines, color=gray!30, dashed] (0,0) grid (6.9,5.9);

\draw[->] (0,0)--(7,0) node[right]{$x$};
\draw[->] (0,0)--(0,6) node[above]{$y$};

\foreach \x in  {1,2,3,4,5,6} 
\draw[shift={(\x,0)},color=black] (0pt,3pt) -- (0pt,-3pt);

\foreach \x in {1,2,3,4,5,6} 
\draw[shift={(\x,0)},color=black] (0pt,0pt) -- (0pt,-3pt) node[below] 
{$\x$};

\foreach \x in  {1,2,3,4,5} 
\draw[shift={(0,\x)},color=black] (-3pt,0pt) -- (3pt,0pt);
\foreach \x in {1,2,3,4,5} 
\draw[shift={(0,\x)},color=black] (0pt,0pt) -- (-3pt,0pt) node[left] 
{$\x$};

\node[circle,inner sep=2pt,fill=gray,label=:{$1$}] at (1,1) {};
\node[circle,inner sep=2pt,fill=red,label=:{$2$}] at (2,1) {};
\node[circle,inner sep=2pt,fill=cyan,label=:{$3$}] at (3,1) {};
\node[circle,inner sep=2pt,fill=red,label=:{$4$}] at (4,1) {};
\node[circle,inner sep=2pt,fill=gray,label=:{$5$}] at (5,1) {};
\node[circle,inner sep=2pt,fill=gray,label=:{$6$}] at (6,1) {};

\node[circle,inner sep=2pt,fill=gray,label=:{$7$}] at (1,2) {};
\node[star,star points=5, ultra thick,inner sep=2pt,fill=cyan,label=:{$8$}] at (2,2) {};
\node[star,star points=5, ultra thick, inner sep=2pt,fill=purple,label=:{${9}$}] at (3,2) {};
\node[circle,inner sep=2pt,fill=cyan,label=:{${10}$}] at (4,2) {};
\node[circle,inner sep=2pt,fill=gray,label=:{${11}$}] at (5,2) {};
\node[circle,inner sep=2pt,fill=gray,label=:{${12}$}] at (6,2) {};

\node[circle,inner sep=2pt,fill=gray,label=:{${13}$}] at (1,3) {};
\node[circle,inner sep=2pt,fill=red,label=:{${14}$}] at (2,3) {};
\node[circle,inner sep=2pt,fill=cyan,label=:{${15}$}] at (3,3) {};
\node[circle,inner sep=2pt,fill=red,label=:{${16}$}] at (4,3) {};
\node[circle,inner sep=2pt,fill=gray,label=:{${17}$}] at (5,3) {};
\node[circle,inner sep=2pt,fill=gray,label=:{${18}$}] at (6,3) {};

\node[circle,inner sep=2pt,fill=gray,label=:{${19}$}] at (1,4) {};
\node[circle,inner sep=2pt,fill=gray,label=:{${20}$}] at (2,4) {};
\node[star,star points=5, ultra thick,inner sep=2pt,fill=gray,label=:{${21}$}] at (3,4) {};
\node[circle,inner sep=2pt,fill=gray,label=:{${22}$}] at (4,4) {};
\node[circle,inner sep=2pt,fill=gray,label=:{${23}$}] at (5,4) {};
\node[circle,inner sep=2pt,fill=gray,label=:{${24}$}] at (6,4) {};

\node[circle,inner sep=2pt,fill=gray,label=:{${25}$}] at (1,5) {};
\node[circle,inner sep=2pt,fill=gray,label=:{${26}$}] at (2,5) {};
\node[circle,inner sep=2pt,fill=gray,label=:{${27}$}] at (3,5) {};
\node[circle,inner sep=2pt,fill=gray,label=:{${28}$}] at (4,5) {};
\node[star,star points=5, ultra thick, inner sep=2pt,fill=gray,label=:{${29}$}] at (5,5) {};
\node[circle,inner sep=2pt,fill=gray,label=:{${30}$}] at (6,5) {};

\end{tikzpicture}
    \caption{A latent space with 30 courses.}
    \label{fig_appendix_latent_space}
\end{figure}
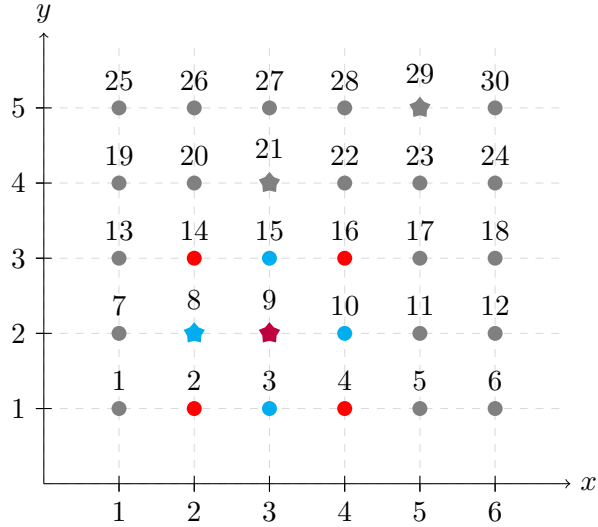

More precisely, we assume that $\courseset{} \subset \mathbb{N}^2$ and that the set of complementarities and substitutabilities are centered around a subset of the popular courses $\courseset{p}$, called centers, i.e., $\courseset{c} \subseteq \courseset{p}$. Then the set of substitutabilities and complementarities around a center point $c \in \courseset{c}$ are defined using the $L_1$ and $L_\infty$ distances as follows:
\begin{itemize}
\item \emph{Set of substitutabilities:}
\begin{equation}
    S_c = \{m \in M | L_1(m, c) \leq r_s\} \\
\end{equation}
\item \emph{Set of complementarities:}
\begin{equation}
    C_c = \{m \in M | L_\infty(m, c) \leq r_c\} \setminus S_c \cup \{c\}
\end{equation}
\end{itemize}

where $r_s, r_c \in \mathbb{N}_{>0}$ are the radius of the set of substitutabilities and the radius of the set of complementarities, respectively. The larger these radii, the more complementarities/substitutabilities are present in the students' preferences. We illustrate an example of this latent space in \Cref{fig_appendix_latent_space}.

\Cref{fig_appendix_latent_space}, depicts a latent space of height 5 and width 6 with the courses $\courseset{} = \{1, \dots, 30\}$, where we assume that both radii are equal to $1$, i.e., $r_s=r_c=1$ and the  set of popular courses is given as $\courseset{p} = \{8, 9, 21, 29\}$. Then, the set of substitutabilities and complementarities defined by the center $\color{purple} 9$ are given by $\color{cyan} S_c =\{9, 3, 8, 10, 15\}$ and $\color{red} C_c =\{9, 2, 4, 14, 16\}$, respectively.

To model the impact of complementarities/substitutabilities on students' preferences, we use piece-wise constant step functions $\psi_c:\{0,\ldots,|C_c|\}\to\mathbb{R}_+$ and $\xi_c:\{0,\ldots,|S_c|\}\to\mathbb{R}_-$ to represent the multiplicative bonuses and penalty factors for a schedule of courses if its items belong to the set of complementarities $C_c$ or the set of substitutabilities $S_c$. We further assume that $\psi_c$ is monotonically non-decreasing and $\xi_c$ is monotonically non-increasing. For example, given a center $c$, $\psi_c(3)$ represents a student's multiplicative bonus for taking three courses from the set of complementarities $C_c$.

With all these building blocks, the utility of a student $ i \in \studentset{}$ for a schedule $x \in \{0,1\}^m$ is defined as: 
\begin{align}
    u_{i}(x) = & \sum_{j \in x} u_{i}(\{j\}) \label{utility:def} \\ 
            +& \sum_{c \in \courseset{c}} \sum_{j \in C_c}  \psi_c (\tau_c(x)) u_{i}(\{j\}) \nonumber \\
            +& \sum_{c \in \courseset{c}} \sum_{j \in S_c}  \xi_c(\kappa_c(x))  u_{i}(\{j\}), \nonumber
\end{align} 
where $u_{i}(\{j\})$ represents the base values of student $i \in \studentset{}$ for course $j$ and  $\tau_c(x)\coloneqq|\{j\in x:j\in C_c\}|$ is the number of courses that belong both to the schedule $x$ and the set of complementarities at center $c$, i.e., $C_c$. Analogously, $\kappa_c(x)\coloneqq|\{j\in x:j\in S_c\}|$ is the number of courses that belong both to the schedule $x$ and the set of substitutabilities at center $c$, i.e., $S_c$. 

This preference generator enables to efficiently encode students' complementarities/substitabilities into a MIP (see electronic companion~\ref{A_MIP}). Thus, it allows us to conduct experiments and compare MLCM with the original CM mechanism in a synthetic setting where we have access to the optimal allocations w.r.t. students' true preferences.

\section{MIP for the Student Preference Generator}\label{A_MIP}
In this section, we provide a succinct MIP formulation for calculating the optimal allocation for a single student $i$ given her preferences generated by our proposed student preference generator from \Cref{A_SPG} and a price vector $p$. This enables us to compute, for any price vector, the allocation under the students' true preferences. First, we define the individual student optimization problem.

\begin{definition}[Individual student utility optimization problem]\label{def:single_student_optimization_problem}
For student $i \in N$, utility function $u_i:\mathcal{
X}\to \mathbb{R}_+$ defined in \Cref{utility:def}, price vector $p\in \mathbb{R}_+^m$ and budget $b_i\in [1, 1+\beta],\, \beta>0$, we define the individual student utility optimization problem as 
\begin{equation}\label{eq:single_student}
\argmax_{\{x\in \Psi_i:\, x\cdot p\le b_i\}}u_i(x)
\end{equation}
In words, a solution of \Cref{eq:single_student} maximizes student $i$'s true utility (as defined by our preference generator in \Cref{utility:def}) amongst all her permissible course schedules $\Psi_i$ that are affordable at prices $p$ when given a budget $b_i$.
\end{definition}

Next, we provide in \Cref{single_student_MIP} the equivalent MIP.

\begin{proposition}[Individual student  MIP]\label{single_student_MIP}
Using the notation from electronic companion~\ref{A_SPG}, and \Cref{def:single_student_optimization_problem} the individual student utility optimization problem defined in \Cref{eq:single_student} can be equivalently formulated as the following MIP:
\begin{alignat}{3}
     \argmax_{x\in \Psi_i} &\sum_{c \in \courseset{c}} \sum_{j \in C_c}  \sum_{\tau = 1 }^{\bar{\tau}} \psi_c (\tau) \cdot G_{c,j,\tau} \cdot u_i(\{j\}) \label{MIP:obj1} \\ \
    & +\sum_{c \in \courseset{c}} \sum_{j \in S_c}  \sum_{\tau = 1 }^{\bar{\tau}} \xi_c (\tau) \cdot J_{c,j,\tau} \cdot u_i(\{j\})\label{MIP:obj2}\\ 
    &  + \sum_{j \in [m]} x_j \cdot u_i(\{j\}) \label{MIP:obj3}
\end{alignat}
\begin{alignat}{3}
     \text{s.t.} &  \nonumber \\
    &  \sum_{\tau = 1}^{\bar{\tau}} \tau \cdot G_{c,j',\tau} \le \sum_{j \in C_c} x_j,  \; \forall \; j' \in C_c, \forall \;c \in \courseset{c} \label{constr:egsvm1} \\
    & \sum_{\tau = 1 }^{\bar{\tau}} G_{c, j, \tau} \le 1, \; \forall \; j \in C_c, \forall \;c \in M_c  \label{constr:egsvm2} \\
    & \sum_{\tau = 1 }^{\bar{\tau}} \tau \cdot J_{c, j', \tau} \ge \sum_{j \in S_c} x_j, \; \forall \; j' \in S_c, \forall \;c \in M_c \label{constr:egsvm3} \\ 
    &  \sum_{\tau = 1 }^{\bar{\tau}} J_{c, j, \tau} \le 1, \; \forall \; j \in S_c, \forall \; c \in M_c \label{constr:egsvmf} \\
    & \sum_{j \in [m]} p_j \cdot x_j \le b_i \label{constr:prices}
\end{alignat}

\noindent where
\begin{itemize}
    \item $\bar{\tau}$ is the maximum number of courses in a schedule.
    \item $\courseset{c}$ is the set of centers.
    \item $C_c$ is the set of complementarities centered around $c$.
    \item $S_c$ is the set of substitutabilities centered around $c$.
    \item $G_{c,j,\tau}$ denotes the binary variable, whether or not the course indexed by $j$, which belongs to the set of complementarities $C_c$ is in $x$, while in total  $\tau$ courses from that set are included in $x$.
    \item $J_{c,j,\tau}$ denotes the binary variable, whether or not the course indexed by $j$, which belongs to the set of substitutabilities $S_c$ is in $x$, while in total $\tau$ courses from that set are included in $x$.
    \item $b_i$ is the budget of student $i$.
\end{itemize}
\end{proposition}

\proof{Proof.}
The first thing we need to show is that the problem is actually a MIP, i.e., all constraints are linear and all variables are either linear or integer. First, note that all variables are either linear or integer by their definition. Moreover, the same is true for all constraints shown, other than the $x \in \Psi_i$ constraint. Next, we show that this constraint can be encoded in a linear way. 

There are 2 reasons a schedule $x$ would not be permissible for a student $i$. The first one would be if it contained any courses that the student is not eligible for. Let $E_i$ be the indexes of those courses for student $i$. The second constraint would be if it contained more than one courses that take place at the same time. Let $H$ be the set of all course hours (also known as \textit{time slots}, e.g. Wednesdays $10$-$11$am), and $T_h$ the set of indexes of courses that take place at time $h\in H$. Then, $x \in \Psi_i$ is equivalent to the following two \emph{linear} constraints:
\begin{align}
    \sum_{j \in E_i} x_j &\le 0,  \label{constr:non_permissible} \\ 
    \sum_{j \in T_h} x_j &\le 1 \; \forall h \in H. \label{constr:same_hour} 
\end{align}
Constraint~\eqref{constr:non_permissible} enforces that student $i$ cannot enroll in courses that are not permissible for her, while constraint~\eqref{constr:same_hour} enforces that she cannot take more than one courses that take place in the same time slot. Now that we have proven that the problem defined above is a valid MIP respecting all constraints, we need to prove that its solution is a utility-maximizing  course schedule for student $i\in N$, i.e, is equivalent to \Cref{eq:single_student}. 

If all binary variables $G_{c,j,\tau}$ and $J_{c,j,\tau}$ are set to the correct value for a given course schedule $x$, then \Cref{MIP:obj1,MIP:obj2,MIP:obj3} encode the utility of course schedule $x$ for a given student $i$, as the only additive terms not zeroed-out are those in \Cref{utility:def}. Thus, to prove the correctness of the MIP, it suffices to prove that those binary variables are correctly set for any valid $x$ and that the budget constraint is satisfied.

Constraint~\eqref{constr:egsvmf} enforces that for any $(c,j) \in M_c \times S_c$ pair at most one of the binary variables $J_{c,j,\tau}$ will be set to $1$. This combined with constraint~\eqref{constr:egsvm3} enforce that if $\tau$ elements from set $S_c$ are included in $x$, then for the $J_{c,j,\tau'}$ variable it must hold that $\tau' \ge \tau$. But since we have a maximization problem and the function $\xi_c(\cdot)$ is monotonically non-increasing, the smallest value of $\tau'$ that satisfies the above constraint will be set to one, thus $\tau' = \tau$. Finally, note that if $\xi_c(\tau') = \xi_c(\tau)$ for some $\tau' \ge \tau$, i.e. $\xi_c(\cdot)$ is at a constant part of its support, it can be the case that the maximizer sets $J_{c,j,\tau'}=1$ instead of the $J_{c,j,\tau'}=1$. However, since $\xi_c(\tau') = \xi_c(\tau)$, this is not a problem, as the objective value is the same in both cases.

Similarly, constraint~\eqref{constr:egsvm2} enforces that for any $(c,j) \in M_c \times C_c$ pair, at most one of the $G_{c,j,\tau'}$ variables will be set to one. Constraint~\eqref{constr:egsvm1} enforces that if $\tau$ of the elements of $C_c$ are included in $x$, then it must hold that $\tau' \le \tau$. But since this is a maximization problem and $\psi_c(\cdot)$ is monotonically non-decreasing, the solver will set exactly the variable $G_{c,j, \tau}$ to one where $\tau' = \tau$, as this maximizes the objective value out of all feasible solutions. If $\psi_c(\tau) = \psi(\tau')$, $G_{c,j_\tau'}$ could be set to one instead of $G_{c,j,\tau}$, but the same argument as above again applies. 

Finally, constraint~\eqref{constr:prices} enforces that a student cannot take any schedule with a price larger than her budget. 
\Halmos
\endproof

\section{Mistake Profile Calibration} 
\label{A_mistake_calibration}
In this section, we present details on our mistake profile calibration.

\paragraph{Results} \Cref{tab:MistakeCalibration} shows detailed results of our calibration procedure. For each row, we multiply the default reporting mistake parameters $f_{b}, f_{a}, \sigma_{b}, \sigma_{a}$ as defined in \Cref{subsection:calibrating} by the common constant $\gamma$. We see that our mistake calibration for the default common mistake constant $\gamma=1$ (i.e, the rows marked in grey) closely matches the metrics reported in \citet{budish2022can}.
\begin{table}[h!]
\robustify\bfseries
\centering
\begin{sc}
\resizebox{1\textwidth}{!}{
\setlength\tabcolsep{4pt}

\begin{tabular}{llrrrrrrrrr}
\toprule
\multicolumn{2}{c}{\textbf{Setting}}&  \multicolumn{3}{c}{\textbf{\#Courses with value}} & \multicolumn{4}{c}{\textbf{\#Adjustments}} & \multicolumn{1}{c}{\textbf{Accuracy}} & \multicolumn{1}{c}{\textbf{If disagreement}}\\
\cmidrule(l{2pt}r{2pt}){1-2}
\cmidrule(l{2pt}r{2pt}){3-5}
\cmidrule(l{2pt}r{2pt}){6-9} 
\#Pop&  $\gamma$ &                          \multicolumn{1}{c}{$>0$} &                           \multicolumn{1}{c}{$(0,50) $} &              \multicolumn{1}{c}{$[50,100]$} &                      \multicolumn{1}{c}{Mean} & \multicolumn{1}{c}{Median} &    \multicolumn{1}{c}{Min} & \multicolumn{1}{c}{Max} & \multicolumn{1}{c}{\textbf{in \%}}  &                    \multicolumn{1}{c}{\textbf{utility difference in \%}} \\
\cmidrule(l{2pt}r{2pt}){1-1}
\cmidrule(l{2pt}r{2pt}){2-2}
\cmidrule(l{2pt}r{2pt}){3-5}
\cmidrule(l{2pt}r{2pt}){6-9}
\cmidrule(l{2pt}r{2pt}){10-10}
\cmidrule(l{2pt}r{2pt}){11-11}
9& $0.5$ &  18.75 $\pm$ \scriptsize 0.02 &  14.35 $\pm$ \scriptsize 0.07 &  4.40 $\pm$ \scriptsize 0.07 &         3.72 $\pm$ \scriptsize 0.12 &         3.0 &   0 &         17 &         93 $\pm$ \scriptsize 1 &          -3.81 $\pm$ \scriptsize 0.41 \\
9& $0.75$ &  15.62 $\pm$ \scriptsize 0.02 &  10.31 $\pm$ \scriptsize 0.08 &  5.30 $\pm$ \scriptsize 0.08 &         2.06 $\pm$ \scriptsize 0.09 &         1.0 &   0 &         12 &         87 $\pm$ \scriptsize 1 &          -9.41 $\pm$ \scriptsize 0.62 \\
9& $0.9$ &  13.75 $\pm$ \scriptsize 0.02 &   7.91 $\pm$ \scriptsize 0.09 &  5.84 $\pm$ \scriptsize 0.08 &         1.37 $\pm$ \scriptsize 0.07 &         1.0 &   0 &   11 &   84 $\pm$ \scriptsize 1 &   -13.10 $\pm$ \scriptsize 0.72 \\
\rowcolor{gray!40}
9& $1.0$ &  12.49 $\pm$ \scriptsize 0.02 &   6.32 $\pm$ \scriptsize 0.09 &  6.17 $\pm$ \scriptsize 0.09 &   0.98 $\pm$ \scriptsize 0.05 &         1.0 &   0 &   10 &   81 $\pm$ \scriptsize 1 &         -15.52 $\pm$ \scriptsize 0.76 \\    
9& $1.1$ &  11.20 $\pm$ \scriptsize 0.02 &   4.71 $\pm$ \scriptsize 0.09 &  6.49 $\pm$ \scriptsize 0.09 &         0.69 $\pm$ \scriptsize 0.04 &   0.0 &   0 &          7 &         78 $\pm$ \scriptsize 1 &         -18.11 $\pm$ \scriptsize 0.77 \\
9& $1.25$ &   9.39 $\pm$ \scriptsize 0.02 &   2.58 $\pm$ \scriptsize 0.08 &  6.81 $\pm$ \scriptsize 0.08 &         0.37 $\pm$ \scriptsize 0.03 &   0.0 &   0 &          6 &         75 $\pm$ \scriptsize 1 &         -21.60 $\pm$ \scriptsize 0.80 \\
9& $1.5$ &  6.20 $\pm$ \scriptsize 0.02 &  0.36 $\pm$ \scriptsize 0.04 &  5.84 $\pm$ \scriptsize 0.04 &  0.10 $\pm$ \scriptsize 0.01 &    0.0 &      0 &      3 &  66 $\pm$ \scriptsize 1 &  -26.78 $\pm$ \scriptsize 0.78 \\
\midrule
6& $0.5$ &  18.75 $\pm$ \scriptsize 0.02 &  13.71 $\pm$ \scriptsize 0.06 &  5.04 $\pm$ \scriptsize 0.06 &  3.83 $\pm$ \scriptsize 0.12 &    3.0 &      0 &     19 &  93 $\pm$ \scriptsize 1 &          -3.11 $\pm$ \scriptsize 0.37 \\
6& $0.75$ &  15.62 $\pm$ \scriptsize 0.02 &  10.01 $\pm$ \scriptsize 0.07 &  5.61 $\pm$ \scriptsize 0.07 &  2.17 $\pm$ \scriptsize 0.09 &    2.0 &      0 &     13 &  87 $\pm$ \scriptsize 1 &          -8.02 $\pm$ \scriptsize 0.57 \\
6& $0.9$ &  13.75 $\pm$ \scriptsize 0.02 &   7.80 $\pm$ \scriptsize 0.08 &  5.95 $\pm$ \scriptsize 0.07 &  1.47 $\pm$ \scriptsize 0.07 &    1.0 &      0 &     10 &  85 $\pm$ \scriptsize 1 &         -11.42 $\pm$ \scriptsize 0.67 \\
\rowcolor{gray!40}
6& $1.0$ &  12.49 $\pm$ \scriptsize 0.02 &   6.33 $\pm$ \scriptsize 0.08 &  6.16 $\pm$ \scriptsize 0.08 &  1.08 $\pm$ \scriptsize 0.06 &    1.0 &      0 &     10 &  83 $\pm$ \scriptsize 1 &   -13.34 $\pm$ \scriptsize 0.71 \\  
6& $1.1$ &  11.20 $\pm$ \scriptsize 0.02 &   4.82 $\pm$ \scriptsize 0.08 &  6.38 $\pm$ \scriptsize 0.08 &  0.73 $\pm$ \scriptsize 0.04 &    0.0 &      0 &     10 &  80 $\pm$ \scriptsize 1 &         -15.48 $\pm$ \scriptsize 0.75 \\
6& $1.25$ &   9.39 $\pm$ \scriptsize 0.02 &   2.72 $\pm$ \scriptsize 0.08 &  6.67 $\pm$ \scriptsize 0.08 &  0.41 $\pm$ \scriptsize 0.03 &    0.0 &      0 &     10 &  76 $\pm$ \scriptsize 1 &         -19.12 $\pm$ \scriptsize 0.77 \\
6& $1.5$ &  6.20 $\pm$ \scriptsize 0.02 &  0.38 $\pm$ \scriptsize 0.04 &  5.82 $\pm$ \scriptsize 0.04 &  0.11 $\pm$ \scriptsize 0.02 &    0.0 &      0 &      3 &  68 $\pm$ \scriptsize 1 &  -23.83 $\pm$ \scriptsize 0.78 \\
\midrule
\multicolumn{2}{l}{\cite{budish2022can}}  &   12.45\hphantom{ $\pm$ \scriptsize 0.02}  &  6.17\hphantom{ $\pm$ \scriptsize 0.02}  &  6.27\hphantom{ $\pm$ \scriptsize 0.02} & 1.08\hphantom{ $\pm$ \scriptsize 0.02}  & 0 &  0 &         10 &         0.84\hphantom{ $\pm$ \scriptsize 0.02}  &          -13.35 $\pm$ \scriptsize 0.41 \\
\bottomrule
\end{tabular}
}
        \vskip 0.1cm
\caption{Mistake profile calibration experiment for several settings defined by the number of popular courses (\textsc{\#Pop}) and the common mistake constant $\gamma$ compared to the experimental findings in \cite{budish2022can}. Our two default settings (i.e., $\gamma=1$) are marked in grey. 2,000 students in total. We show the number of courses with reported value in 3 distinct intervals, the mean, median, minimum and maximum number of adjustments in the students' reports, the accuracy of their reports as determined by asking CQs and the median scaled utility difference between the two schedules in a CQ in case of disagreement between the CQ answer and the reported preferences. Shown are average results and a 95\% CI.}
\label{tab:MistakeCalibration}
\end{sc}
\end{table}

\section{Experiment Details}\label{sec:app_Experiment Details}
In this section, we present all details of our experiments from \Cref{Section_efficiency_results}.

\subsection{Compute Infrastructure}\label{subsec:app:compute_infrastructure}
All experiments were conducted on a compute cluster running Debian GNU/Linux 10 with Intel Xeon E5-2650 v4 2.20GHz processors with 24 cores and 128GB RAM and Intel E5 v2 2.80GHz processors with 20 cores and 128GB RAM and Python 3.8.10.

\subsection{Hyperparameter Optimization}\label{subsec:app_hpo}
In this section, we provide details on our exact hyperparameter optimization (HPO) methodology and the ranges that we used. \

We optimized the MVNN hyperparameters (HPs) using a different seed than all other experiments in the paper. Specifically, we first created a simulated course allocation instance with a supply ratio of $1.25$, $9$ popular courses and $100$ students, using our simulation framework as described in \Cref{sec_SPG}.
Then, we calculated an A-CEEI based on the GUI reports of the students, with their mistakes calibrated to match the reports of \citet{budish2022can} as detailed in electronic companion~\ref{A_mistake_calibration}.
For each HP configuration we tested, we trained an MVNN on the same GUI reports of the students as described in \Cref{alg_mixed_training}, and iteratively generated $10$ CQs using our OBIS algorithm as detailed in \Cref{subsec:query_generation}, with the same GUI prices used to estimate the attainability of different schedules.
Finally, for each HP configuration, we trained the MVNN of each student on her GUI reports plus the corresponding CQs of that configuration, and measured its generalization performance on a validation set consisting of $1000$ schedule-true student utility pairs.  
The number of students used to evaluate each configuration was set to 10. 
Finally, we selected the set of HPs that performed the best on this validation set with respect to the coefficient of determination ($R^2$). The full range of HPs tested and the winning configuration are shown in in \Cref{table_HPO_ranges}. 
Note that we performed HPO only for one setting and one distribution of students' preferences, and used those hyperparameters for all other experiments in this paper.
This further highlights the robustness of our approach, and its applicability to a large variety of settings.  

\begin{table}
\centering
\resizebox{0.7\textwidth}{!}{
\begin{tabular}{llc}
\toprule
 \multicolumn{1}{l}{\textbf{Hyperparameter}} &  \multicolumn{1}{l}{\textbf{HPO-Range}} &  \multicolumn{1}{l}{\textbf{Winning Configuration}}\\
\midrule
Hidden Layers            & {[}1,2,3{]}   & 1                      \\
Neurons per Hidden Layer             & {[}10, 20, 30{]} & 20 \\
Cardinal Learning Rate  & (1e-4, 1e-2)      & 1e-2                  \\
Cardinal Epochs            & {[}100, 200, 300{]}  & 100                          \\
Cardinal L2-Regularization & (1e-6, 1e-1)          & 1e-3             \\
Ordinal Learning Rate  & (1e-4, 5e-2)          & 1e-2              \\
Ordinal Epochs            & {[}5, 10, 20, 50, 100{]}  & 10                          \\
Ordinal L2-Regularization & (0, 1e-5)           & 0            \\
\bottomrule
\end{tabular}
}
\vskip 0.1cm
    \caption{HPO ranges and winning configuration}
    \label{table_HPO_ranges}
\end{table}

\section{Welfare Results for Other Settings} \label{A_welfare_results}
In this section, we provide additional welfare results for the following five settings defined by a supply ratio (SR) and a number of popular courses (\textsc{\#Pop}):
\begin{itemize}[topsep=6pt,partopsep=6pt, parsep=6pt]
\item $\text{SR}=1.1$ and $\textsc{\#Pop}=9$ (see \Cref{table_cqs_sr_1.1_pop9}),
\item $\text{SR}=1.50$ and $\textsc{\#Pop}=9$ (see \Cref{table_cqs_sr_1.5_pop9}),
\item $\text{SR}=1.1$ and $\textsc{\#Pop}=6$ (see \Cref{table_cqs_sr_1.1_pop_6}),
\item $\text{SR}=1.25$ and $\textsc{\#Pop}=6$ (see \Cref{table_cqs_sr_1.25_pop_6}),
\item $\text{SR}=1.50$ and $\textsc{\#Pop}=6$ (see \Cref{table_cqs_sr_1.5_pop_6}),
\end{itemize}
where we use the same experimental setup as described in the main paper in \Cref{subsec:experiment_setup}.

From \Cref{table_cqs_sr_1.5_pop9}, \Cref{table_cqs_sr_1.5_pop_6}, and \Cref{table_cqs_sr_1.25_pop_6}, we see that the results (both average and minimum student utility) are better for $\text{SR}=1.5$ and worse for $6$ popular courses but qualitatively similar compared to the results presented in the main paper in \Cref{subsec:experimental_results}.

Namely, for $\text{SR}=1.5$ and $9$ popular courses \textsc{MLCM (10 OBIS CQs)} increases average utility from $79.0\%$ to $87.7\%$, an $8.7$ percentage points increase (which is a $11.0$\% relative increase) and minimum utility from $40.7\%$ to $52.4\%$ , an $11.7$ percentage points increase (which is a $28.7$\% relative increase). At the most extreme setting of $\text{SR}=1.1$ and $6$ popular courses, \textsc{MLCM (10 OBIS CQs)} increases average utility compared to CM from $84.1\%$ to $89.6\%$, an increase of $5.5$ percentage points (a $6.3$\% relative increase) and minimum utility from $50.1\%$ to $58.8\%$ (a $17.4$\% relative increase). For all settings, as the number of CQs increases, the performance of \textsc{MLCM} improves further.

\begin{table}[h!]
	\robustify\bfseries
	\centering
	\begin{sc}
	\resizebox{.8\textwidth}{!}{
	\setlength\tabcolsep{4pt}
\begin{tabular}{llrrr}
\toprule
& & \multicolumn{2}{c}{\textbf{Student Utility (in \%)}}  & \multicolumn{1}{c}{\textbf{Time}}\\
\cmidrule(l{2pt}r{2pt}){3-4}
\textbf{Mechanism} & \textbf{Parametrization} & \multicolumn{1}{c}{\textbf{Average}} & \multicolumn{1}{c}{\textbf{Minimum}}  & \multicolumn{1}{c}{\textbf{(in min)}} \\
\cmidrule(l{2pt}r{2pt}){1-2}
\cmidrule(l{2pt}r{2pt}){3-4}
\cmidrule(l{2pt}r{2pt}){5-5}
CM\textsuperscript{*} & Full Preferences & 100.0 $\pm$ \scriptsize 0.0 & 74.2 $\pm$ \scriptsize 0.4 &  77.4 \\
CM & No Mistakes & 98.9 $\pm$ \scriptsize 0.1 & 73.8 $\pm$ \scriptsize 0.4 &  67.3 \\
\cmidrule(l{2pt}r{2pt}){1-2}
\cmidrule(l{2pt}r{2pt}){3-4}
\cmidrule(l{2pt}r{2pt}){5-5}
RSD & - & 75.1 $\pm$ \scriptsize 0.2 & 22.1 $\pm$ \scriptsize 0.6 &   0.0 \\
\rowcolor{gray!40}
CM & - & 79.5 $\pm$ \scriptsize 0.2 & 41.8 $\pm$ \scriptsize 0.5 &  25.5 \\
\cmidrule(l{2pt}r{2pt}){1-2}
\cmidrule(l{2pt}r{2pt}){3-4}
\cmidrule(l{2pt}r{2pt}){5-5}
MLCM & \hphantom{0}1 OBIS CQ & 80.3 $\pm$ \scriptsize 0.2 & 42.7 $\pm$ \scriptsize 0.5 &  12.0 \\
& \hphantom{0}5 OBIS CQs & 85.2 $\pm$ \scriptsize 0.2 & 49.5 $\pm$ \scriptsize 0.5 &   9.1 \\
& 10 OBIS CQs & 87.5 $\pm$ \scriptsize 0.2 & 51.9 $\pm$ \scriptsize 0.5 &  10.7 \\
& 15 OBIS CQs & 89.0 $\pm$ \scriptsize 0.2 & 53.9 $\pm$ \scriptsize 0.5 &  10.5 \\
& 20 OBIS CQs & 90.1 $\pm$ \scriptsize 0.2 & 54.9 $\pm$ \scriptsize 0.5 &  10.9 \\
\cmidrule(l{2pt}r{2pt}){2-2}
\cmidrule(l{2pt}r{2pt}){3-4}
\cmidrule(l{2pt}r{2pt}){5-5}
& 10 random CQs & 80.2 $\pm$ \scriptsize 0.2 & 43.0 $\pm$ \scriptsize 0.5 &  10.9 \\
& 20 random CQs & 81.2 $\pm$ \scriptsize 0.2 & 44.2 $\pm$ \scriptsize 0.5 &  11.0 \\
\cmidrule(l{2pt}r{2pt}){2-2}
\cmidrule(l{2pt}r{2pt}){3-4}
\cmidrule(l{2pt}r{2pt}){5-5}
& 10 na\"{i}ve CQs & 85.7 $\pm$ \scriptsize 0.2 & 45.8 $\pm$ \scriptsize 0.6 &  10.9 \\
& 20 na\"{i}ve CQs & 87.3 $\pm$ \scriptsize 0.2 & 46.7 $\pm$ \scriptsize 0.5 &  10.9 \\
\cmidrule(l{2pt}r{2pt}){1-2}
\cmidrule(l{2pt}r{2pt}){3-4}
\cmidrule(l{2pt}r{2pt}){5-5}
MLCM-Projected & 10 OBIS CQs & 87.5 $\pm$ \scriptsize 0.2 & 52.4 $\pm$ \scriptsize 0.5 & 100.7 \\
& 20 OBIS CQs & 90.0 $\pm$ \scriptsize 0.2 & 54.8 $\pm$ \scriptsize 0.5 &  95.4 \\
\bottomrule
\end{tabular}
}
\end{sc}
\vskip 0.1cm
\caption{Comparison of different mechanisms for a supply ratio of 1.1, 9 popular courses, and default parameterization for reporting mistakes.  Standard CM is highlighted in grey. Shown are averages over 500 runs. We normalize average and minimum utility by the average utility of \textsc{CM\textsuperscript{*}} and also show their 95\% CIs.}
\label{table_cqs_sr_1.1_pop9}
\vspace{-.6cm}
\end{table}

\begin{table}[h!]
	\robustify\bfseries
	\centering
	\begin{sc}
	\resizebox{.8\textwidth}{!}{
	\setlength\tabcolsep{4pt}

\begin{tabular}{llrrc}
\toprule
& & \multicolumn{2}{c}{\textbf{Student Utility (in \%)}}  & \multicolumn{1}{c}{\textbf{Time}}\\
\cmidrule(l{2pt}r{2pt}){3-4}
\textbf{Mechanism} & \textbf{Parametrization} & \multicolumn{1}{c}{\textbf{Average}} & \multicolumn{1}{c}{\textbf{Minimum}}  & \multicolumn{1}{c}{\textbf{(in min)}} \\
\cmidrule(l{2pt}r{2pt}){1-2}
\cmidrule(l{2pt}r{2pt}){3-4}
\cmidrule(l{2pt}r{2pt}){5-5}
CM\textsuperscript{*} & Full Preferences & 100.0 $\pm$ \scriptsize 0.0 & 71.4 $\pm$ \scriptsize 0.4 & 47.5 \\
CM & No Mistakes & 98.5 $\pm$ \scriptsize 0.1 & 70.9 $\pm$ \scriptsize 0.4 & 41.3 \\
\cmidrule(l{2pt}r{2pt}){1-2}
\cmidrule(l{2pt}r{2pt}){3-4}
\cmidrule(l{2pt}r{2pt}){5-5}
RSD & - & 77.7 $\pm$ \scriptsize 0.2 & 36.8 $\pm$ \scriptsize 0.5 & 0.0 \\
\rowcolor{gray!40}
CM & - & 79.0 $\pm$ \scriptsize 0.2 & 40.7 $\pm$ \scriptsize 0.5 & 21.7 \\
\cmidrule(l{2pt}r{2pt}){1-2}
\cmidrule(l{2pt}r{2pt}){3-4}
\cmidrule(l{2pt}r{2pt}){5-5}
MLCM & \hphantom{0}1 OBIS CQ & 79.9 $\pm$ \scriptsize 0.2 & 41.7 $\pm$ \scriptsize 0.5 & 10.2 \\
& \hphantom{0}5 OBIS CQs & 85.1 $\pm$ \scriptsize 0.2 & 49.1 $\pm$ \scriptsize 0.5 & 9.4 \\
& 10 OBIS CQs & 87.7 $\pm$ \scriptsize 0.2 & 52.4 $\pm$ \scriptsize 0.5 & 11.3 \\
& 15 OBIS CQs & 89.2 $\pm$ \scriptsize 0.2 & 53.9 $\pm$ \scriptsize 0.5 & 8.3 \\
& 20 OBIS CQs & 90.2 $\pm$ \scriptsize 0.2 & 55.0 $\pm$ \scriptsize 0.5 & 10.6 \\
\cmidrule(l{2pt}r{2pt}){2-2}
\cmidrule(l{2pt}r{2pt}){3-4}
\cmidrule(l{2pt}r{2pt}){5-5}
& 10 random CQs & 79.5 $\pm$ \scriptsize 0.2 & 41.6 $\pm$ \scriptsize 0.5 & 9.4 \\
& 20 random CQs & 80.4 $\pm$ \scriptsize 0.2 & 43.4 $\pm$ \scriptsize 0.5 & 10.2 \\
\cmidrule(l{2pt}r{2pt}){2-2}
\cmidrule(l{2pt}r{2pt}){3-4}
\cmidrule(l{2pt}r{2pt}){5-5}
& 10 na\"{i}ve CQs & 87.7 $\pm$ \scriptsize 0.2 & 46.7 $\pm$ \scriptsize 0.6 & 9.0 \\
& 20 na\"{i}ve CQs & 90.2 $\pm$ \scriptsize 0.2 & 47.1 $\pm$ \scriptsize 0.6 & 8.0 \\
\cmidrule(l{2pt}r{2pt}){1-2}
\cmidrule(l{2pt}r{2pt}){3-4}
\cmidrule(l{2pt}r{2pt}){5-5}
MLCM-Projected & 10 OBIS CQs & 87.5 $\pm$ \scriptsize 0.2 & 52.2 $\pm$ \scriptsize 0.5 & 80.4 \\
& 20 OBIS CQs & 90.2 $\pm$ \scriptsize 0.2 & 55.0 $\pm$ \scriptsize 0.5 & 80.3 \\
\bottomrule
\end{tabular}

}
\end{sc}
\vskip 0.1cm
\caption{Comparison of different mechanisms for a supply ratio of 1.5, 9 popular courses, and default parameterization for reporting mistakes.  Standard CM is highlighted in grey. Shown are averages over 500 runs. We normalize average and minimum utility by the average utility of \textsc{CM\textsuperscript{*}} and also show their 95\% CIs.}
\label{table_cqs_sr_1.5_pop9}
\vspace{-.6cm}
\end{table}

\begin{table}[h!]
	\robustify\bfseries
	\centering
	\begin{sc}
    \resizebox{.8\textwidth}{!}{
	\setlength\tabcolsep{4pt}
\begin{tabular}{llrrc}
\toprule
& & \multicolumn{2}{c}{\textbf{Student Utility (in \%)}}  & \multicolumn{1}{c}{\textbf{Time}}\\
\cmidrule(l{2pt}r{2pt}){3-4}
\textbf{Mechanism} & \textbf{Parametrization} & \multicolumn{1}{c}{\textbf{Average}} & \multicolumn{1}{c}{\textbf{Minimum}}  & \multicolumn{1}{c}{\textbf{(in min)}} \\
\cmidrule(l{2pt}r{2pt}){1-2}
\cmidrule(l{2pt}r{2pt}){3-4}
\cmidrule(l{2pt}r{2pt}){5-5}
CM\textsuperscript{*} & Full Preferences & 100.0 $\pm$ \scriptsize 0.0 & 78.4 $\pm$ \scriptsize 0.3 & 143.6 \\
CM & No Mistakes & 99.3 $\pm$ \scriptsize 0.1 & 78.2 $\pm$ \scriptsize 0.3 & 126.4 \\
\cmidrule(l{2pt}r{2pt}){1-2}
\cmidrule(l{2pt}r{2pt}){3-4}
\cmidrule(l{2pt}r{2pt}){5-5}
RSD & - & 79.4 $\pm$ \scriptsize 0.3 & 24.8 $\pm$ \scriptsize 0.6 & 0.0 \\
\rowcolor{gray!40}
CM & - & 84.1 $\pm$ \scriptsize 0.2 & 50.1 $\pm$ \scriptsize 0.6 & 41.6 \\
\cmidrule(l{2pt}r{2pt}){1-2}
\cmidrule(l{2pt}r{2pt}){3-4}
\cmidrule(l{2pt}r{2pt}){5-5}
MLCM & \hphantom{0}1 OBIS CQ & 84.1 $\pm$ \scriptsize 0.2 & 50.5 $\pm$ \scriptsize 0.6 & 22.0 \\
& \hphantom{0}5 OBIS CQs & 87.9 $\pm$ \scriptsize 0.2 & 56.2 $\pm$ \scriptsize 0.6 & 17.2 \\
& 10 OBIS CQs & 89.6 $\pm$ \scriptsize 0.2 & 58.8 $\pm$ \scriptsize 0.5 & 16.6 \\
& 15 OBIS CQs & 90.7 $\pm$ \scriptsize 0.2 & 59.4 $\pm$ \scriptsize 0.6 & 17.1 \\
& 20 OBIS CQs & 91.5 $\pm$ \scriptsize 0.2 & 60.4 $\pm$ \scriptsize 0.6 & 16.4 \\
\cmidrule(l{2pt}r{2pt}){2-2}
\cmidrule(l{2pt}r{2pt}){3-4}
\cmidrule(l{2pt}r{2pt}){5-5}
& 10 random CQs & 83.1 $\pm$ \scriptsize 0.2 & 49.5 $\pm$ \scriptsize 0.6 & 18.8 \\
& 20 random CQs & 83.1 $\pm$ \scriptsize 0.2 & 50.0 $\pm$ \scriptsize 0.6 & 19.8 \\
\cmidrule(l{2pt}r{2pt}){2-2}
\cmidrule(l{2pt}r{2pt}){3-4}
\cmidrule(l{2pt}r{2pt}){5-5}
& 10 na\"{i}ve CQs & 85.9 $\pm$ \scriptsize 0.2 & 50.8 $\pm$ \scriptsize 0.6 & 18.5 \\
& 20 na\"{i}ve CQs & 86.4 $\pm$ \scriptsize 0.2 & 51.2 $\pm$ \scriptsize 0.6 & 18.8 \\
\cmidrule(l{2pt}r{2pt}){1-2}
\cmidrule(l{2pt}r{2pt}){3-4}
\cmidrule(l{2pt}r{2pt}){5-5}
MLCM-Projected & 10 OBIS CQs & 89.5 $\pm$ \scriptsize 0.2 & 58.5 $\pm$ \scriptsize 0.5 & 135.1 \\
& 20 OBIS CQs & 91.2 $\pm$ \scriptsize 0.2 & 59.9 $\pm$ \scriptsize 0.6 & 100.6 \\
\bottomrule
\end{tabular}

}
\end{sc}
\vskip 0.1cm
\caption{Comparison of different mechanisms for a supply ratio of 1.1, 6 popular courses, and default parameterization for reporting mistakes.  Standard CM is highlighted in grey. Shown are averages over 500 runs. We normalize average and minimum utility by the average utility of \textsc{CM\textsuperscript{*}} and also show their 95\% CIs.}
\label{table_cqs_sr_1.1_pop_6}
\vspace{-0.6cm}
\end{table}

\begin{table}[h!]
	\robustify\bfseries
	\centering
	\begin{sc}
    \resizebox{.85\textwidth}{!}{
	\setlength\tabcolsep{4pt}
\begin{tabular}{llrrc}
\toprule
& & \multicolumn{2}{c}{\textbf{Student Utility (in \%)}}  & \multicolumn{1}{c}{\textbf{Time }}\\
\cmidrule(l{2pt}r{2pt}){3-4}
\textbf{Mechanism} & \textbf{Parametrization} & \multicolumn{1}{c}{\textbf{Average}} & \multicolumn{1}{c}{\textbf{Minimum}}  & \multicolumn{1}{c}{\textbf{(in min)}} \\
\cmidrule(l{2pt}r{2pt}){1-2}
\cmidrule(l{2pt}r{2pt}){3-4}
\cmidrule(l{2pt}r{2pt}){5-5}
CM\textsuperscript{*} & Full Preferences & 100.0 $\pm$ \scriptsize 0.0 & 77.5 $\pm$ \scriptsize 0.3 &  95.0 \\
CM & No Mistakes & 98.8 $\pm$ \scriptsize 0.1 & 76.9 $\pm$ \scriptsize 0.3 &  93.5 \\
\cmidrule(l{2pt}r{2pt}){1-2}
\cmidrule(l{2pt}r{2pt}){3-4}
\cmidrule(l{2pt}r{2pt}){5-5}
RSD & - & 80.2 $\pm$ \scriptsize 0.3 & 31.3 $\pm$ \scriptsize 0.5 &   0.0 \\
\rowcolor{gray!40}
CM & - & 83.8 $\pm$ \scriptsize 0.2 & 49.7 $\pm$ \scriptsize 0.6 &  39.9 \\
\cmidrule(l{2pt}r{2pt}){1-2}
\cmidrule(l{2pt}r{2pt}){3-4}
\cmidrule(l{2pt}r{2pt}){5-5}
MLCM & \hphantom{0}1 OBIS CQ & 83.8 $\pm$ \scriptsize 0.2 & 49.7 $\pm$ \scriptsize 0.6 &  17.2 \\
& \hphantom{0}5 OBIS CQs & 87.7 $\pm$ \scriptsize 0.2 & 56.2 $\pm$ \scriptsize 0.5 &  13.6 \\
& 10 OBIS CQs & 89.4 $\pm$ \scriptsize 0.2 & 57.9 $\pm$ \scriptsize 0.6 &  13.7 \\
& 15 OBIS CQs & 90.4 $\pm$ \scriptsize 0.2 & 59.3 $\pm$ \scriptsize 0.5 &  13.3 \\
& 20 OBIS CQs & 91.2 $\pm$ \scriptsize 0.2 & 59.8 $\pm$ \scriptsize 0.6 &  12.9 \\
\cmidrule(l{2pt}r{2pt}){2-2}
\cmidrule(l{2pt}r{2pt}){3-4}
\cmidrule(l{2pt}r{2pt}){5-5}
& 10 random CQs & 82.9 $\pm$ \scriptsize 0.2 & 49.3 $\pm$ \scriptsize 0.5 &  15.7 \\
& 20 random CQs & 83.0 $\pm$ \scriptsize 0.2 & 49.9 $\pm$ \scriptsize 0.6 &  15.2 \\
\cmidrule(l{2pt}r{2pt}){2-2}
\cmidrule(l{2pt}r{2pt}){3-4}
\cmidrule(l{2pt}r{2pt}){5-5}
& 10 na\"{i}ve CQs & 86.3 $\pm$ \scriptsize 0.2 & 50.7 $\pm$ \scriptsize 0.6 &  15.6 \\
& 20 na\"{i}ve CQs & 87.0 $\pm$ \scriptsize 0.2 & 51.2 $\pm$ \scriptsize 0.6 &  14.7 \\
\cmidrule(l{2pt}r{2pt}){1-2}
\cmidrule(l{2pt}r{2pt}){3-4}
\cmidrule(l{2pt}r{2pt}){5-5}
MLCM-Projected & 10 OBIS CQs & 89.4 $\pm$ \scriptsize 0.2 & 57.8 $\pm$ \scriptsize 0.6 & 137.0 \\
& 20 OBIS CQs & 91.0 $\pm$ \scriptsize 0.2 & 59.8 $\pm$ \scriptsize 0.6 & 129.2 \\
\bottomrule
\end{tabular}

}
\end{sc}
\vskip 0.1cm
\caption{Comparison of different mechanisms for a supply ratio of 1.25, 6 popular courses, and default parameterization for reporting mistakes.  Standard CM is highlighted in grey. Shown are averages over 500 runs. We normalize average and minimum utility by the average utility of \textsc{CM\textsuperscript{*}} and also show their 95\% CIs.}
\label{table_cqs_sr_1.25_pop_6}
\vspace{-0.6cm}
\end{table}

\begin{table}[h!]
	\robustify\bfseries
	\centering
	\begin{sc}
	\resizebox{.85\textwidth}{!}{
	\setlength\tabcolsep{4pt}
\begin{tabular}{llrrc}
\toprule
& & \multicolumn{2}{c}{\textbf{Student Utility (in \%)}}  & \multicolumn{1}{c}{\textbf{Time}}\\
\cmidrule(l{2pt}r{2pt}){3-4}
\textbf{Mechanism} & \textbf{Parametrization} & \multicolumn{1}{c}{\textbf{Average}} & \multicolumn{1}{c}{\textbf{Minimum}}  & \multicolumn{1}{c}{\textbf{(in min)}} \\
\cmidrule(l{2pt}r{2pt}){1-2}
\cmidrule(l{2pt}r{2pt}){3-4}
\cmidrule(l{2pt}r{2pt}){5-5}
CM\textsuperscript{*} & Full Preferences & 100.0 $\pm$ \scriptsize 0.0 & 76.5 $\pm$ \scriptsize 0.4 &  83.0 \\
CM & No Mistakes & 98.5 $\pm$ \scriptsize 0.1 & 75.4 $\pm$ \scriptsize 0.4 &  61.4 \\
\cmidrule(l{2pt}r{2pt}){1-2}
\cmidrule(l{2pt}r{2pt}){3-4}
\cmidrule(l{2pt}r{2pt}){5-5}
RSD & - & 80.6 $\pm$ \scriptsize 0.3 & 34.5 $\pm$ \scriptsize 0.6 &   0.0 \\
\rowcolor{gray!40}
CM & - & 83.6 $\pm$ \scriptsize 0.2 & 49.1 $\pm$ \scriptsize 0.6 &  34.6 \\
\cmidrule(l{2pt}r{2pt}){1-2}
\cmidrule(l{2pt}r{2pt}){3-4}
\cmidrule(l{2pt}r{2pt}){5-5}
MLCM & \hphantom{0}1 OBIS CQ & 83.8 $\pm$ \scriptsize 0.2 & 49.4 $\pm$ \scriptsize 0.6 &  15.2 \\
& \hphantom{0}5 OBIS CQs & 87.9 $\pm$ \scriptsize 0.2 & 55.8 $\pm$ \scriptsize 0.6 &  11.4 \\
& 10 OBIS CQs & 89.7 $\pm$ \scriptsize 0.2 & 58.3 $\pm$ \scriptsize 0.5 &  11.9 \\
& 15 OBIS CQs & 90.7 $\pm$ \scriptsize 0.2 & 59.7 $\pm$ \scriptsize 0.5 &  12.7 \\
& 20 OBIS CQs & 91.4 $\pm$ \scriptsize 0.2 & 60.6 $\pm$ \scriptsize 0.5 &  11.0 \\
\cmidrule(l{2pt}r{2pt}){2-2}
\cmidrule(l{2pt}r{2pt}){3-4}
\cmidrule(l{2pt}r{2pt}){5-5}
& 10 random CQs & 82.8 $\pm$ \scriptsize 0.2 & 48.7 $\pm$ \scriptsize 0.5 &  13.5 \\
& 20 random CQs & 83.0 $\pm$ \scriptsize 0.2 & 49.6 $\pm$ \scriptsize 0.6 &  12.4 \\
\cmidrule(l{2pt}r{2pt}){2-2}
\cmidrule(l{2pt}r{2pt}){3-4}
\cmidrule(l{2pt}r{2pt}){5-5}
& 10 na\"{i}ve CQs & 87.2 $\pm$ \scriptsize 0.2 & 51.6 $\pm$ \scriptsize 0.6 &  11.7 \\
& 20 na\"{i}ve CQs & 88.5 $\pm$ \scriptsize 0.2 & 52.1 $\pm$ \scriptsize 0.6 &  11.7 \\
\cmidrule(l{2pt}r{2pt}){1-2}
\cmidrule(l{2pt}r{2pt}){3-4}
\cmidrule(l{2pt}r{2pt}){5-5}
MLCM-Projected & 10 OBIS CQs & 89.6 $\pm$ \scriptsize 0.2 & 58.4 $\pm$ \scriptsize 0.6 & 119.3 \\
& 20 OBIS CQs & 91.4 $\pm$ \scriptsize 0.2 & 60.5 $\pm$ \scriptsize 0.5 & 119.7 \\
\bottomrule
\end{tabular}

}
\end{sc}
\vskip 0.1cm
\caption{Comparison of different mechanisms for a supply ratio of 1.5, 6 popular courses, and default parameterization for reporting mistakes.  Standard CM is highlighted in grey. Shown are averages over 500 runs. We normalize average and minimum utility by the average utility of \textsc{CM\textsuperscript{*}} and also show their 95\% CIs.}
\label{table_cqs_sr_1.5_pop_6}
\vspace{-.6cm}
\end{table}

\newpage
\clearpage

\section{Mistake Robustness Experiments}
\label{A_mistake_robustness_results}
\begin{figure}[h!]
    \vskip -0.5cm
    \centering
    \includegraphics[width=0.95\columnwidth]{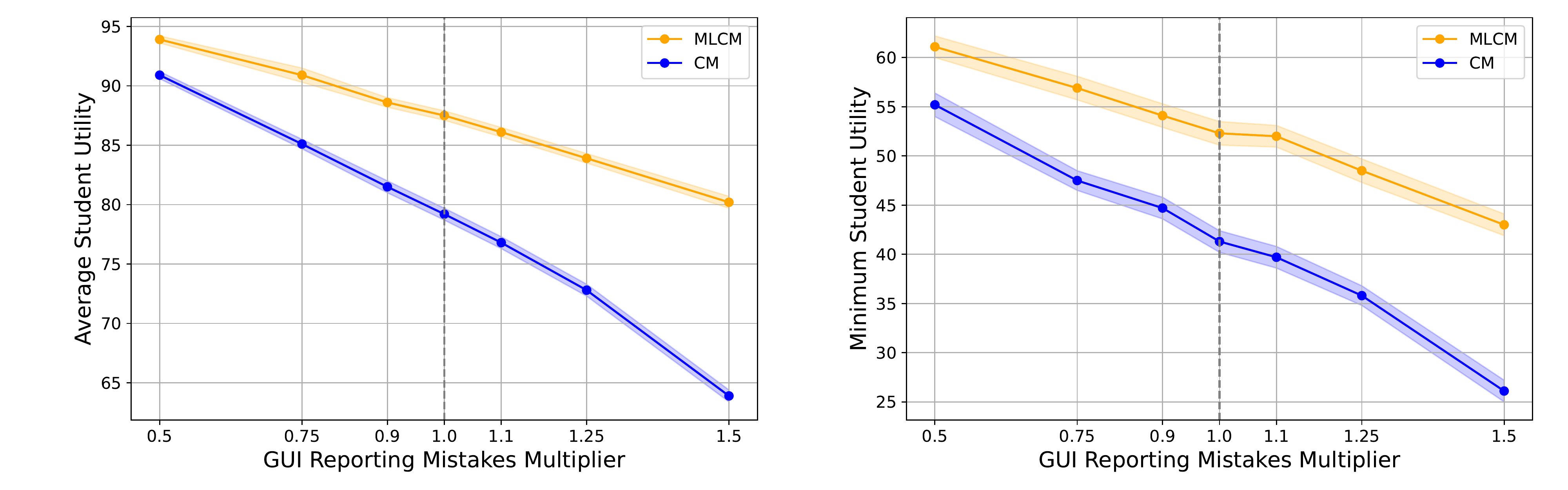}
    \vskip -0.25cm
    \caption{Reporting mistakes robustness experiment for a supply ratio of 1.1 and 9 popular courses. Shown are average results in \% for the final allocation over 100 runs including 95\% CI.}
    \label{fig:noise_robustness_sr_1.1_pop9}
\end{figure}

\begin{figure}[h!]
    \vskip -0.45cm
    \centering
    \includegraphics[width=0.95\columnwidth]{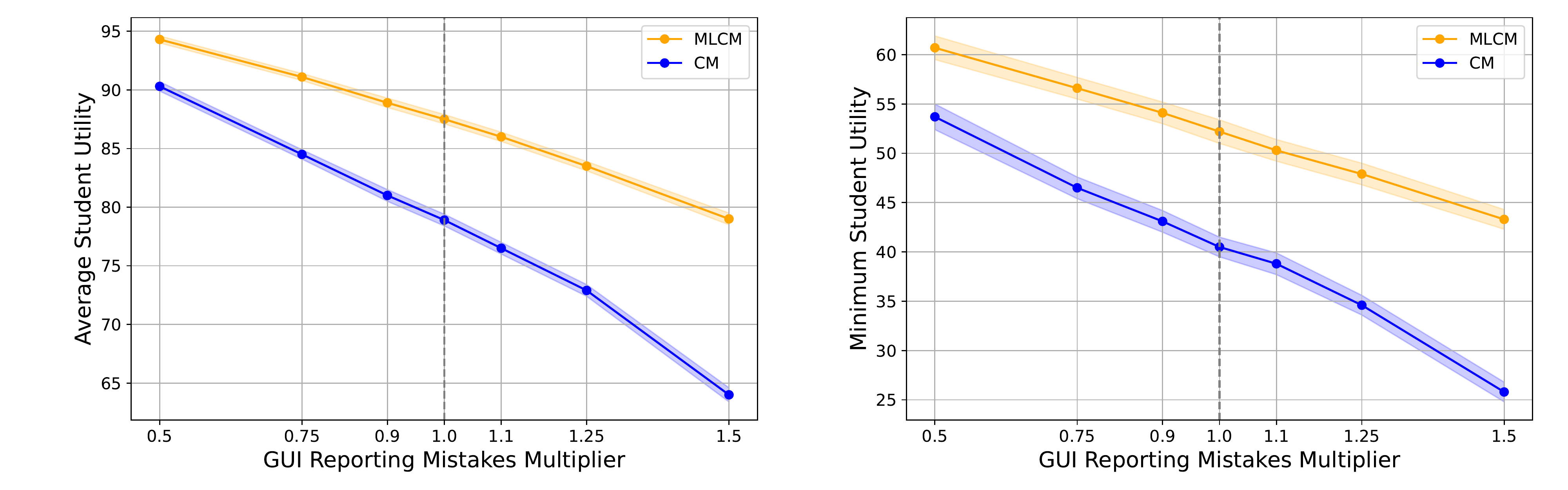}
    \vskip -0.25cm
    \caption{Reporting mistakes robustness experiment for a supply ratio of 1.5 and 9 popular courses. Shown are average results in \% for the final allocation over 100 runs including 95\% CI.}
    \label{fig:noise_robustness_sr_1.5_pop9}
\end{figure}

\begin{figure}[h!]
    \vskip -0.45cm
    \centering
    \includegraphics[width=0.95\columnwidth]{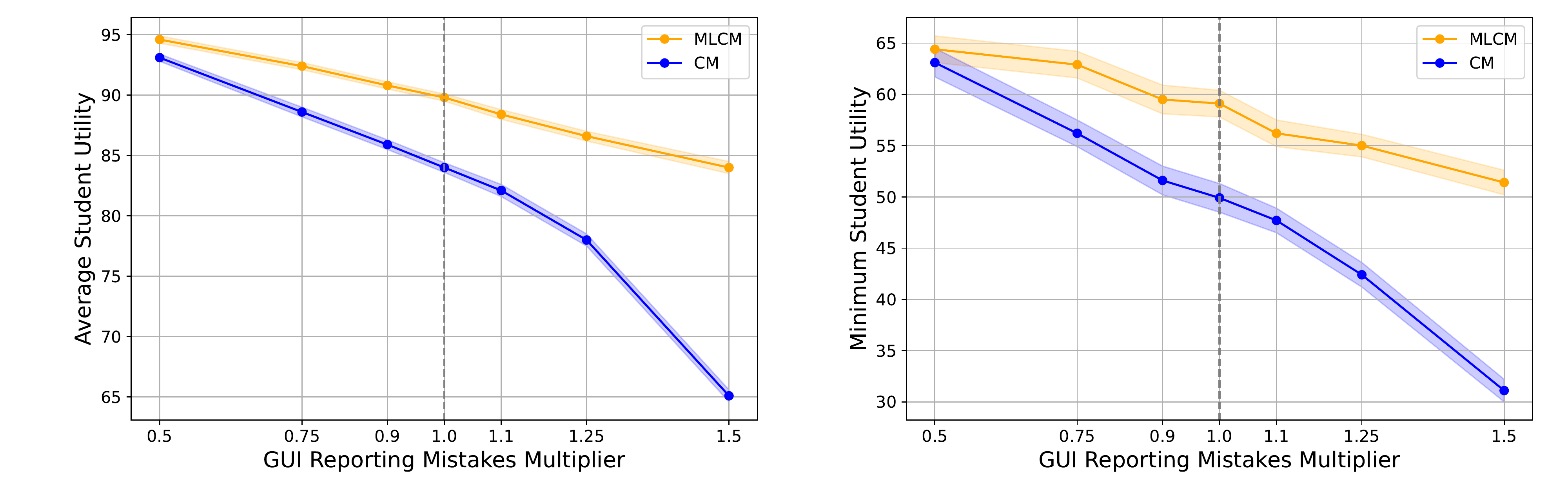}
    \vskip -0.25cm
    \caption{Reporting mistakes robustness experiment for a supply ratio of 1.1 and 6 popular courses. Shown are average results in \% for the final allocation over 100 runs including 95\% CI.}
    \label{fig:noise_robustness_sr_1.1_pop6}
\end{figure}

\begin{figure}[h!]
    \vskip -0.45cm
    \centering
    \includegraphics[width=0.95\columnwidth]{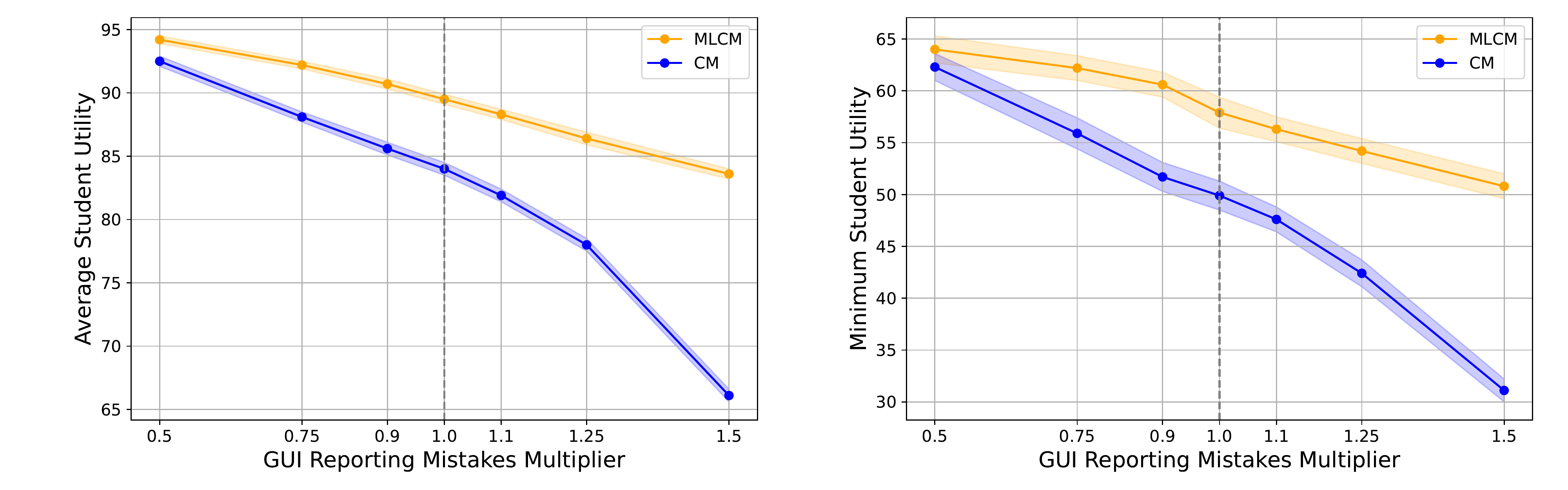}
    \vskip -0.3cm
    \caption{Reporting mistakes robustness experiment for a supply ratio of 1.25 and 6 popular courses. Shown are average results in \% for the final allocation over 100 runs including 95\% CI.}
    \label{fig:noise_robustness_sr_1.25_pop6}
\end{figure}

\begin{figure}[h!]
    \vskip -0.45cm
    \centering
    \includegraphics[width=0.95\columnwidth]{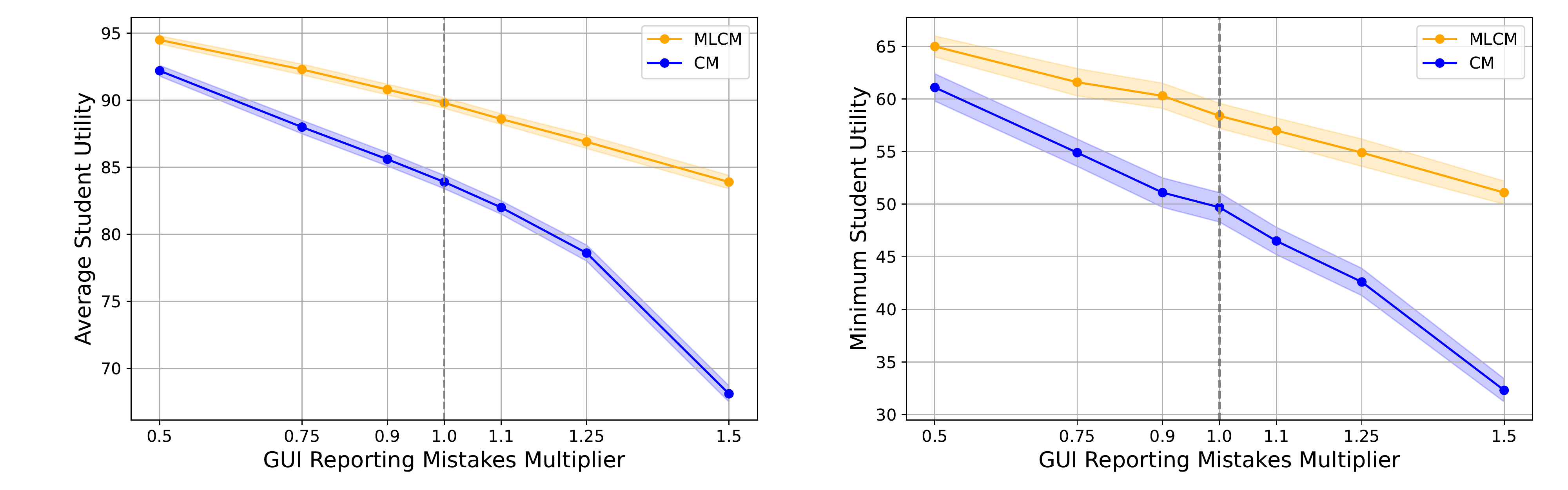}
    \vskip -0.25cm
    \caption{Reporting mistakes robustness experiment for a supply ratio of 1.5 and 6 popular courses. Shown are average results in \% for the final allocation over 100 runs including 95\% CI.}
    \label{fig:noise_robustness_sr_1.5_pop6}
\end{figure}

In this section, we present five further reporting mistake robustness experiments for the following choices of supply ratios and number of popular courses, respectively:
\begin{itemize}[topsep=6pt,partopsep=6pt, parsep=6pt]
\item A supply ratio of 1.1 and 9 popular courses (see \Cref{fig:noise_robustness_sr_1.1_pop9})
\item A supply ratio of 1.5 and 9 popular courses (see \Cref{fig:noise_robustness_sr_1.5_pop9})
\item A supply ratio of 1.1 and 6 popular courses (see \Cref{fig:noise_robustness_sr_1.1_pop6})
\item A supply ratio of 1.25 and 6 popular courses (see \Cref{fig:noise_robustness_sr_1.25_pop6})
\item A supply ratio of 1.5 and 6 popular courses (see \Cref{fig:noise_robustness_sr_1.5_pop6})
\end{itemize}

\paragraph{Results} Similarly to the results presented in the main paper in \Cref{fig:noise_robustness} (supply ratio of 1.25 and 9 popular courses), we see that, as $\gamma$ increases, the performance of both, CM and MLCM, monotonically decreases. This should not come as a surprise, as for values of $\gamma$ larger than $1$ the students make more mistakes when reporting their preferences to the GUI. MLCM significantly outperforms CM for all $\gamma \in [0.75, 1.5]$. Moreover, as $\gamma$ increases, the relative performance gap of the two mechanisms gets larger. These results further highlight the robustness of our design to changes in the mistake profile of the students.

\newpage
\clearpage

\section{Should Individual Students Opt Into the ML Feature?}
\label{A_unilateral_switch}
In this section, we first provide additional results for a supply ratio of $1.1$ and $1.5$ for the experiment measuring the expected utility gain of a student if she were the only one
to opt into MLCM, as described in \Cref{subsec:Should Individual Students Opt Into MLCM?}. We present those results in \Cref{tab: appendix_table_first_student_to_switch}. 
Furthermore, we perform a similar test, measuring the expected gain of a student opting into MLCM, if \textit{every} other student also opted in. We present those results in \Cref{tab:app_all_students_opt_in}. 

\paragraph{No Other Student to Opt into the ML Feature}
Similarly to the results presented in the main paper (see \Cref{table_first_student_to_switch} for a supply ratio of 1.25), \Cref{tab: appendix_table_first_student_to_switch} shows that for all SRs tested, the \emph{expected relative gain} from opting in is at least $9.5\%$ (across all settings). Furthermore, the student prefers the ``MLCM schedule'' in at least $71.9\%$ of the cases, while she prefers the ``CM schedule'' in at most $17.1\%$ of the cases.
As the number of comparison queries (CQs) the student answers increases, the benefit from opting into MLCM's ML feature as the first student becomes even larger. Finally, the improvement is larger for more popular courses.
\begin{table}[t!]
	\robustify\bfseries
	\centering
	\begin{sc}
	\resizebox{1\columnwidth}{!}{
	\setlength\tabcolsep{4pt}

\begin{tabular}{ccc r  r  r r r r}
\toprule
\multicolumn{3}{c}{\textbf{Setting}} & \multicolumn{3}{c}{\textbf{Preferred Mechanism}} & \multicolumn{3}{c}{\textbf{Gain from Opting Into MLCM}}  \\
\cmidrule(l{2pt}r{2pt}){1-3}
\cmidrule(l{2pt}r{2pt}){4-6}
\cmidrule(l{2pt}r{2pt}){7-9}
                    \multicolumn{1}{c}{\textbf{SR}} & \multicolumn{1}{c}{\textbf{\#PoP}} & \multicolumn{1}{c}{\textbf{\#CQs}}& \multicolumn{1}{c}{\textbf{MLCM}} & \multicolumn{1}{c}{\textbf{CM}} & \multicolumn{1}{c}{\textbf{Indiff.}} & \multicolumn{1}{c}{\textbf{Expected}} & \multicolumn{1}{c}{\textbf{if pref MLCM}} &  \multicolumn{1}{c}{\textbf{if pref CM}} \\
\cmidrule(l{2pt}r{2pt}){1-3}
\cmidrule(l{2pt}r{2pt}){4-6}
\cmidrule(l{2pt}r{2pt}){7-9}
1.1 & 6 & 10  & 73.75\% & 14.86\% & 11.39\% & 10.54\% & 15.88\% & -7.78\% \\
1.1 & 6 & 15  & 78.60\% & 11.90\% & 9.50\%  & 12.58\% & 17.11\% & -7.28\% \\
1.1 & 6 & 20  & 81.74\% & 10.23\% & 8.03\%  & 13.93\% & 17.96\% & -7.03\% \\
\cmidrule(l{2pt}r{2pt}){1-3}
\cmidrule(l{2pt}r{2pt}){4-6}
\cmidrule(l{2pt}r{2pt}){7-9}
1.25 & 6 & 10 &    71.92\% &   17.07\% &     11.01\% &       9.61\% &  15.58\% & -9.17\% \\
1.25 & 6 & 15 &    75.99\% &   14.99\% &      9.02\% &      11.25\% &  16.65\% & -9.19\% \\
1.25 & 6 & 20 &    78.26\% &   14.05\% &      7.69\% &      12.38\% &  17.53\% & -9.32\% \\
\cmidrule(l{2pt}r{2pt}){1-3}
\cmidrule(l{2pt}r{2pt}){4-6}
\cmidrule(l{2pt}r{2pt}){7-9}
1.5 & 6 & 10  & 71.99\% & 16.55\% & 11.46\% & 9.50\% & 15.43\% & -9.38\% \\
1.5 & 6 & 15  & 76.91\% & 13.96\% & 9.13\%  & 11.40\% & 16.61\% & -9.53\% \\
1.5 & 6 & 20  & 79.47\% & 12.75\% & 7.78\%  & 12.41\% & 17.17\% & -9.24\% \\
\cmidrule(l{2pt}r{2pt}){1-3}
\cmidrule(l{2pt}r{2pt}){4-6}
\cmidrule(l{2pt}r{2pt}){7-9}
1.1 & 9 & 10  & 78.04\% & 10.19\% & 11.77\% & 15.33\% & 20.81\% & -8.81\% \\
1.1 & 9 & 15  & 82.67\% & 8.50\%  & 8.83\%  & 17.87\% & 22.47\% & -8.37\% \\
1.1 & 9 & 20  & 85.35\% & 7.43\%  & 7.22\%  & 19.74\% & 23.89\% & -8.65\% \\
\cmidrule(l{2pt}r{2pt}){1-3}
\cmidrule(l{2pt}r{2pt}){4-6}
\cmidrule(l{2pt}r{2pt}){7-9}
1.25 & 9 & 10  & 78.49\% & 10.39\% & 11.12\% & 14.53\% & 19.72\% & -8.94\% \\
1.25 & 9 & 15  & 82.95\% & 8.72\%  & 8.33\%  & 17.24\% & 21.71\% & -8.50\% \\
1.25 & 9 & 20  & 85.22\% & 8.02\%  & 6.76\%  & 18.69\% & 22.74\% & -8.13\% \\
\cmidrule(l{2pt}r{2pt}){1-3}
\cmidrule(l{2pt}r{2pt}){4-6}
\cmidrule(l{2pt}r{2pt}){7-9}
1.5 & 9 & 10  & 78.90\% & 9.74\%  & 11.36\% & 14.19\% & 19.18\% & -9.41\% \\
1.5 & 9 & 15  & 83.89\% & 7.71\%  & 8.40\%  & 16.71\% & 20.84\% & -9.66\% \\
1.5 & 9 & 20  & 85.58\% & 7.60\%  & 6.82\%  & 18.18\% & 22.11\% & -9.34\% \\
\bottomrule
\end{tabular}

}
\end{sc}
\vskip 0.1cm
\caption{Expected gain of opting into MLCM's ML feature when no other student opts in. Shown are average results across 10,000 students per setting (\textsc{SR},\textsc{\#PoP},\textsc{\#CQs}). CIs for all metrics are $\approx0$.}
\label{tab: appendix_table_first_student_to_switch}
\vskip -0.6 cm
\end{table}

\paragraph{All Other Students to Opt into the ML Feature}
Now suppose that all students have decided to opt into MLCM's ML feature, except for one. How much would that last student benefit if she decided to also ``opt in''? As before, we answer this question by running both MLCM and CM twice -- once where all but one student have opted in, and once where \textit{all} students have opted in.\footnote{As before, we report Stage 1 results but now use the fixed price vector that would result if all students chose to opt in.}
Our analysis encompasses 100 instances, with 100 students in each instance, totaling 10,000 students. We report averages over those 10,000 students.
\Cref{tab:app_all_students_opt_in} shows those results for all three supply ratios of $1.1$, $1.25$ and $1.5$.
We see that the \emph{expected relative gain} from opting in is at least $7.8\%$ (across all settings). Furthermore, the student prefers the ``MLCM schedule'' in at least $66.2$\% of the cases, while she prefers the ``CM schedule'' in at most $22.5$\% of the cases.
As the student answers more CQs, the benefit from opting into MLCM's ML feature becomes even larger. Finally, the improvement is larger for more popular courses.
\begin{table}[t!]   
	\robustify\bfseries
	\centering
	\begin{sc}
	\resizebox{1\columnwidth}{!}{
	\setlength\tabcolsep{4pt}

\begin{tabular}{ccc r  r  r r r r}
\toprule
\multicolumn{3}{c}{\textbf{Setting}} & \multicolumn{3}{c}{\textbf{Preferred Mechanism}} & \multicolumn{3}{c}{\textbf{Gain from Opting Into MLCM}}  \\
\cmidrule(l{2pt}r{2pt}){1-3}
\cmidrule(l{2pt}r{2pt}){4-6}
\cmidrule(l{2pt}r{2pt}){7-9}
                    \multicolumn{1}{c}{\textbf{SR}} & \multicolumn{1}{c}{\textbf{\#PoP}} & \multicolumn{1}{c}{\textbf{\#CQs}}& \multicolumn{1}{c}{\textbf{MLCM}} & \multicolumn{1}{c}{\textbf{CM}} & \multicolumn{1}{c}{\textbf{Indiff.}} & \multicolumn{1}{c}{\textbf{Expected}} & \multicolumn{1}{c}{\textbf{if pref MLCM}} &  \multicolumn{1}{c}{\textbf{if pref CM}} \\
\cmidrule(l{2pt}r{2pt}){1-3}
\cmidrule(l{2pt}r{2pt}){4-6}
\cmidrule(l{2pt}r{2pt}){7-9}
1.1 & 9 &  10 &    72.33\% &   16.82\% &     10.85\% &      12.75\% &  19.81\% & -9.31\% \\
1.1 & 9 & 15 &    76.51\% &   15.18\% &      8.31\% &      14.76\% &  21.25\% & -9.76\% \\
1.1 & 9 & 20 &    79.09\% &   13.87\% &      7.04\% &      16.28\% &  22.30\% & -9.83\% \\
\cmidrule(l{2pt}r{2pt}){1-3}
\cmidrule(l{2pt}r{2pt}){4-6}
\cmidrule(l{2pt}r{2pt}){7-9}
1.25 & 9 & 10 &    74.94\% &   14.04\% &     11.02\% &      13.33\% &  19.61\% & -9.76\% \\
1.25 & 9 & 15 &    78.71\% &    12.8\% &      8.49\% &      15.37\% &  21.03\% & -9.32\% \\
1.25 & 9 & 20 &    80.63\% &   12.45\% &      6.92\% &      16.50\% &  21.92\% & -9.15\% \\
\cmidrule(l{2pt}r{2pt}){1-3}
\cmidrule(l{2pt}r{2pt}){4-6}
\cmidrule(l{2pt}r{2pt}){7-9}
1.5 & 9 & 10 &    77.11\% &   11.47\% &     11.42\% &      13.78\% &  19.26\% & -9.31\% \\
1.5 & 9 & 15 &    81.27\% &   10.22\% &      8.51\% &      15.90\% &  20.73\% & -9.30\% \\
1.5 & 9 & 20 &    83.48\% &    9.57\% &      6.95\% &      17.34\% &  21.86\% & -9.19\% \\
\cmidrule(l{2pt}r{2pt}){1-3}
\cmidrule(l{2pt}r{2pt}){4-6}
\cmidrule(l{2pt}r{2pt}){7-9}
1.1 & 6 & 10 &     66.8\% &   22.46\% &     10.74\% &       8.02\% &  14.89\% & -8.56\% \\
1.1 & 6 & 15 &    69.68\% &   21.62\% &      8.70\% &       9.24\% &  16.07\% & -8.96\% \\
1.1 & 6 & 20 &    73.22\% &   19.08\% &      7.70\% &      10.21\% &  16.21\% & -8.63\% \\
\cmidrule(l{2pt}r{2pt}){1-3}
\cmidrule(l{2pt}r{2pt}){4-6}
\cmidrule(l{2pt}r{2pt}){7-9}
1.25 & 6 & 10 &    66.16\% &    23.4\% &     10.44\% &       7.84\% &  15.20\% & -9.37\% \\
1.25 & 6 & 15 &    69.63\% &   21.75\% &      8.62\% &       8.92\% &  15.74\% & -9.25\% \\
1.25 & 6 & 20 &     72.1\% &   21.01\% &      6.89\% &      10.00\% &  16.62\% & -9.46\% \\
\cmidrule(l{2pt}r{2pt}){1-3}
\cmidrule(l{2pt}r{2pt}){4-6}
\cmidrule(l{2pt}r{2pt}){7-9}
1.5 & 6 & 10 &     68.9\% &   20.07\% &     11.03\% &       8.73\% &  15.43\% & -9.35\% \\
1.5 & 6 & 15 &    72.54\% &   18.47\% &      8.99\% &      10.06\% &  16.39\% & -9.84\% \\
1.5 & 6 & 20 &    75.32\% &   17.16\% &      7.52\% &      10.87\% &  16.63\% & -9.60\% \\
\bottomrule
\end{tabular}

}
\end{sc}
\vskip 0.1cm
\caption{Expected gain of opting into MLCM's ML feature when all other students also opt in. Shown are average results across 10,000 students per setting (\textsc{SR},\textsc{\#PoP},\textsc{\#CQs}). CIs for all metrics are $\approx0$.}
\label{tab:app_all_students_opt_in}
\vskip -0.5 cm
\end{table}

\section{Additive Preferences}
\label{sec:app:Results_for_Additive Preferences}
In this section, we repeat all of our experiments for the special case of students having \emph{additive} true preferences, i.e., we use our student preference generator (see \Cref{sec:StudentPreferenceGenerator} and electronic companion~\ref{A_SPG}) to generate additive/linear true students' utility functions. Again, we calibrated the mistake profile of the students so that both their accuracy and the reported utility difference in case of disagreements match those determined in the lab experiment of \citet{budish2022can} (see \Cref{tab:app:noise_calibration_additive_preferences}). 

\begin{table}[h!]
	\robustify\bfseries
	\centering
	\begin{sc}
	\resizebox{1\columnwidth}{!}{
	\setlength\tabcolsep{4pt}

\begin{tabular}{llrrrrrrrrr}
\toprule
 \multicolumn{2}{c}{\textbf{Setting}} & \multicolumn{3}{c}{\textbf{\#Courses with value}} & \multicolumn{4}{c}{\textbf{\#Adjustments}} & \multicolumn{1}{c}{\textbf{Accuracy}} & \multicolumn{1}{c}{\textbf{If disagreement}}\\
 \cmidrule(l{2pt}r{2pt}){1-2}
\cmidrule(l{2pt}r{2pt}){3-5}
\cmidrule(l{2pt}r{2pt}){6-9} 
 \#Pop & $\gamma$&                         $> 0$ &                           $(0,50)$ &              $[50,100]$ &                      Mean & Median &    Min & Max &    \multicolumn{1}{c}{\textbf{in \%}}                         & \textbf{utility difference in \%} \\
\cmidrule(l{2pt}r{2pt}){1-2}
\cmidrule(l{2pt}r{2pt}){3-5}
\cmidrule(l{2pt}r{2pt}){6-9}
\cmidrule(l{2pt}r{2pt}){10-10}
\cmidrule(l{2pt}r{2pt}){11-11}
9& 0.5& 17.20 $\pm$ \scriptsize 0.02 & 9.96 $\pm$ \scriptsize 0.10 & 7.24 $\pm$ \scriptsize 0.09 & 0.00 $\pm$ \scriptsize 0.00 &    0.0 &      0 &      0 & 95 $\pm$ \scriptsize 1 &  -1.47 $\pm$ \scriptsize 0.31 \\
9& 0.75& 13.39 $\pm$ \scriptsize 0.03 & 5.66 $\pm$ \scriptsize 0.11 & 7.73 $\pm$ \scriptsize 0.11 & 0.00 $\pm$ \scriptsize 0.00 &    0.0 &      0 &      0 & 91 $\pm$ \scriptsize 1 &  -6.44 $\pm$ \scriptsize 0.73 \\
9& 0.9& 11.03 $\pm$ \scriptsize 0.01 & 3.00 $\pm$ \scriptsize 0.11 & 8.03 $\pm$ \scriptsize 0.11 & 0.00 $\pm$ \scriptsize 0.00 &    0.0 &      0 &      0 & 87 $\pm$ \scriptsize 1 & -11.11 $\pm$ \scriptsize 0.89 \\
\ccell 9& \ccell1&  \ccell 9.49 $\pm$ \scriptsize 0.03 & \ccell 1.53 $\pm$ \scriptsize 0.09 & \ccell 7.96 $\pm$ \scriptsize 0.09 & \ccell 0.00 $\pm$ \scriptsize 0.00 & \ccell    0.0 &   \ccell    0 &  \ccell    0 &  \ccell 84 $\pm$ \scriptsize 1 & \ccell -14.03 $\pm$ \scriptsize 1.00 \\
9& 1.1& 7.92 $\pm$ \scriptsize 0.02 & 0.50 $\pm$ \scriptsize 0.06 & 7.42 $\pm$ \scriptsize 0.06 & 0.00 $\pm$ \scriptsize 0.00 &    0.0 &      0 &      0 & 78 $\pm$ \scriptsize 1 & -17.70 $\pm$ \scriptsize 1.07 \\
9& 1.25& 5.62 $\pm$ \scriptsize 0.03 & 0.04 $\pm$ \scriptsize 0.01 & 5.58 $\pm$ \scriptsize 0.03 & 0.00 $\pm$ \scriptsize 0.00 &    0.0 &      0 &      0 & 69 $\pm$ \scriptsize 2 & -22.18 $\pm$ \scriptsize 1.07 \\
9& 1.5& 1.75 $\pm$ \scriptsize 0.03 & 0.00 $\pm$ \scriptsize 0.00 & 1.75 $\pm$ \scriptsize 0.03 & 0.00 $\pm$ \scriptsize 0.00 &    0.0 &      0 &      0 & 59 $\pm$ \scriptsize 2 & -17.01 $\pm$ \scriptsize 1.36 \\
\midrule
\multicolumn{2}{c}{\citet{budish2022can}}  &   12.45\hphantom{ $\pm$ \scriptsize 0.00}  &  6.17\hphantom{ $\pm$ \scriptsize 0.00}  &  6.27\hphantom{ $\pm$ \scriptsize 0.00} & 1.08\hphantom{ $\pm$ \scriptsize 0.00}  & 0 &  0 &         10 &         84\hphantom{ $\pm$ \scriptsize 0}  &          -13.35 $\pm$ \scriptsize 0.41 \\\bottomrule 
\end{tabular}

}
\vskip 0.1cm
\caption{Mistake profile calibration experiment for \emph{additive} preferences for several settings defined by the number of popular courses (\textsc{\#Pop}) and the common mistake constant $\gamma$ compared to the experimental findings in \cite{budish2022can}. Our default settings (i.e., $\gamma=1$) is marked in grey. 1000 students in total. We show the number of courses with reported value in 3 distinct intervals, the mean, median, minimum and maximum number of adjustments in the students' reports, the accuracy of their reports as determined by asking CQs and the median scaled utility difference between the two schedules in a CQ in case of disagreement between the CQ answer and the reported preferences. Shown are average results and a 95\% CI.}
\label{tab:app:noise_calibration_additive_preferences}
\end{sc}
\end{table}

For these new preferences we present:
\begin{itemize}
    \item the main welfare results for supply ratios $1.1$, $1.25$ and $1.5$ 
    (\Cref{tab:app:welfare_SR1.1_Pop9,tab:app:welfare_SR1.25_Pop9,tab:app:welfare_SR1.5_Pop9}), 
    \item the reporting mistakes robustness study for supply ratios of $1.1$, $1.25$ and $1.5$ (\Cref{fig:app_noise_robustness_additive_sr1.1_pop9,fig:app_noise_robustness_additive_sr1.25_pop9,fig:app_noise_robustness_additive_sr1.5_pop9}),  
    \item the experiment measuring the expected gain of opting into MLCM's ML feature when no/all other students do so (\Cref{tab:app:table_first_student_to_switch_additive,tab:app:table_last_student_to_switch_additive}).
\end{itemize}

Overall our results in the following three subsections show that MLCM achieves qualitatively the same results when the true students' utilities are restricted to be additive. These results further highlight the robustness of our design, and its adaptability to different environments. 

\subsection{Welfare Results for Additive Preferences}
In this subsection, we present the main welfare results when students' true utility functions are restricted to be \emph{additive/linear}. Please see \Cref{subsec:experiment_setup} for details on the experiment setup.

We normalize all results by the average utility of \textsc{CM\textsuperscript{*} (Full Preferences)} after Stage 1. 
Note that in this setting with additive preferences, the mechanisms \textsc{CM\textsuperscript{*} (Full Preferences)} and \textsc{CM (No Mistakes)} completely coincide. 
In Table \ref{tab:app:welfare_SR1.25_Pop9}, we present results for $\text{SR}=1.25$ (which is very close to Wharton's SR; see \cite{budish2022can}) and $9$ popular courses. The results are qualitatively very similar for the other two supply ratios. 
We see that \textsc{MLCM (10 OBIS CQs)} significantly outperforms \textsc{CM}, both in average and minimum student utility. In particular, \textsc{MLCM (10 OBIS CQs)} increases average utility from $85.5\%$ to $93.2\%$, which is an increase of $7.7$ percentage points (a $9$\% relative increase increase) and minimum student utility from $54.3\%$ to $67.6\%$, an increase of $13.3$ percentage points (a $24.5$\% relative increase). As the number of CQs increases, the performance of \textsc{MLCM} improves further.
Interestingly, as opposed to the results in \Cref{subsec:experimental_results} for preferences that match the reports of \citet{budish2022can}, for additive preferences the relative performance gain of MLCM over CM increases as the supply ratio decreases. 


\begin{table}[h!]
	\robustify\bfseries
	\centering
	\begin{sc}
	\resizebox{0.7\columnwidth}{!}{
	\setlength\tabcolsep{4pt}
\begin{tabular}{llrrc}
    \toprule
    & & \multicolumn{2}{c}{\textbf{Student Utility (in \%)}}  & \multicolumn{1}{c}{\textbf{Time}}\\
    \cmidrule(l{2pt}r{2pt}){3-4}
    \textbf{Mechanism} & \textbf{Parametrization} & \multicolumn{1}{c}{\textbf{Average}} & \multicolumn{1}{c}{\textbf{Minimum}}  & \multicolumn{1}{c}{\textbf{(in min)}} \\
    \cmidrule(l{2pt}r{2pt}){1-2}
    \cmidrule(l{2pt}r{2pt}){3-4}
    \cmidrule(l{2pt}r{2pt}){5-5}
    $\text{CM}^*$ & Full Preferences & 100.0 $\pm$ \scriptsize 0.0 & 83.4 $\pm$ \scriptsize 0.3 &  4.6 \\
    \cmidrule(l{2pt}r{2pt}){1-2}
    \cmidrule(l{2pt}r{2pt}){3-4}
    \cmidrule(l{2pt}r{2pt}){5-5}
    \rowcolor{gray!40}
    CM & - & 83.6 $\pm$ \scriptsize 0.1 & 51.7 $\pm$ \scriptsize 0.5 &  1.4 \\
    \cmidrule(l{2pt}r{2pt}){1-2}
    \cmidrule(l{2pt}r{2pt}){3-4}
    \cmidrule(l{2pt}r{2pt}){5-5}
    MLCM & \hphantom{0}1 OBIS CQ & 87.6 $\pm$ \scriptsize 0.1 & 59.1 $\pm$ \scriptsize 0.4 & 17.6 \\
    & \hphantom{0}5 OBIS CQs & 90.8 $\pm$ \scriptsize 0.1 & 63.7 $\pm$ \scriptsize 0.4 & 14.1 \\
    & 10 OBIS CQs & 92.4 $\pm$ \scriptsize 0.1 & 65.7 $\pm$ \scriptsize 0.5 & 12.7 \\
    & 15 OBIS CQs & 93.5 $\pm$ \scriptsize 0.1 & 67.3 $\pm$ \scriptsize 0.4 & 12.3 \\
    & 20 OBIS CQs & 94.3 $\pm$ \scriptsize 0.1 & 68.2 $\pm$ \scriptsize 0.4 & 13.2 \\
    \bottomrule
\end{tabular}
}
\vskip 0.1cm
\caption{Comparison of CM and MLCM for \emph{additive} preferences, a supply ratio of 1.1, 9 popular courses, and default parameterization for reporting mistakes.  Standard CM is highlighted in grey. Shown are averages over 500 runs. We normalize average and minimum utility by the average utility of \textsc{CM\textsuperscript{*}} and also show their 95\% CIs.}
\label{tab:app:welfare_SR1.1_Pop9}
\end{sc}
\end{table}

\begin{table}[h!]
	\robustify\bfseries
	\centering
	\begin{sc}
	\resizebox{0.7\columnwidth}{!}{
	\setlength\tabcolsep{4pt}

\begin{tabular}{llrrc}
    \toprule
    & & \multicolumn{2}{c}{\textbf{Student Utility (in \%)}}  & \multicolumn{1}{c}{\textbf{Time}}\\
    \cmidrule(l{2pt}r{2pt}){3-4}
    \textbf{Mechanism} & \textbf{Parametrization} & \multicolumn{1}{c}{\textbf{Average}} & \multicolumn{1}{c}{\textbf{Minimum}}  & \multicolumn{1}{c}{\textbf{(in min)}} \\
    \cmidrule(l{2pt}r{2pt}){1-2}
    \cmidrule(l{2pt}r{2pt}){3-4}
    \cmidrule(l{2pt}r{2pt}){5-5}
    $\text{CM}^*$ & Full Preferences & 100.0 $\pm$ \scriptsize 0.0 & 84.2 $\pm$ \scriptsize 0.3 &  3.1 \\
    \cmidrule(l{2pt}r{2pt}){1-2}
    \cmidrule(l{2pt}r{2pt}){3-4}
    \cmidrule(l{2pt}r{2pt}){5-5}
    \rowcolor{gray!40}
    CM & - & 85.5 $\pm$ \scriptsize 0.1 & 54.3 $\pm$ \scriptsize 0.5 &  1.3 \\
    \cmidrule(l{2pt}r{2pt}){1-2}
    \cmidrule(l{2pt}r{2pt}){3-4}
    \cmidrule(l{2pt}r{2pt}){5-5}
    MLCM & \hphantom{0}1 OBIS CQ & 88.4 $\pm$ \scriptsize 0.1 & 60.1 $\pm$ \scriptsize 0.5 & 13.9 \\
    & \hphantom{0}5 OBIS CQs & 91.5 $\pm$ \scriptsize 0.1 & 65.2 $\pm$ \scriptsize 0.4 & 11.6 \\
    & 10 OBIS CQs & 93.2 $\pm$ \scriptsize 0.1 & 67.6 $\pm$ \scriptsize 0.5 & 10.3 \\
    & 15 OBIS CQs & 94.2 $\pm$ \scriptsize 0.1 & 69.5 $\pm$ \scriptsize 0.5 & 10.7 \\
    & 20 OBIS CQs & 94.9 $\pm$ \scriptsize 0.1 & 70.3 $\pm$ \scriptsize 0.4 & 10.8 \\
    \bottomrule
\end{tabular}

}
\vskip 0.1cm
\caption{Comparison of CM and MLCM for \emph{additive} preferences, a supply ratio of 1.25, 9 popular courses, and default parameterization for reporting mistakes.  Standard CM is highlighted in grey. Shown are averages over 500 runs. We normalize average and minimum utility by the average utility of \textsc{CM\textsuperscript{*}} and also show their 95\% CIs.}
\label{tab:app:welfare_SR1.25_Pop9}
\end{sc}
\end{table}

\begin{table}[h!]
	\robustify\bfseries
	\centering
	\begin{sc}
	\resizebox{0.7\columnwidth}{!}{
	\setlength\tabcolsep{4pt}
\begin{tabular}{llrrc}
    \toprule
    & & \multicolumn{2}{c}{\textbf{Student Utility (in \%)}}  & \multicolumn{1}{c}{\textbf{Time}}\\
    \cmidrule(l{2pt}r{2pt}){3-4}
    \textbf{Mechanism} & \textbf{Parametrization} & \multicolumn{1}{c}{\textbf{Average}} & \multicolumn{1}{c}{\textbf{Minimum}}  & \multicolumn{1}{c}{\textbf{(in min)}} \\
    \cmidrule(l{2pt}r{2pt}){1-2}
    \cmidrule(l{2pt}r{2pt}){3-4}
    \cmidrule(l{2pt}r{2pt}){5-5}
    $\text{CM}^*$ & Full Preferences & 100.0 $\pm$ \scriptsize 0.0 & 84.2 $\pm$ \scriptsize 0.2 &  2.7 \\
    \cmidrule(l{2pt}r{2pt}){1-2}
    \cmidrule(l{2pt}r{2pt}){3-4}
    \cmidrule(l{2pt}r{2pt}){5-5}
    \rowcolor{gray!40}
    CM & - & 88.1 $\pm$ \scriptsize 0.1 & 57.3 $\pm$ \scriptsize 0.5 &  1.2 \\
    \cmidrule(l{2pt}r{2pt}){1-2}
    \cmidrule(l{2pt}r{2pt}){3-4}
    \cmidrule(l{2pt}r{2pt}){5-5}
    MLCM & \hphantom{0}1 OBIS CQ & 89.9 $\pm$ \scriptsize 0.1 & 60.8 $\pm$ \scriptsize 0.5 & 12.3 \\
    & \hphantom{0}5 OBIS CQs & 92.6 $\pm$ \scriptsize 0.1 & 66.6 $\pm$ \scriptsize 0.4 & 10.6 \\
    & 10 OBIS CQs & 94.1 $\pm$ \scriptsize 0.1 & 68.8 $\pm$ \scriptsize 0.4 &  9.7 \\
    & 15 OBIS CQs & 95.0 $\pm$ \scriptsize 0.1 & 70.6 $\pm$ \scriptsize 0.4 &  9.8 \\
    & 20 OBIS CQs & 95.7 $\pm$ \scriptsize 0.1 & 71.3 $\pm$ \scriptsize 0.4 &  9.9 \\
    \bottomrule
\end{tabular}

}
\vskip 0.1cm
\caption{Comparison of CM and MLCM for \emph{additive} preferences, a supply ratio of 1.5, 9 popular courses, and default parameterization for reporting mistakes.  Standard CM is highlighted in grey. Shown are averages over 500 runs. We normalize average and minimum utility by the average utility of \textsc{CM\textsuperscript{*}} and also show their 95\% CIs.}
\label{tab:app:welfare_SR1.5_Pop9}
\end{sc}
\end{table}

\newpage
\clearpage

\subsection{Reporting Mistakes Robustness Study for Additive Preferences}
\begin{figure}[h!]
    \centering
    \vskip -0.49cm
    \includegraphics[width=0.95\columnwidth]{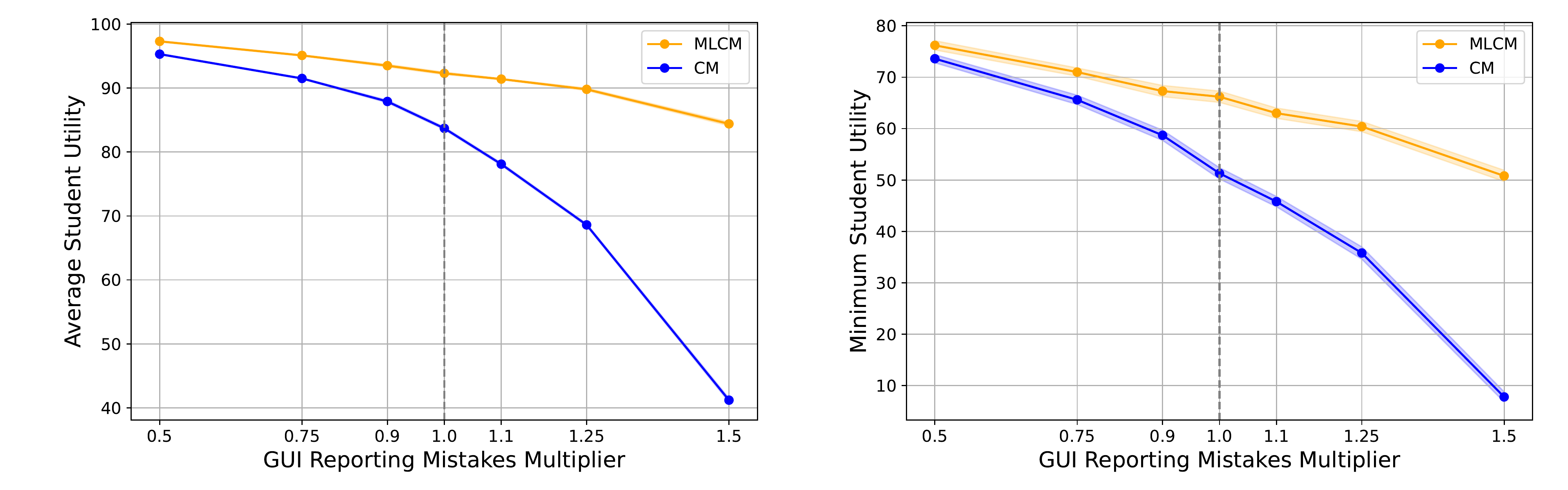}
    \vskip -0.3cm
    \caption{Reporting mistakes robustness experiment for \emph{additive} preferences for a supply ratio of 1.1 and 9 popular courses. Shown are average results in \% for the final allocation over 100 runs including 95\% CI.}
    \label{fig:app_noise_robustness_additive_sr1.1_pop9}
\end{figure}

\begin{figure}[h!]
    \centering
    \vskip -1.09cm
    \includegraphics[width=0.95\columnwidth]{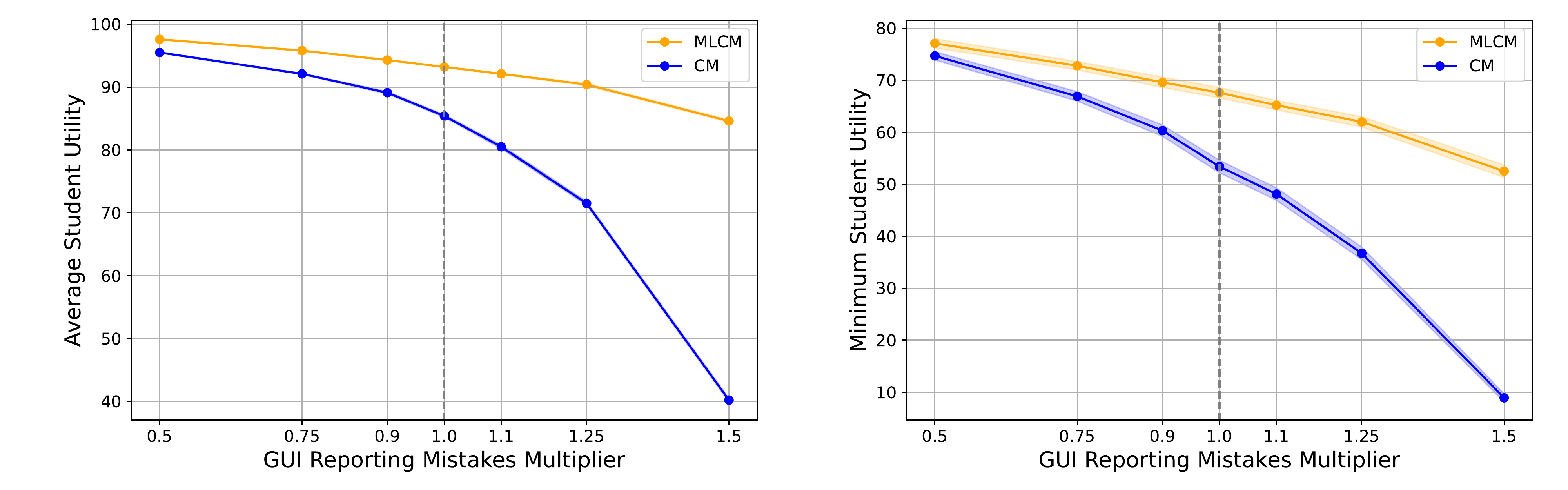}
    \vskip -0.3cm
    \caption{Reporting mistakes robustness experiment for \emph{additive} preferences for a supply ratio of 1.25 and 9 popular courses. Shown are average results in \% for the final allocation over 100 runs including 95\% CI.}
    \label{fig:app_noise_robustness_additive_sr1.25_pop9}
\end{figure}

\begin{figure}[h!]
    \centering
    \vskip -1.09cm
    \includegraphics[width=0.95\columnwidth]{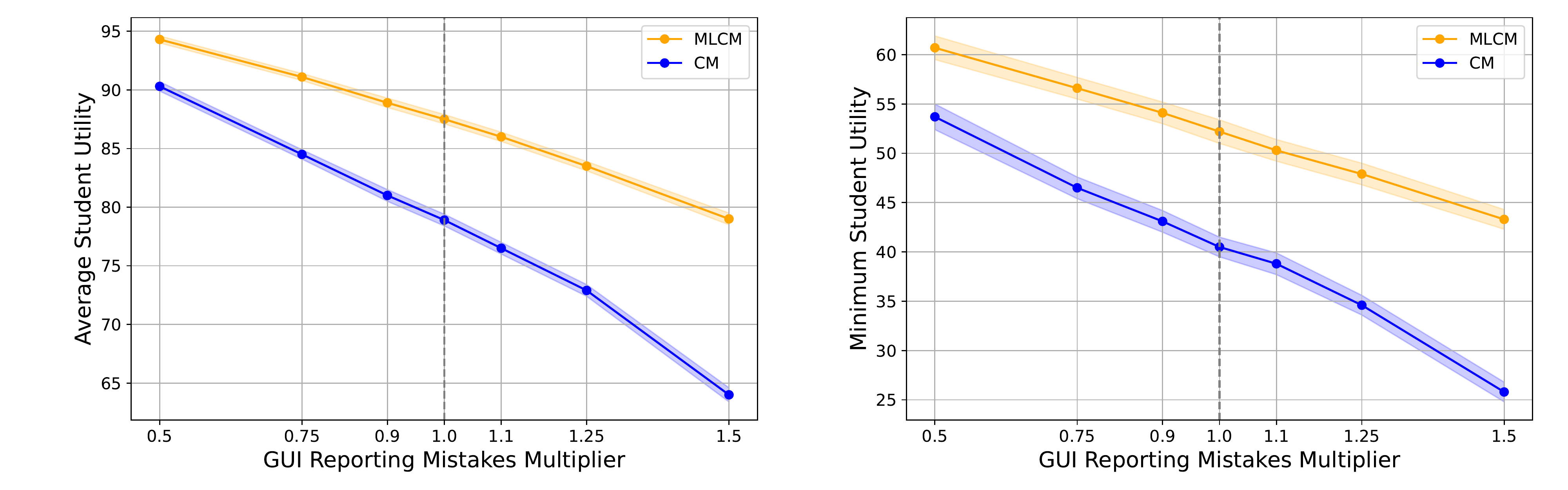}
    \vskip -0.3cm
    \caption{Reporting mistakes robustness experiment for \emph{additive} preferences for a supply ratio of 1.5 and 9 popular courses. Shown are average results in \% for the final allocation over 100 runs including 95\% CI.}
    \vskip -2cm
    \label{fig:app_noise_robustness_additive_sr1.5_pop9}
\end{figure}

In this subsection, we present in \Cref{fig:app_noise_robustness_additive_sr1.1_pop9,fig:app_noise_robustness_additive_sr1.25_pop9,fig:app_noise_robustness_additive_sr1.5_pop9} the reporting mistakes robustness study when students' true utility functions are restricted to be \emph{additive/linear}. Please see \Cref{subsec:Reporting Mistakes Robustness Study} for details on the experiment setup.

Similar to the results in \Cref{subsec:Reporting Mistakes Robustness Study}, as $\gamma$ increases, the performance of both CM and MLCM monotonically decreases. MLCM significantly outperforms CM for all $\gamma \in [0.5, 1.5]$. As $\gamma$ increases, the relative performance gap of the two mechanisms gets significantly larger. Those results could be further improved by retuning MLCM's hyperparameters for each value of $\gamma$.


\newpage
\clearpage
\subsection{Should Individual Students Opt Into MLCM?}
In this section, we present the experimental results measuring a students expected gain from opting into MLCM when students' true utility functions are \emph{additive/linear}. 
Our analysis encompasses 100 instances, with 100 students in each instance, totaling 10,000 students. We report averages over those 10,000 students.
Please see \Cref{subsec:Should Individual Students Opt Into MLCM?} for more details on the experiment setup.

\paragraph{No Other Student to Opt into the ML Feature}\Cref{tab:app:table_first_student_to_switch_additive} shows the results when all other students opt out of MLCM's ML feature. 
The \emph{expected relative gain} from opting in is at least $8.9\%$. Furthermore, the student prefers the ``MLCM schedule'' in at least $76$\% of the cases, while she prefers the ``CM schedule'' in at most $14.1$\% of the cases. As the number of CQs the student answers increases, the benefit from opting into MLCM's ML feature becomes even larger. 

\begin{table}[h!]
\vskip -0.49cm
	\robustify\bfseries
	\centering
	\begin{sc}
	\resizebox{0.85\columnwidth}{!}{
	\setlength\tabcolsep{4pt}
\begin{tabular}{ccc r  r  r r r r}
\toprule
\multicolumn{3}{c}{\textbf{Setting}} & \multicolumn{3}{c}{\textbf{Preferred Mechanism}} & \multicolumn{3}{c}{\textbf{Gain from Opting Into MLCM}}  \\
\cmidrule(l{2pt}r{2pt}){1-3}
\cmidrule(l{2pt}r{2pt}){4-6}
\cmidrule(l{2pt}r{2pt}){7-9}
\multicolumn{1}{c}{\textbf{SR}} & \multicolumn{1}{c}{\textbf{\#PoP}} & \multicolumn{1}{c}{\textbf{\#CQs}}& \multicolumn{1}{c}{\textbf{MLCM}} & \multicolumn{1}{c}{\textbf{CM}} & \multicolumn{1}{c}{\textbf{Indiff.}} & \multicolumn{1}{c}{\textbf{Expected}} & \multicolumn{1}{c}{\textbf{if pref MLCM}} &  \multicolumn{1}{c}{\textbf{if pref CM}} \\
\cmidrule(l{2pt}r{2pt}){1-3}
\cmidrule(l{2pt}r{2pt}){4-6}
\cmidrule(l{2pt}r{2pt}){7-9}
1.1 & 9 & 10 &    81.13\% &   13.09\% &      5.78\% &      12.74\% &  16.80\% & -6.63\% \\
1.1 & 9 & 15 &    84.66\% &   10.98\% &      4.36\% &      14.51\% &  18.05\% & -6.83\% \\
1.1 & 9 & 20 &    86.47\% &    9.57\% &      3.96\% &      15.71\% &  18.90\% & -6.45\% \\
\cmidrule(l{2pt}r{2pt}){1-3}
\cmidrule(l{2pt}r{2pt}){4-6}
\cmidrule(l{2pt}r{2pt}){7-9}
1.25 & 9 & 10 &    78.68\% &   13.92\% &      7.40\% &      11.07\% &  15.27\% & -6.69\% \\
1.25 & 9 & 15 &    82.55\% &   11.77\% &      5.68\% &      12.58\% &  16.15\% & -6.42\% \\
1.25 & 9 & 20 &    85.14\% &   10.34\% &      4.52\% &      13.69\% &  16.81\% & -5.94\% \\
\cmidrule(l{2pt}r{2pt}){1-3}
\cmidrule(l{2pt}r{2pt}){4-6}
\cmidrule(l{2pt}r{2pt}){7-9}
1.5 & 9 & 10 &    76.09\% &   14.13\% &      9.78\% &       8.87\% &  12.81\% & -6.28\% \\
1.5 & 9 & 15 &    80.38\% &   12.14\% &      7.48\% &      10.28\% &  13.67\% & -5.90\% \\
1.5 & 9 & 20 &    83.22\% &   10.07\% &      6.71\% &      11.15\% &  14.08\% & -5.77\% \\
\bottomrule
\end{tabular}
}
\vskip 0.1cm
\caption{Expected gain of opting into MLCM's ML feature when no other student opts in for \emph{additive} preferences. Shown are average results across 10,000 students per setting (\textsc{SR},\textsc{\#PoP},\textsc{\#CQs}). CIs are $\approx0$.}
\label{tab:app:table_first_student_to_switch_additive}
\end{sc}
\vskip -0.5cm
\end{table}

\paragraph{All Other Students to Opt into the ML Feature}\Cref{tab:app:table_last_student_to_switch_additive} shows the results when all other students opt into MLCM's ML feature.
The \emph{expected relative gain} from opting in is at least $7.6\%$. Furthermore, the student prefers the ``MLCM schedule'' in at least $71.9$\% of the cases, while she prefers the ``CM schedule'' in at most $18.9$\% of the cases. As the number of CQs the student answers increases, the benefit from opting into MLCM's ML feature becomes even larger. 

\begin{table}[h!]
\vskip -0.4cm
	\robustify\bfseries
	\centering
	\begin{sc}
	\resizebox{0.85\columnwidth}{!}{
	\setlength\tabcolsep{4pt}
\begin{tabular}{ccc r  r  r r r r}
\toprule
\multicolumn{3}{c}{\textbf{Setting}} & \multicolumn{3}{c}{\textbf{Preferred Mechanism}} & \multicolumn{3}{c}{\textbf{Gain from Opting Into MLCM}}  \\
\cmidrule(l{2pt}r{2pt}){1-3}
\cmidrule(l{2pt}r{2pt}){4-6}
\cmidrule(l{2pt}r{2pt}){7-9}
\multicolumn{1}{c}{\textbf{SR}} & \multicolumn{1}{c}{\textbf{\#PoP}} & \multicolumn{1}{c}{\textbf{\#CQs}}& \multicolumn{1}{c}{\textbf{MLCM}} & \multicolumn{1}{c}{\textbf{CM}} & \multicolumn{1}{c}{\textbf{Indiff.}} & \multicolumn{1}{c}{\textbf{Expected}} & \multicolumn{1}{c}{\textbf{if pref MLCM}} &  \multicolumn{1}{c}{\textbf{if pref CM}} \\
\cmidrule(l{2pt}r{2pt}){1-3}
\cmidrule(l{2pt}r{2pt}){4-6}
\cmidrule(l{2pt}r{2pt}){7-9}
1.1 & 9 & 10 &     75.7\% &   18.14\% &      6.16\% &      10.17\% &  15.01\% & -6.49\% \\
1.1 & 9 & 15 &    79.88\% &   15.51\% &      4.61\% &      11.68\% &  15.86\% & -6.47\% \\
1.1 & 9 & 20 &    81.11\% &   14.73\% &      4.16\% &      12.25\% &  16.21\% & -6.12\% \\
\cmidrule(l{2pt}r{2pt}){1-3}
\cmidrule(l{2pt}r{2pt}){4-6}
\cmidrule(l{2pt}r{2pt}){7-9}
1.25 & 9 & 10 &    73.57\% &   18.92\% &      7.51\% &       9.07\% &  13.93\% & -6.37\% \\
1.25 & 9 & 15 &    76.75\% &   17.25\% &      6.00\% &       9.89\% &  14.30\% & -6.37\% \\
1.25 & 9 & 20 &    78.72\% &   16.65\% &      4.63\% &      10.55\% &  14.64\% & -5.91\% \\
\cmidrule(l{2pt}r{2pt}){1-3}
\cmidrule(l{2pt}r{2pt}){4-6}
\cmidrule(l{2pt}r{2pt}){7-9}
1.5 & 9 & 10 &    71.85\% &   18.27\% &      9.88\% &       7.59\% &  12.19\% & -6.59\% \\
1.5 & 9 & 15 &     74.7\% &   17.79\% &      7.51\% &       8.50\% &  12.76\% & -5.93\% \\
1.5 & 9 & 20 &     77.2\% &   16.36\% &      6.44\% &       9.07\% &  12.95\% & -5.86\% \\
\bottomrule
\end{tabular}

}
\vskip 0.1cm
\caption{Expected gain of opting into MLCM's ML feature when all other students also opt in for \emph{additive} preferences. Shown are average results across 10,000 students per setting (\textsc{SR},\textsc{\#PoP},\textsc{\#CQs}). CIs are $\approx0$.}
\label{tab:app:table_last_student_to_switch_additive}
\end{sc}
\end{table}

\newpage
\clearpage

\section{Accelerating A-CEEI computation for MVNNs} \label{subsec:app_speedup}
In this section, we detail the technical optimization that made computing an A-CEEI, using the algorithm described in \citet{rubinstein2023practical}, faster for the preferences learned by MLCM and represented by MVNNs compared to the CM GUI preferences, as shown in \Cref{subsec:MLCM_practicability}.

The main idea of the algorithm described in \citet{rubinstein2023practical} is that, for every price vector being examined, the algorithm is allowed to make small changes to the budgets of individual students in order to reduce the clearing error. 
Those changes however can invert the budget ordering, i.e., the priority, of different students. 
In order for a budget change to be permissible, it must hold that for every student $i$, that student is not envious of the schedule that any other student $i'$ with a lower budget priority (with respect to the original budgets) receives. 
Formally, let $x_i$ be the (tentative) schedule allocated to student $i$ and $x_{i'}$ be the tentative schedule allocated to student $i'$, where $i'$ is a student with lower budget priority compared to $i$. Then, it must hold that: 
\begin{gather}
    \nexists S \subseteq  x_{i'}: \widehat{u}_i(x_i) < \widehat{u}_i(S), \label{eq:eftb_constr_general}
\end{gather}
where $\widehat{u}_i$ is student $i$'s utility function, either with respect to her MVNN model, or with respect to her CM GUI reports. 
This is a constraint that cannot be encoded at a MIP level, but instead for every price vector and for every budget change being examined, this constraint needs to be individually checked, for all students. 
MVNNs however can only represent monotone value functions, i.e., $\mathcal{M}(x_i) \ge \mathcal{M}(S) \; \forall \; S \subseteq x$. 
Consequently, Constraint~\ref{eq:eftb_constr_general} simplifies to:
\begin{gather}
    \widehat{u}_i(x_i) < \widehat{u}_i(x_{i'}), \label{eq:eftb_constr_mvnn}
\end{gather}
Thus, in case MVNNs are used, MLCM, when examining a price vector and budget change, does not need to explicitly check that all constraints prescribed by \ref{eq:eftb_constr_general} are respected, but only those of the form of \ref{eq:eftb_constr_mvnn}.
Note that for students requesting schedules consisting of at most $5$ courses, this optimization can yield up to a $32$-fold decrease in the number of constraints that need to be explicitly checked.
This decrease in the number of constraints in practice yields a significant enough speed improvement that MLCM only requires between $30$\% to $60$\% of the time that CM requires to calculate an A-CEEI, as discussed in \Cref{subsec:MLCM_practicability}.

\section{Projecting the learned preferences back to the GUI language} \label{subsec:app_projection}
In this section, we detail how we project the learned student's preferences back to the CM GUI language described in \Cref{sec:Course-Match-Introduction}.

Let $\mathcal{M}_i$ be the ML model representing student $i$'s utility function. 
As discussed in $\Cref{sec:MachineLearningInstantiation}$, this is a regression model, meaning it can predict student $i$'s utility for any course schedule. 
Projecting $\mathcal{M}_i$ back to CM's GUI language is a two step process. 
First, we create a very large dataset of schedule-ML utility prediction pairs, using the trained model $\mathcal{M}_i$. 
Then, we extend the schedules in that dataset by generating and appending their polynomial interaction features of degree $2$. 
For $m$ courses, this extended representation has exactly $m + {m \choose 2}$ coefficients, i.e., one additional coefficient for each adjustment that the student could report. 
For example, if there are only $2$ courses, $a$ and $b$, 
then $[a,b]$ would be converted to $[a, b , ab]$ in the extended representation. 
Finally, for each student $i$, we train a constrained linear regression with $m + {m \choose 2}$ coefficients on her dataset of extended schedules-ML utility prediction pairs. 
The first $m$ coefficients of the constrained linear regression represent the student's learned base vales for the $m$ courses, and are restricted to be in the range $[0, 100]$, as in the GUI language. 
The next ${m \choose 2}$ coefficients represent the student's adjustments, and are restricted to be in the range $[-200, + 200]$, as would be the case for the GUI language. 
This process is formally described in \Cref{alg:gui_projection}. 
We create the dataset consisting of pairs of points in their extended representation and their predicted values by $\mathcal{M}_i$ in Lines~\ref{alg:extension_start} to \ref{alg:extension_end}. 
In Lines~\ref{alg:init_regression} to \ref{alg:end_regression} we initialize a constrained linear regression model and train it on that dataset. 
For the results in this paper, we generate $1000$ schedules for each student and train the constrained linear regression model using projected gradient descent.

\begin{algorithm}[t!]
\caption{Projection of Learned Student Preferences to CM GUI Language}
\label{alg_projection}
\textbf{Input}: ML model $\mathcal{M}_i$, number of schedules to sample $k$ \\
\textbf{Output}: Projection of $\mathcal{M}_i$ to the CM GUI language

\begin{algorithmic}[1] 
\STATE $\mathcal{D}_{\mathcal{M}_i} \gets \{\}$ \label{alg:extension_start}
\FOR{$j = 1$ to $k$} 
    \STATE $x \sim U \{0,1\}^{m}$
    \STATE $x' \gets \text{PolynomialFeatures}(x,2)$
    \STATE $\mathcal{D}_{\mathcal{M}_i} \gets \mathcal{D}_{\mathcal{M}_i} \cup (x', \mathcal{M}_i(x)) $ 
\ENDFOR \label{alg:extension_end} 
\STATE $l \gets $ Constrained linear regression model with $m + {m \choose 2}$ coefficients \label{alg:init_regression}
\STATE Restrict first $m$ coefficients to range $[0, 100]$
\STATE Restrict next ${m \choose 2}$ coefficients to range $[-200, 200]$
\STATE $l \gets \textsc{Train}(l, \mathcal{D}_{\mathcal{M}_i})$ \label{alg:end_regression}
\STATE \textbf{return} Trained regression model $l$ with learned student preferences 
\end{algorithmic}
\label{alg:gui_projection}
\end{algorithm}

\section{Price Approximation Sensitivity Experiments} \label{subsec:app_price_sensitivity}
In this section, we study MLCM's sensitivity to the quality of the approximate prices used in MLCM's phase $4$. 

\begin{figure}[h!]
    \vskip -0.49cm
    \centering
    \includegraphics[width=0.9\columnwidth]{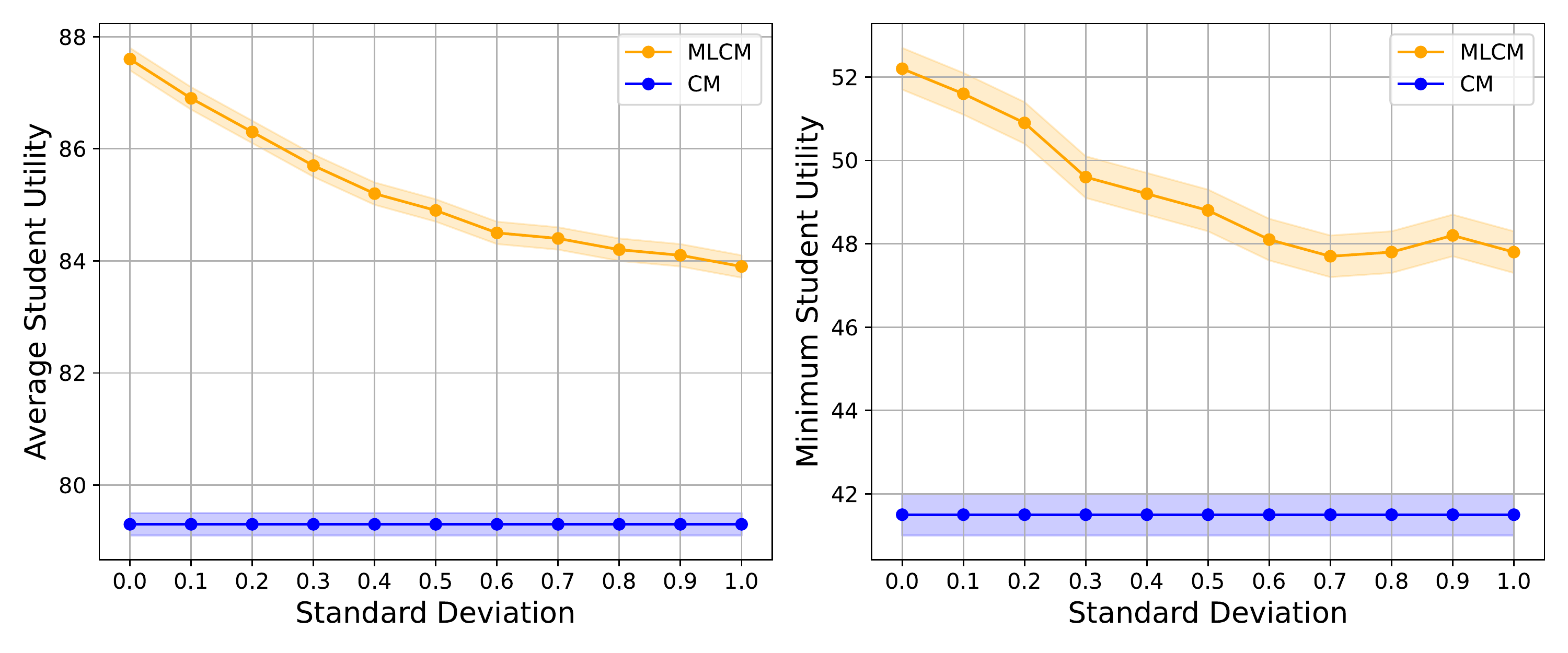}
    \vskip -0.2cm
    \caption{Approximate price sensitivity experiment for a supply ratio of 1.25, 9 popular courses and normally distributed, additive noise in the price approximation. Shown are average results in \% for the final allocation over 500 runs including 95\% CIs.}
    \label{fig:app_price_sensitivity_sr_1.25_pop_9_normal}
\end{figure}

\begin{figure}[h!]
\vskip -0.2cm
    \centering
    \includegraphics[width=0.9\columnwidth]{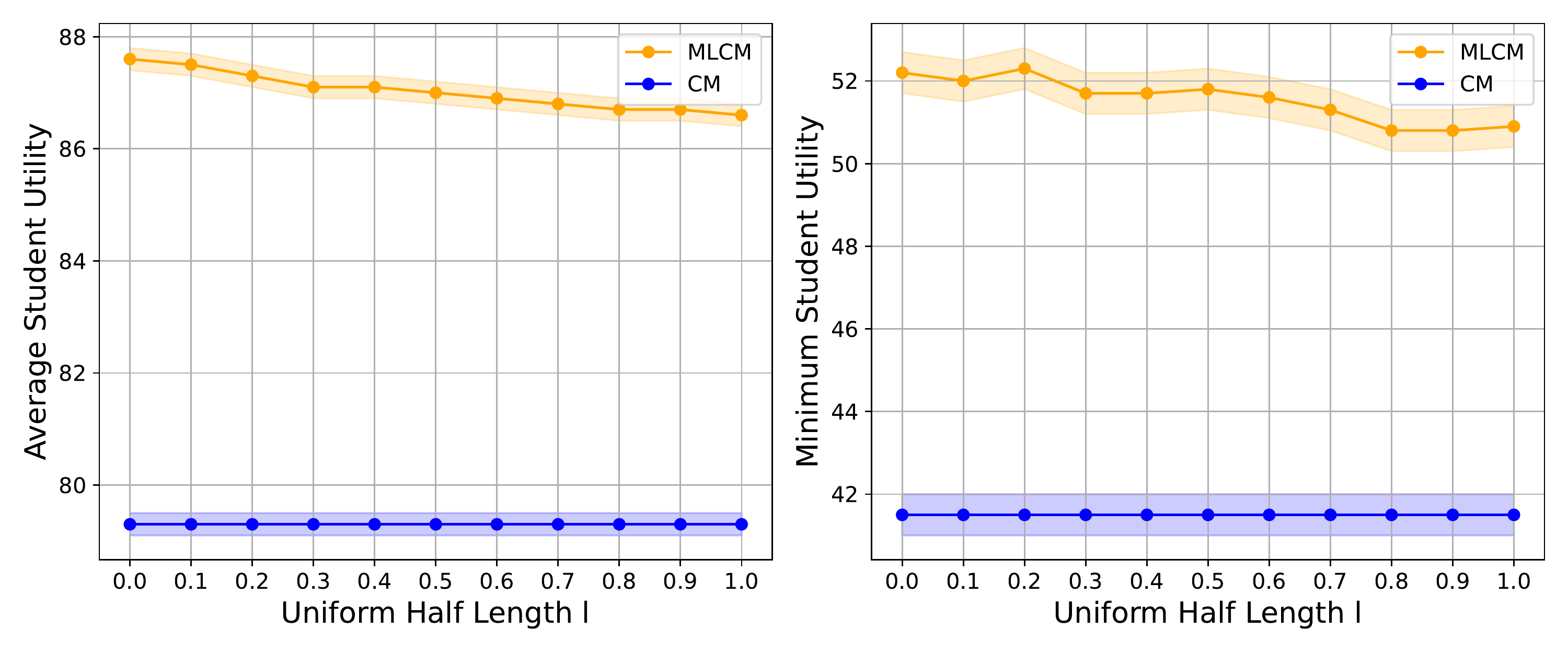}
    \vskip -0.2cm
    \caption{Approximate price sensitivity experiment for a supply ratio of 1.25, 9 popular courses and uniformly distributed, multiplicative noise in the price approximation. Shown are average results in \% for the final allocation over 500 runs including 95\% CIs.}
    \label{fig:app_price_sensitivity_sr_1.25_pop_9_uniform}
\end{figure}

For this test, we use again our MLCM mechanism as described in \Cref{subsec:mlcm_details} and perform a welfare experiment very similar to the one described in \Cref{subsec:experimental_results}.
The difference is that now, after generating the approximate prices in MLCM's Phase $3$, we perturb them by adding some amount of noise to them, and use the resulting prices as input to our OBIS algorithm to generate the CQs. 
We test with two different types of noise. 
The first one is an additive noise that follows a uniform distribution, for which we vary its standard deviation. 
The second one is a multiplicative noise that follows a uniform distribution, for which we vary its uniform range. 
Formally, let ${p}$ be the prices generated in MLCM's Phase $3$, and $p'$ be the noisy estimate of those prices.  
For the normally distributed, additive noise we have: 
\begin{gather*}
    p_j' = p + \epsilon, \epsilon \sim N(0, \sigma)  
\end{gather*}
And for the uniformly distributed, multiplicative noise we have:
\begin{gather*}
    p_j' = p (1 + \epsilon), \epsilon \sim U[-l, +l],
\end{gather*}
where $2l$ is the interval length of the uniform distribution.  
In both cases, we restrict the prices to be non-negative and less or equal to the lowest student budget (note that the difference between the lowest and highest student budget is only $4$\%).  
We report average and minimum student utility as a function of the amount of noise in the approximate prices, with each student answering $10$ CQs in MLCM, similarly to the noise robustness experiment described in \Cref{subsec:Reporting Mistakes Robustness Study}. 
\Cref{fig:app_price_sensitivity_sr_1.25_pop_9_normal,fig:app_price_sensitivity_sr_1.25_pop_9_uniform} show these results for the normally and uniformly distributed noise respectively, for the most realistic setting of an SR of $1.25$ and $9$ popular courses. 

\begin{figure}[h!]
\vskip -0.2cm
    \centering
    \includegraphics[width=0.9\columnwidth]{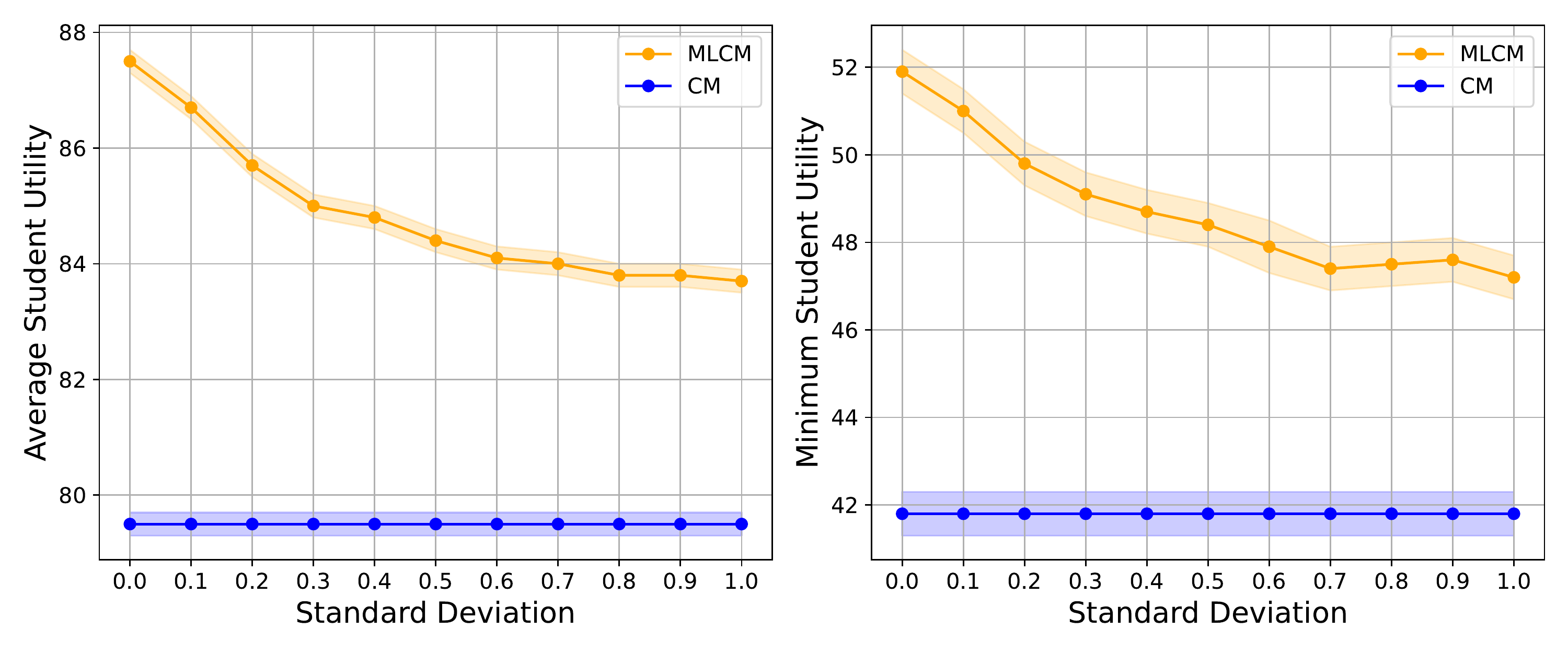}
    \vskip -0.2cm
    \caption{Approximate price sensitivity experiment for a supply ratio of 1.1, 9 popular courses and normally distributed, additive noise in the price approximation. Shown are average results in \% for the final allocation over 500 runs including 95\% CIs.}
    \label{fig:app_price_sensitivity_sr_1.1_pop_9_normal}
\end{figure}

\begin{figure}[h!]
    \vskip -0.5cm
    \centering
    \includegraphics[width=0.9\columnwidth]{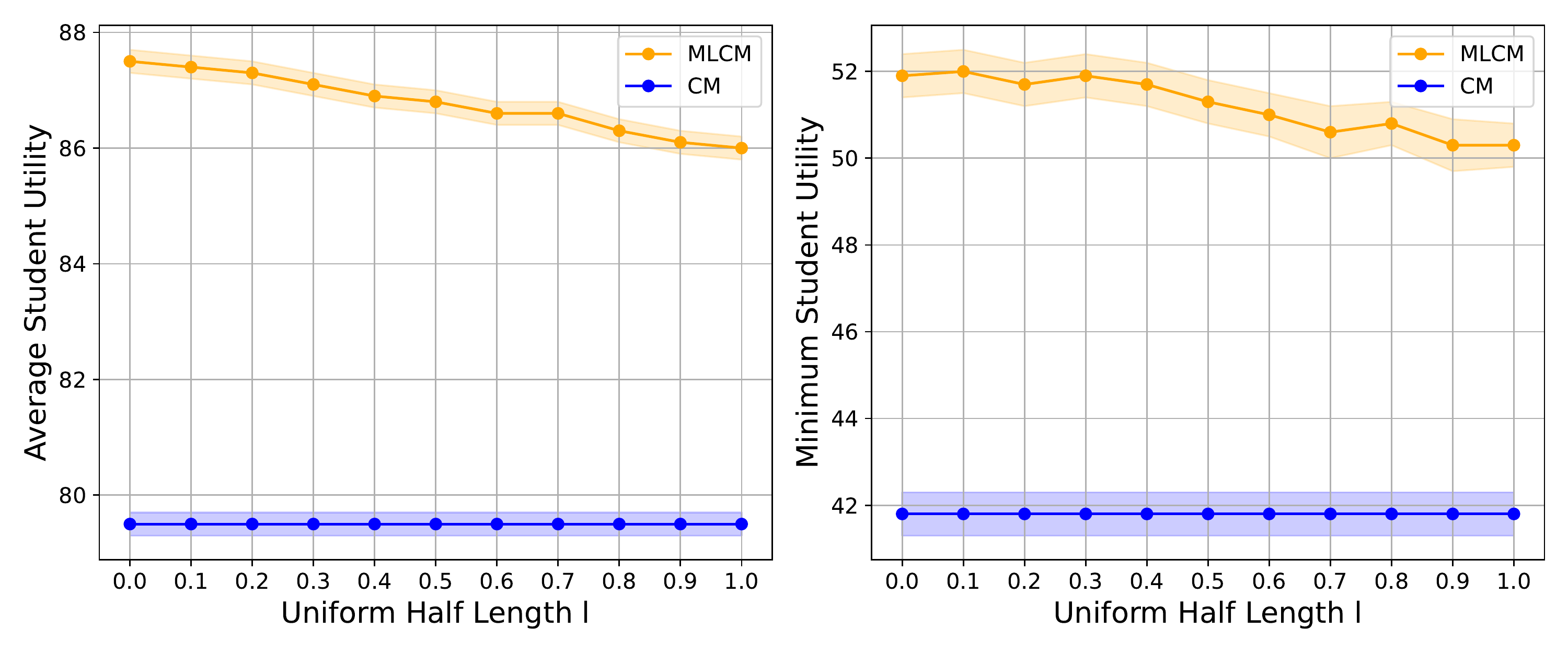}
    \vskip -0.2cm
    \caption{Approximate price sensitivity experiment for a supply ratio of 1.1, 9 popular courses and uniformly distributed, multiplicative noise in the price approximation. Shown are average results in \% for the final allocation over 500 runs including 95\% CIs.}
    \label{fig:app_price_sensitivity_sr_1.1_pop_9_uniform}
\end{figure}

\begin{figure}[h!]
    \vskip -0.2cm
    \centering
    \includegraphics[width=0.9\columnwidth]{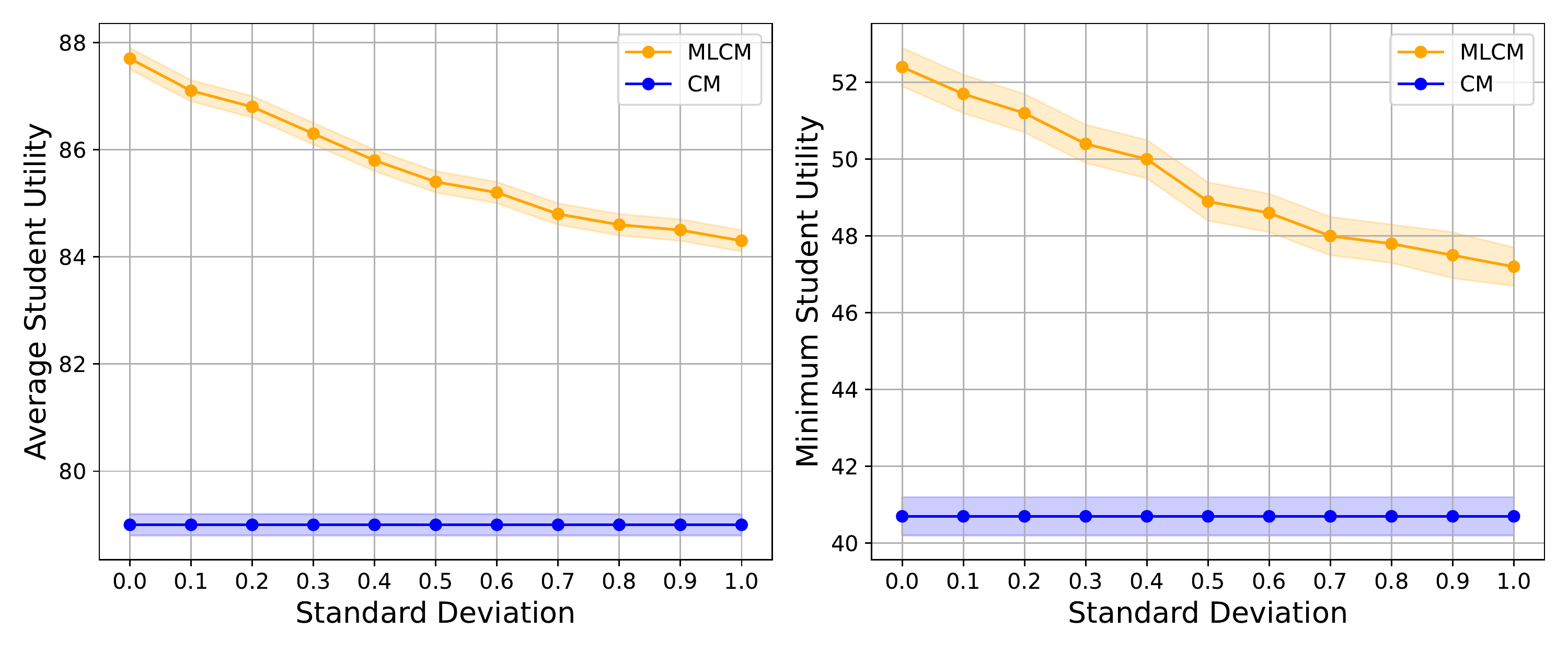}
    \vskip -0.2cm
    \caption{Approximate price sensitivity experiment for a supply ratio of 1.5, 9 popular courses and normally distributed, additive noise in the price approximation. Shown are average results in \% for the final allocation over 500 runs including 95\% CIs.}
    \label{fig:app_price_sensitivity_sr_1.5_pop_9_normal}
\end{figure}

\begin{figure}[h!]
\vskip -0.2cm
    \centering
    \includegraphics[width=0.9\columnwidth]{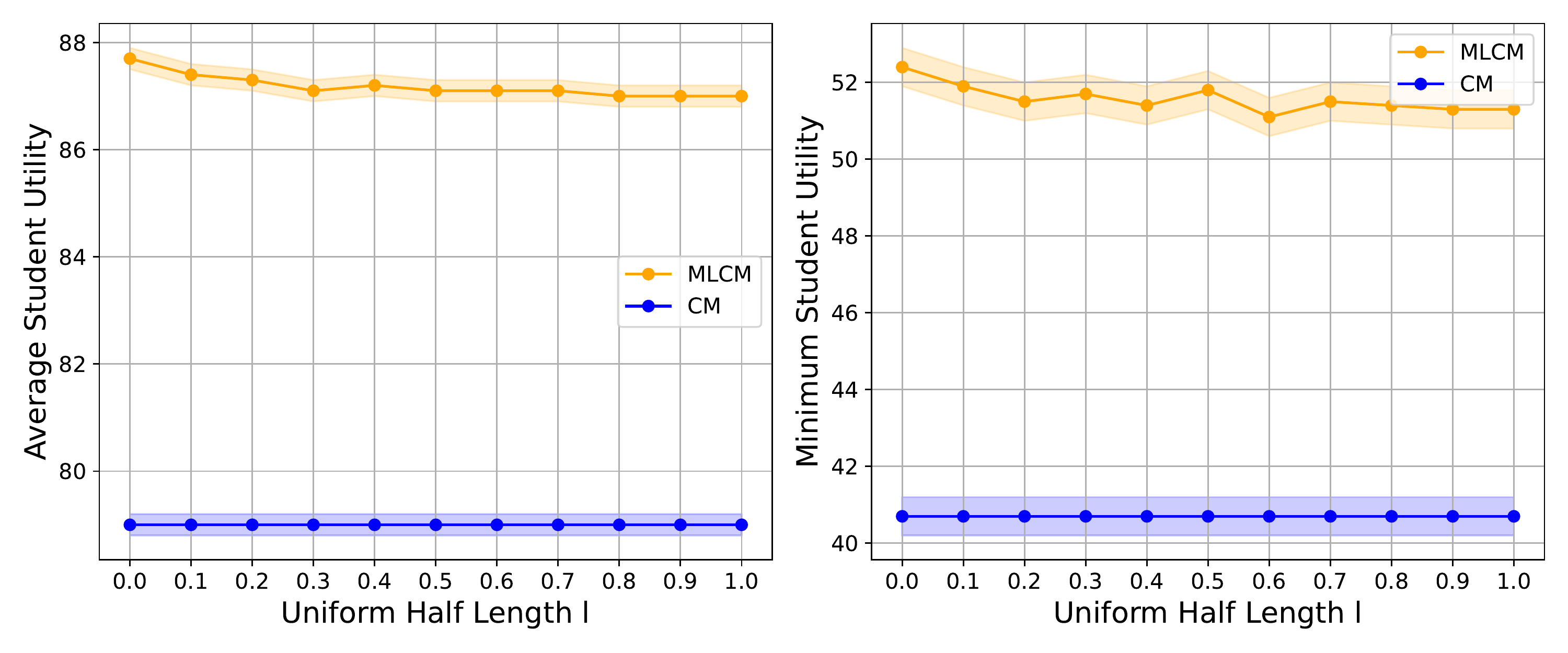}
    \vskip -0.2cm
    \caption{Approximate price sensitivity experiment for a supply ratio of 1.5, 9 popular courses and uniformly distributed, multiplicative noise in the price approximation. Shown are average results in \% for the final allocation over 500 runs including 95\% CIs.}
    \label{fig:app_price_sensitivity_sr_1.5_pop_9_uniform}
\end{figure}

\begin{figure}[h!]
\vskip -0.5cm
    \centering
    \includegraphics[width=0.9\columnwidth]{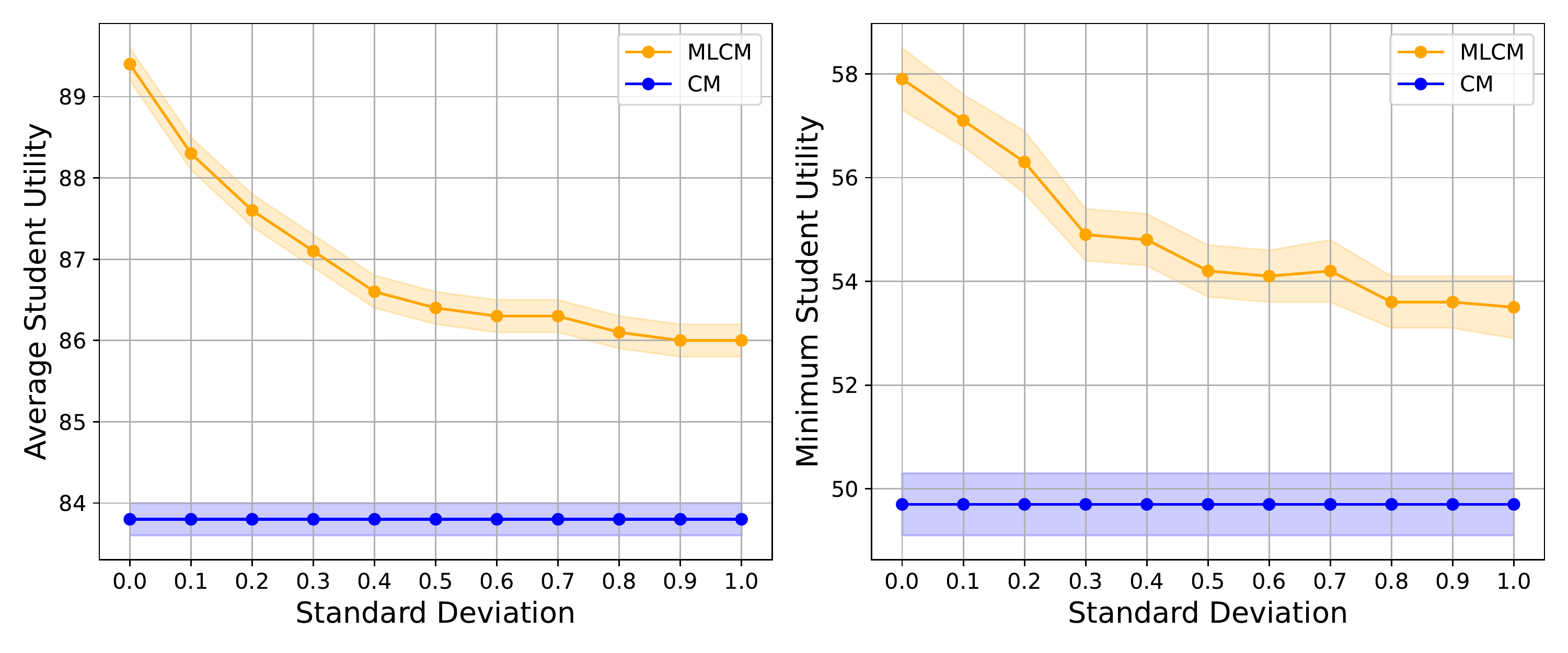}
    \vskip -0.2cm
    \caption{Approximate price sensitivity experiment for a supply ratio of 1.25, 6 popular courses and normally distributed, additive noise in the price approximation. Shown are average results in \% for the final allocation over 500 runs including 95\% CIs.}
    \label{fig:app_price_sensitivity_sr_1.25_pop_6_normal}
\end{figure}

\begin{figure}[h!]
\vskip -0.2cm
    \centering
    \includegraphics[width=0.9\columnwidth]{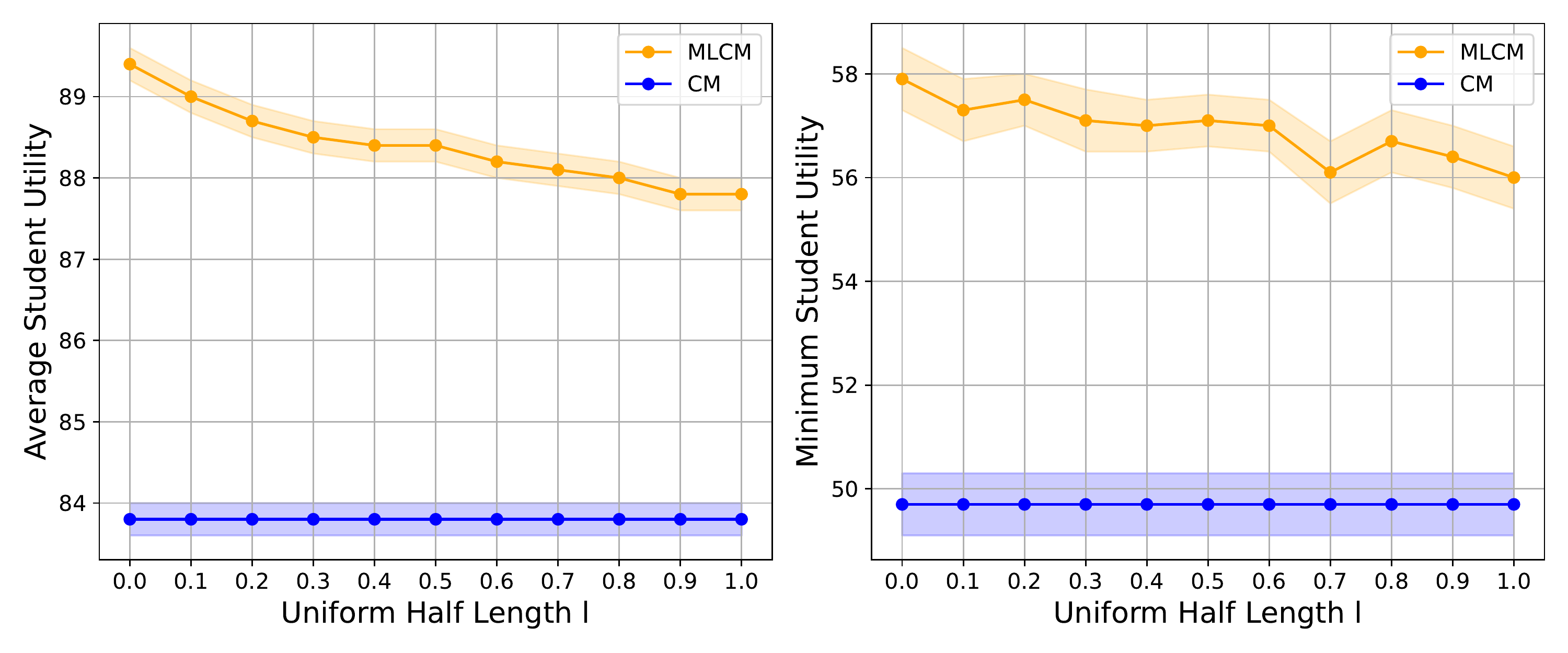}
    \vskip -0.2cm
    \caption{Approximate price sensitivity experiment for a supply ratio of 1.25, 6 popular courses and uniformly distributed, multiplicative noise in the price approximation. Shown are average results in \% for the final allocation over 500 runs including 95\% CIs.}
    \label{fig:app_price_sensitivity_sr_1.25_pop_6_uniform}
\end{figure}

\begin{figure}[h!]
\vskip -0.2cm
    \centering
    \includegraphics[width=0.9\columnwidth]{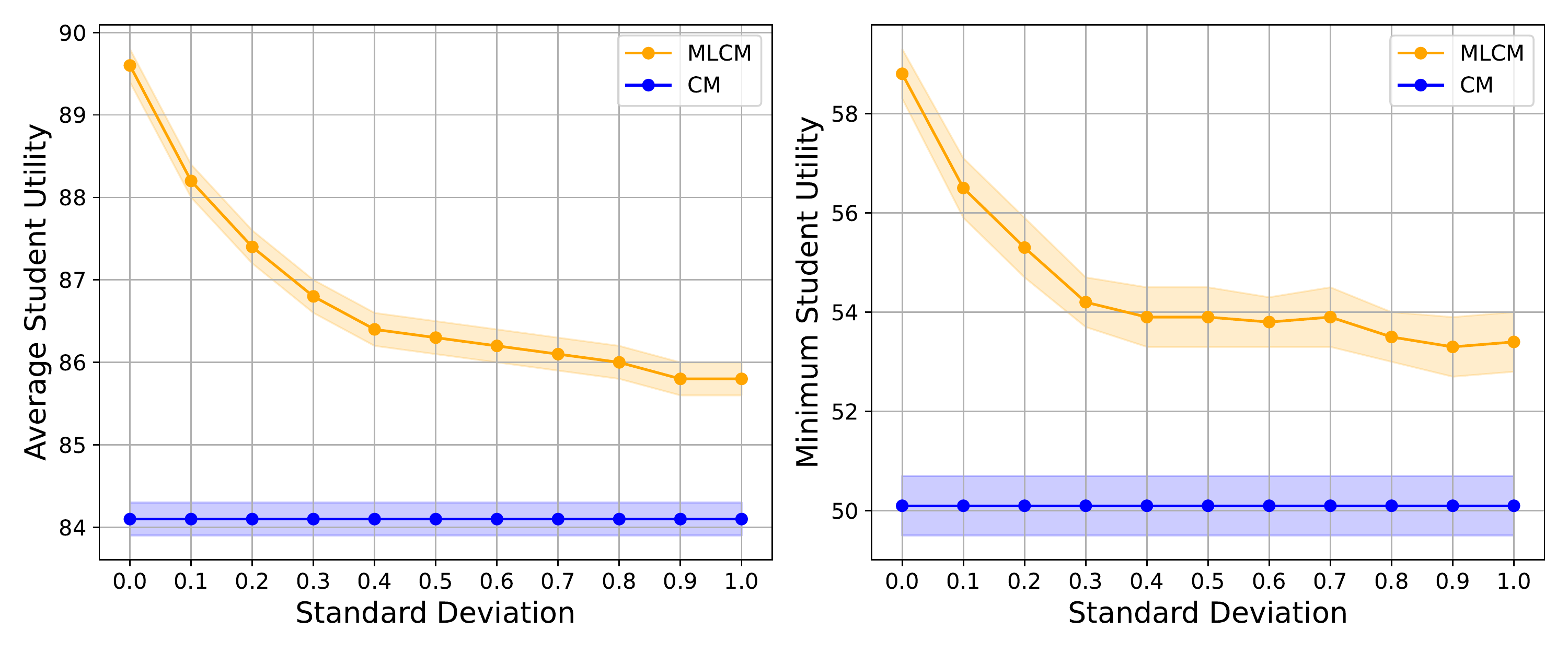}
    \vskip -0.2cm
    \caption{Approximate price sensitivity experiment for a supply ratio of 1.1, 6 popular courses and normally distributed, additive noise in the price approximation. Shown are average results in \% for the final allocation over 500 runs including 95\% CIs.}
    \label{fig:app_price_sensitivity_sr_1.1_pop_6_normal}
\end{figure}

\begin{figure}[h!]
\vskip -0.5cm
    \centering
    \includegraphics[width=0.9\columnwidth]{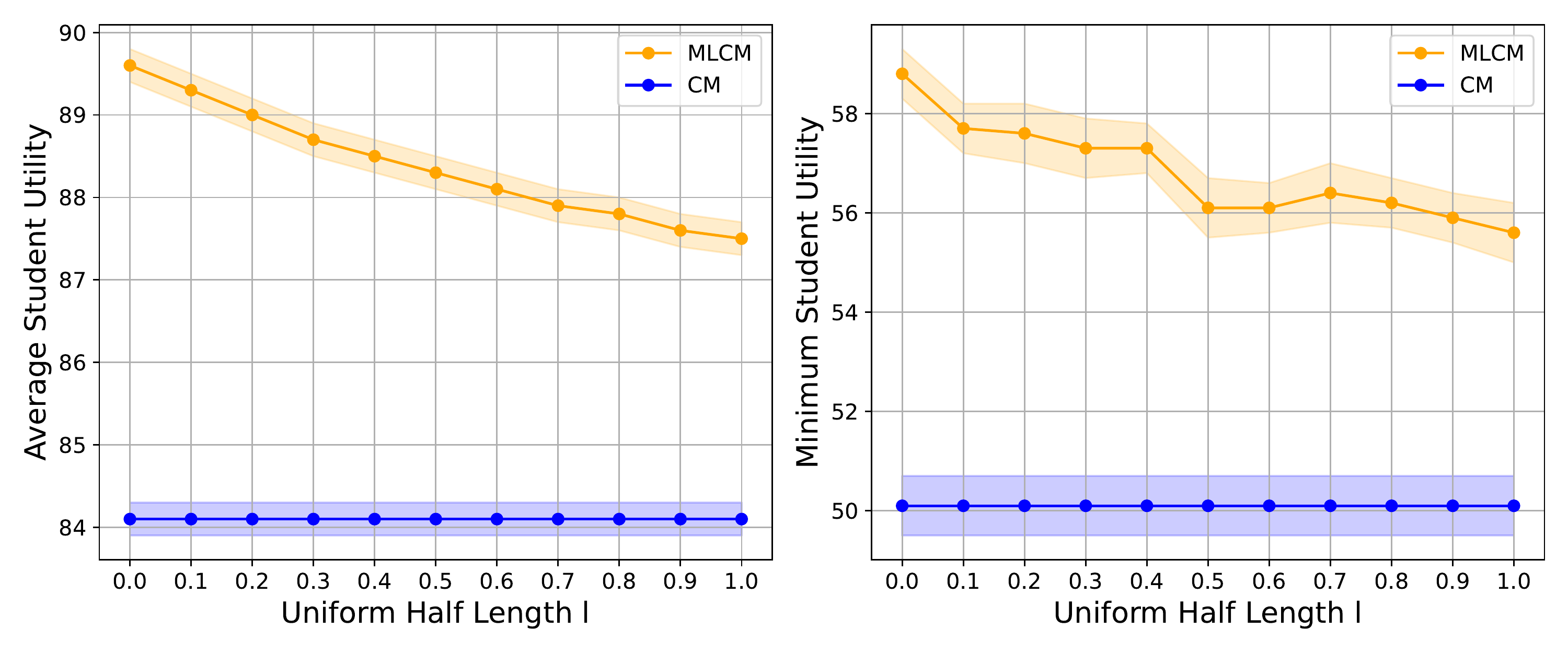}
    \vskip -0.2cm
    \caption{Approximate price sensitivity experiment for a supply ratio of 1.1, 6 popular courses and uniformly distributed, multiplicative noise in the price approximation. Shown are average results in \% for the final allocation over 500 runs including 95\% CIs.}
    \label{fig:app_price_sensitivity_sr_1.1_pop_6_uniform}
\end{figure}

\begin{figure}[h!]
\vskip -0.2cm
    \centering
    \includegraphics[width=0.9\columnwidth]{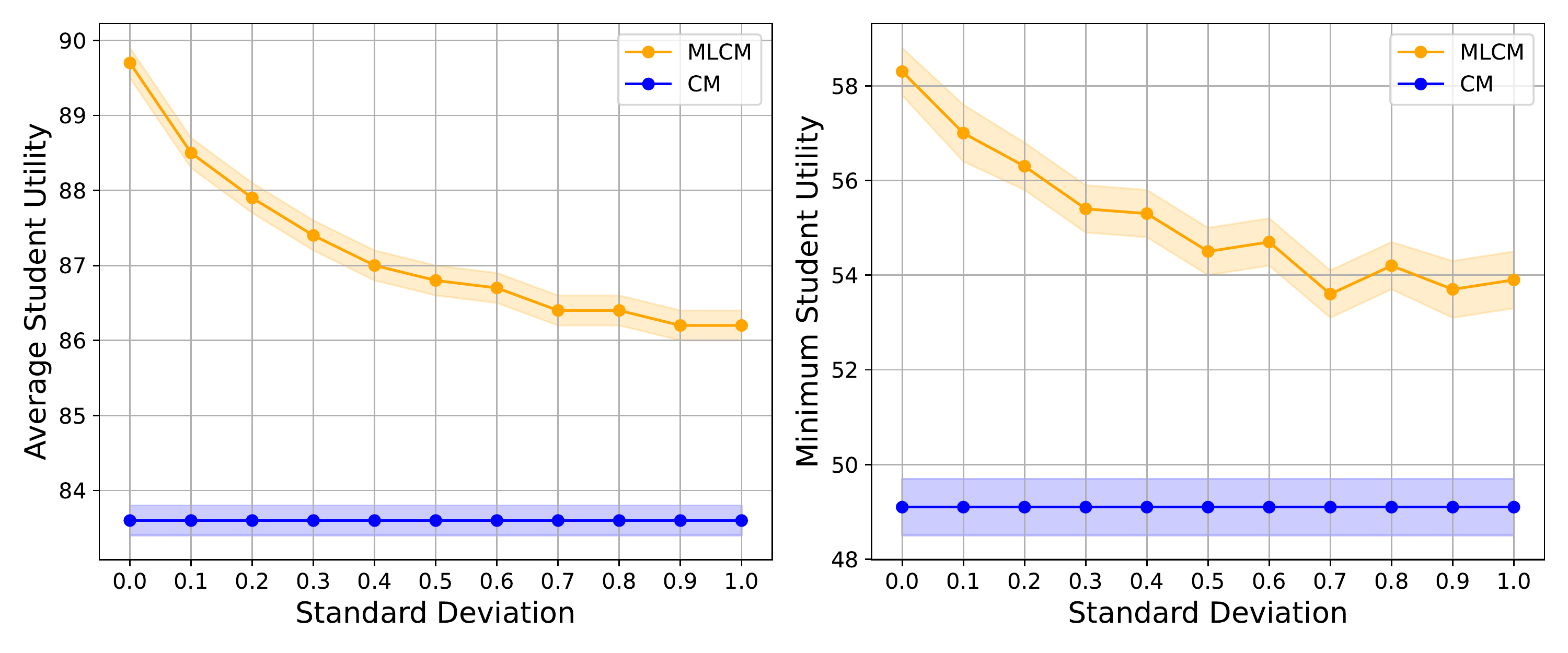}
    \vskip -0.2cm
    \caption{Approximate price sensitivity experiment for a supply ratio of 1.5, 6 popular courses and normally distributed, additive noise in the price approximation. Shown are average results in \% for the final allocation over 500 runs including 95\% CIs.}
    \label{fig:app_price_sensitivity_sr_1.5_pop_6_normal}
\end{figure}

\begin{figure}[h!]
\vskip -0.2cm
    \centering
    \includegraphics[width=0.9\columnwidth]{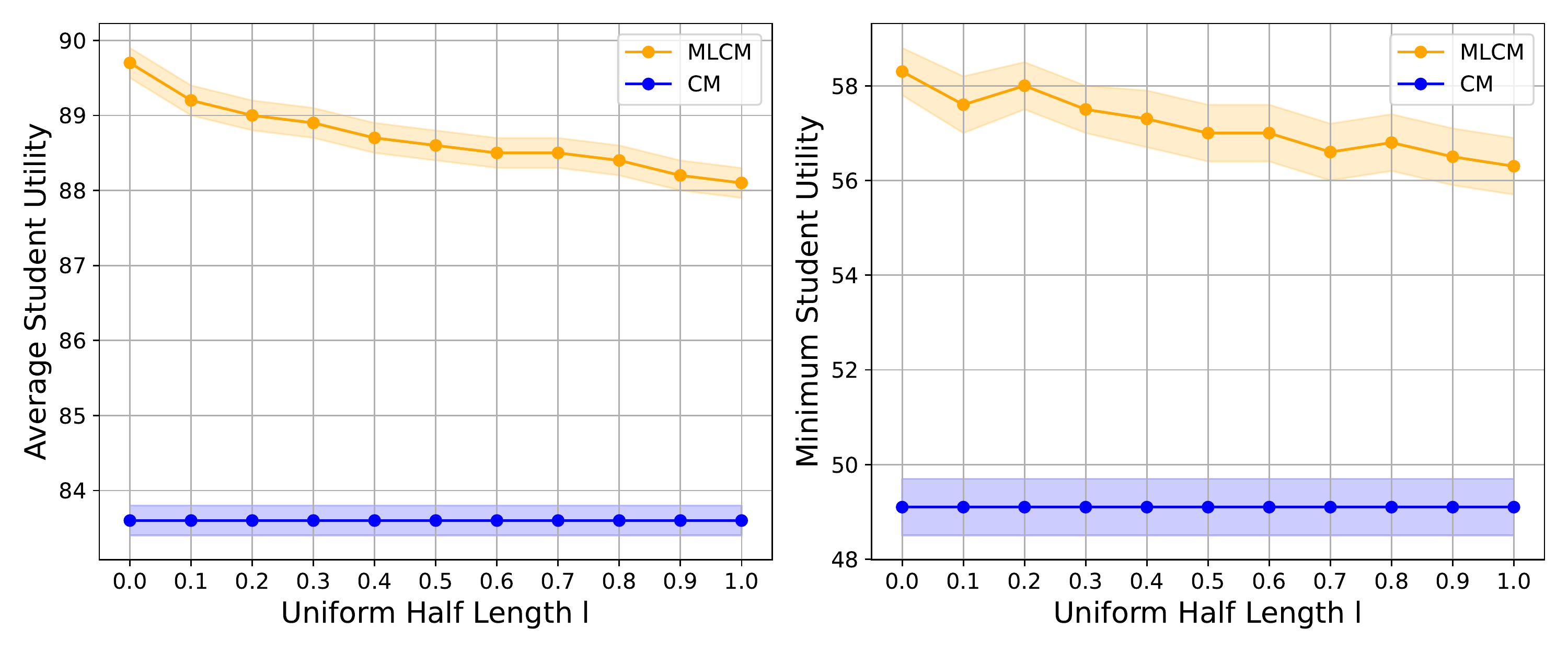}
    \vskip -0.2cm
    \caption{Approximate price sensitivity experiment for a supply ratio of 1.5, 6 popular courses and uniformly distributed, multiplicative noise in the price approximation. Shown are average results in \% for the final allocation over 500 runs including 95\% CIs.}
    \label{fig:app_price_sensitivity_sr_1.5_pop_6_uniform}
\end{figure}

For both types of noise, we can see that MLCM's performance, both in terms of average and minimum student utility, degrades as we increase the noise in the approximate prices used to generate the students' CQs. 
However, these same figures also show how robust MLCM is to the approximation error in these prices. 
For all noise configurations tested, MLCM significantly outperforms CM. 
Namely, when using a multiplicative, uniform noise with $l = 1$, which implies that the price of each course used in OBIS is sampled uniformly at random in the range between $0$ and two times its most accurate estimate, MLCM still causes a relative increase compared to CM in average and minimum student utility of $9.2$\% and $22.7$\% respectively. 
The results for all other settings are qualitatively identical and presented in Figures~\ref{fig:app_price_sensitivity_sr_1.1_pop_9_normal} to~\ref{fig:app_price_sensitivity_sr_1.5_pop_6_uniform}.

\newpage
\clearpage

\section{CQ Mistake Robustness Study}\label{subsec:app:CQ_robustness}
\begin{figure}[h!]
    \centering
    \includegraphics[width=0.9\columnwidth]{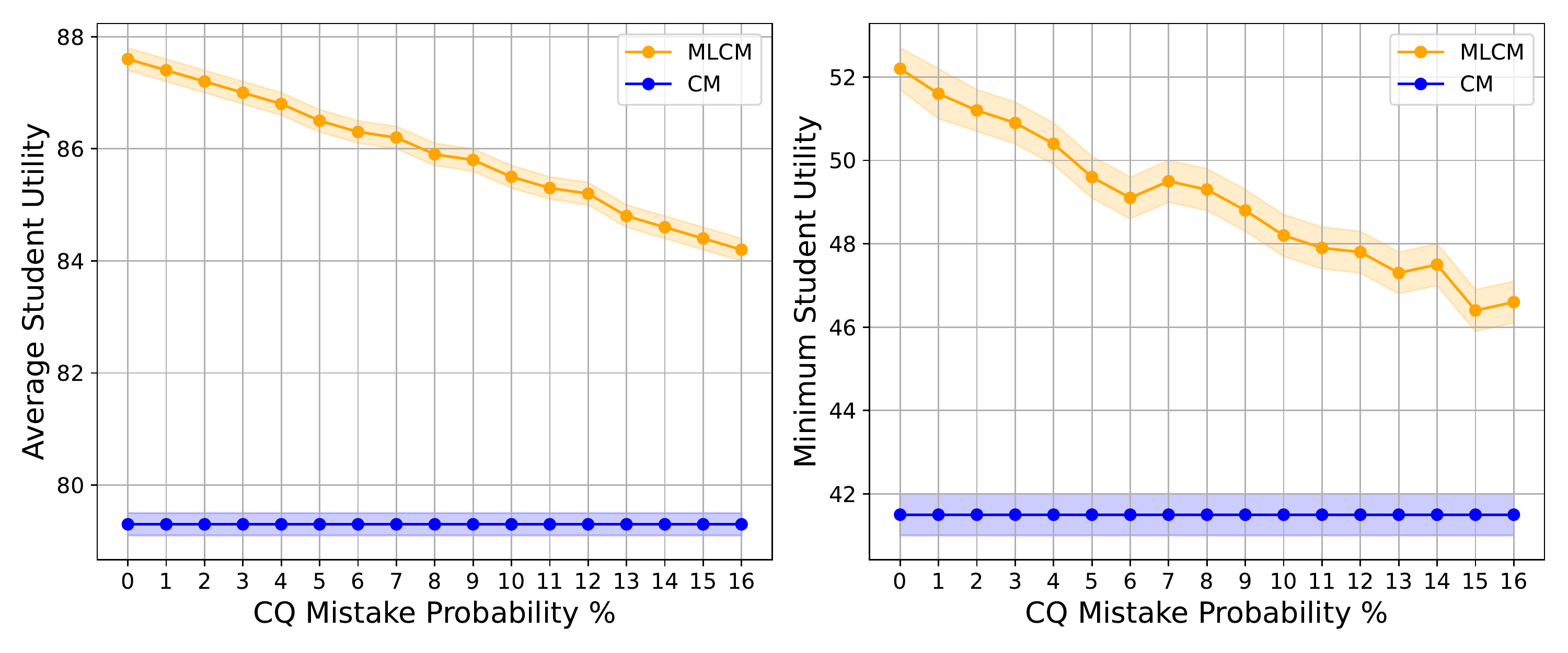}
    \caption{CQ reporting mistakes robustness experiment for a supply ratio of 1.25 and 9 popular courses. Shown are average results in \% for the final allocation over 500 runs including 95\% CI.}
    \label{fig:cq_accuracy_pop_9}
\end{figure}

\begin{figure}[h!]
\vskip -1cm
    \centering
    \includegraphics[width=0.9\columnwidth]{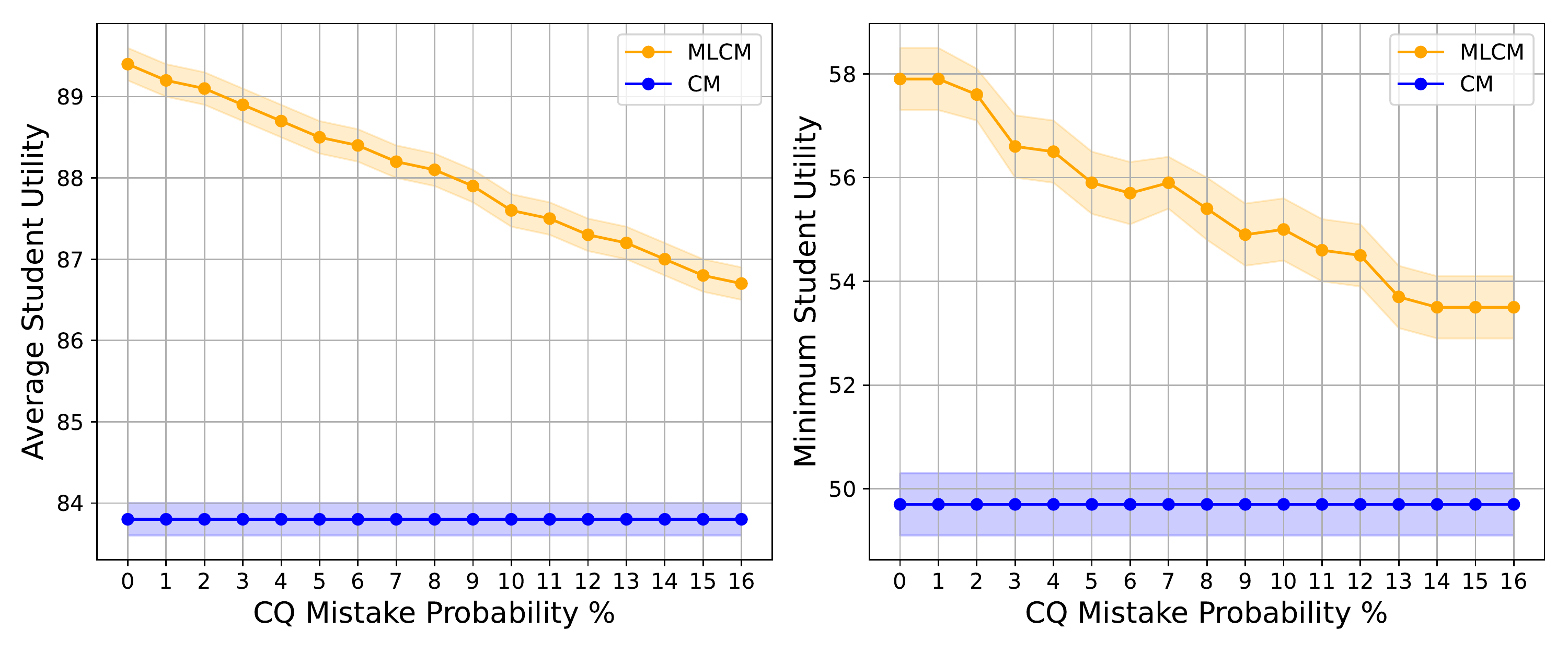}
    \caption{CQ reporting mistakes robustness experiment for a supply ratio of 1.25 and 9 popular courses. Shown are average results in \% for the final allocation over 500 runs including 95\% CI.}
    \label{fig:cq_accuracy_pop_6}
\end{figure}

In this section, we experimentally evaluate MLCM's robustness to the students' mistakes when answering CQs.
For this, we keep the setting fixed (to the most realistic SR of $1.25$ and $9$ or $6$ popular courses) and introduce a probability $p_M$ with which each student, when presented with a CQ, will provide a wrong response. 
For each value of $p_M$, we run the same 500 instances for MLCM, with each student answering $10$ CQs, and compare against CM. 
We test for values of $p_M$ in the range $[0, 0.16]$. 
Note that we consider this range extremely large, as $p_M = 0.16$ means that the students' accuracy when answering CQs is lower than their accuracy when using the cognitively more complicated GUI reporting language, as measured in the lab experiment of \citet{budish2022can}.
Additionally, in that lab experiment the authors treated the students' answers to CQs as ground truth,
arguing that it is a reasonable assumption for students to be able to report which of two schedules they prefer when asked so directly.

\Cref{fig:cq_accuracy_pop_9,fig:cq_accuracy_pop_6} show the results of this experiment. 
We see that MLCM's performance decreases as the students' mistake probability when answering CQs increases. 
However, MLCM significantly outperforms CM for all $p_M \in [0, 0.16]$. 
Note that $p_M = 0.16$ corresponds to students making even more mistakes in their CQ replies than in the original GUI language. 
Despite this, MLCM with 10 CQs per student still achieves a relative performance increase over CM for average and minimum student utility of $6.2$\% and $13.1$\% respective for $9$ popular courses, and a relative increase of $3.5$\% and $7.7$\% for average and minimum student utility respectively. 
Overall, these results highlight MLCM's robustness to the students' accuracy when answering CQs.


\section{Statistical Significance Tests} \label{A_significance_tests}
To test whether our results from Section \ref{Section_efficiency_results} are statistically significant, we perform a \textit{multivariate analysis of variance (MANOVA)} test, using as the independent variable the mechanism and as dependent variables (a) average student utility, and (b) the minimum student utility. We use as significance level for this test a value of $0.05$. The MANOVA test determined that there is a statistically significant treatment effect between the different mechanisms. Given this, we then performed a Post-Hoc tukey test for all pairs of mechanisms. For those tests, we used one dependent variable (i.e., either average or minimum student utility), with a significance level of $0.025$ where the null hypothesis is that there is no treatment effect between \textsc{Group1} and \textsc{Group2}, i.e., $\mathcal{H}_0:\mu_{Group1}= \mu_{Group2}$. The results for each setting are provided in \Cref{tab:tukey_1,tab:tukey_2,tab:tukey_3,tab:tukey_4,tab:tukey_5,tab:tukey_6,tab:tukey_7,tab:tukey_8,tab:tukey_9,tab:tukey_10,tab:tukey_11,tab:tukey_12,tab:tukey_13,tab:tukey_14,tab:tukey_15,tab:tukey_16,tab:tukey_17,tab:tukey_18}.

\begin{table}[h!]
        \vskip 0.5cm
	\robustify\bfseries
	\centering
	\begin{sc}
	\resizebox{0.75\columnwidth}{!}{
	\setlength\tabcolsep{4pt}
\begin{tabular}{llrrrrr}
\toprule
       group1 &        group2 &  meandiff &  p-adj &   lower &   upper &  reject \\
\midrule
CM &   MLCM (10 OBIS CQs) &    0.0797 & 0.0000 &  0.0754 &  0.0840 &    True \\
CM &   MLCM (20 OBIS CQs) &    0.1061 & 0.0000 &  0.1017 &  0.1104 &    True \\
CM &  MLCM (20 na\"{i}ve CQs) &    0.0777 & 0.0000 &  0.0734 &  0.0820 &    True \\
CM & MLCM (20 random CQs) &    0.0172 & 0.0000 &  0.0129 &  0.0215 &    True \\
CM &                  RSD &   -0.0436 & 0.0000 & -0.0480 & -0.0393 &    True \\
MLCM (10 OBIS CQs) &   MLCM (20 OBIS CQs) &    0.0263 & 0.0000 &  0.0220 &  0.0307 &    True \\
MLCM (10 OBIS CQs) &  MLCM (20 na\"{i}ve CQs) &   -0.0020 & 0.6886 & -0.0064 &  0.0023 &   False \\
MLCM (10 OBIS CQs) & MLCM (20 random CQs) &   -0.0625 & 0.0000 & -0.0668 & -0.0582 &    True \\
MLCM (10 OBIS CQs) &                  RSD &   -0.1234 & 0.0000 & -0.1277 & -0.1190 &    True \\
MLCM (20 OBIS CQs) &  MLCM (20 na\"{i}ve CQs) &   -0.0284 & 0.0000 & -0.0327 & -0.0241 &    True \\
MLCM (20 OBIS CQs) & MLCM (20 random CQs) &   -0.0888 & 0.0000 & -0.0932 & -0.0845 &    True \\
MLCM (20 OBIS CQs) &                  RSD &   -0.1497 & 0.0000 & -0.1540 & -0.1454 &    True \\
MLCM (20 na\"{i}ve CQs) & MLCM (20 random CQs) &   -0.0605 & 0.0000 & -0.0648 & -0.0561 &    True \\
MLCM (20 na\"{i}ve CQs) &                  RSD &   -0.1213 & 0.0000 & -0.1256 & -0.1170 &    True \\
MLCM (20 random CQs) &                  RSD &   -0.0609 & 0.0000 & -0.0652 & -0.0565 &    True \\
\bottomrule
\end{tabular}

}
\end{sc}
\vskip 0.1cm
\caption{Post-hoc tukey test for average student utility. Supply ratio 1.1, 9 popular courses. Significance level for \textsc{Reject} column was set to $0.025$.}
\label{tab:tukey_1}
\vskip -0.5cm
\end{table}

\begin{table}[h!]
        \vskip 0.5cm
	\robustify\bfseries
	\centering
	\begin{sc}
	\resizebox{0.75\columnwidth}{!}{
	\setlength\tabcolsep{4pt}
\begin{tabular}{llrrrrr}
\toprule
       group1 &        group2 &  meandiff &  p-adj &   lower &   upper &  reject \\
\midrule
CM &   MLCM (10 OBIS CQs) &    0.1002 &    0.0 &  0.0881 &  0.1123 &    True \\
CM &   MLCM (20 OBIS CQs) &    0.1305 &    0.0 &  0.1184 &  0.1426 &    True \\
CM &  MLCM (20 na\"{i}ve CQs) &    0.0485 &    0.0 &  0.0364 &  0.0605 &    True \\
CM & MLCM (20 random CQs) &    0.0239 &    0.0 &  0.0119 &  0.0360 &    True \\
CM &                  RSD &   -0.1973 &    0.0 & -0.2093 & -0.1852 &    True \\
  MLCM (10 OBIS CQs) &   MLCM (20 OBIS CQs) &    0.0303 &    0.0 &  0.0182 &  0.0424 &    True \\
  MLCM (10 OBIS CQs) &  MLCM (20 na\"{i}ve CQs) &   -0.0518 &    0.0 & -0.0638 & -0.0397 &    True \\
  MLCM (10 OBIS CQs) & MLCM (20 random CQs) &   -0.0763 &    0.0 & -0.0883 & -0.0642 &    True \\
  MLCM (10 OBIS CQs) &                  RSD &   -0.2975 &    0.0 & -0.3095 & -0.2854 &    True \\
  MLCM (20 OBIS CQs) &  MLCM (20 na\"{i}ve CQs) &   -0.0821 &    0.0 & -0.0941 & -0.0700 &    True \\
  MLCM (20 OBIS CQs) & MLCM (20 random CQs) &   -0.1066 &    0.0 & -0.1187 & -0.0945 &    True \\
  MLCM (20 OBIS CQs) &                  RSD &   -0.3278 &    0.0 & -0.3399 & -0.3157 &    True \\
 MLCM (20 na\"{i}ve CQs) & MLCM (20 random CQs) &   -0.0245 &    0.0 & -0.0366 & -0.0124 &    True \\
 MLCM (20 na\"{i}ve CQs) &                  RSD &   -0.2457 &    0.0 & -0.2578 & -0.2336 &    True \\
MLCM (20 random CQs) &                  RSD &   -0.2212 &    0.0 & -0.2333 & -0.2091 &    True \\
\bottomrule
\end{tabular}

}
\end{sc}
\vskip 0.1cm
\caption{Post-hoc tukey test for minimum student utility. Supply ratio 1.1, 9 popular courses. Significance level for \textsc{Reject} column was set to $0.025$.}
\label{tab:tukey_2}
\vskip -0.5cm
\end{table}

\begin{table}[h!]
	\robustify\bfseries
	\centering
	\begin{sc}
	\resizebox{0.75\columnwidth}{!}{
	\setlength\tabcolsep{4pt}
\begin{tabular}{llrrrrr}
\toprule
       group1 &        group2 &  meandiff &  p-adj &   lower &   upper &  reject \\
\midrule
CM &   MLCM (10 OBIS CQs) &    0.0836 &    0.0 &  0.0794 &  0.0878 &    True \\
CM &   MLCM (20 OBIS CQs) &    0.1094 &    0.0 &  0.1052 &  0.1136 &    True \\
CM &  MLCM (20 na\"{i}ve CQs) &    0.0955 &    0.0 &  0.0913 &  0.0997 &    True \\
CM & MLCM (20 random CQs) &    0.0162 &    0.0 &  0.0120 &  0.0204 &    True \\
CM &                  RSD &   -0.0276 &    0.0 & -0.0318 & -0.0234 &    True \\
MLCM (10 OBIS CQs) &   MLCM (20 OBIS CQs) &    0.0258 &    0.0 &  0.0216 &  0.0300 &    True \\
MLCM (10 OBIS CQs) &  MLCM (20 na\"{i}ve CQs) &    0.0119 &    0.0 &  0.0077 &  0.0161 &    True \\
MLCM (10 OBIS CQs) & MLCM (20 random CQs) &   -0.0674 &    0.0 & -0.0716 & -0.0632 &    True \\
MLCM (10 OBIS CQs) &                  RSD &   -0.1112 &    0.0 & -0.1154 & -0.1070 &    True \\
MLCM (20 OBIS CQs) &  MLCM (20 na\"{i}ve CQs) &   -0.0139 &    0.0 & -0.0181 & -0.0096 &    True \\
MLCM (20 OBIS CQs) & MLCM (20 random CQs) &   -0.0932 &    0.0 & -0.0974 & -0.0889 &    True \\
MLCM (20 OBIS CQs) &                  RSD &   -0.1370 &    0.0 & -0.1412 & -0.1328 &    True \\
MLCM (20 na\"{i}ve CQs) & MLCM (20 random CQs) &   -0.0793 &    0.0 & -0.0835 & -0.0751 &    True \\
MLCM (20 na\"{i}ve CQs) &                  RSD &   -0.1231 &    0.0 & -0.1273 & -0.1189 &    True \\
MLCM (20 random CQs) &                  RSD &   -0.0438 &    0.0 & -0.0480 & -0.0396 &    True \\
\bottomrule
\end{tabular}

}
\end{sc}
\vskip 0.1cm
\caption{Post-hoc tukey test for average student utility. Supply ratio 1.25, 9 popular courses. Significance level for \textsc{Reject} column was set to $0.025$.}
\label{tab:tukey_3}
\vskip -0.5cm
\end{table}

\begin{table}[h!]
	\robustify\bfseries
	\centering
	\begin{sc}
	\resizebox{0.75\columnwidth}{!}{
	\setlength\tabcolsep{4pt}
\begin{tabular}{llrrrrr}
\toprule
       group1 &        group2 &  meandiff &  p-adj &   lower &   upper &  reject \\
\midrule
CM &   MLCM (10 OBIS CQs) &    0.1073 &    0.0 &  0.0952 &  0.1193 &    True \\
CM &   MLCM (20 OBIS CQs) &    0.1364 &    0.0 &  0.1243 &  0.1484 &    True \\
CM &  MLCM (20 na\"{i}ve CQs) &    0.0621 &    0.0 &  0.0500 &  0.0741 &    True \\
CM & MLCM (20 random CQs) &    0.0256 &    0.0 &  0.0136 &  0.0376 &    True \\
CM &                  RSD &   -0.1120 &    0.0 & -0.1240 & -0.1000 &    True \\
MLCM (10 OBIS CQs) &   MLCM (20 OBIS CQs) &    0.0291 &    0.0 &  0.0171 &  0.0411 &    True \\
MLCM (10 OBIS CQs) &  MLCM (20 na\"{i}ve CQs) &   -0.0452 &    0.0 & -0.0572 & -0.0332 &    True \\
MLCM (10 OBIS CQs) & MLCM (20 random CQs) &   -0.0817 &    0.0 & -0.0937 & -0.0697 &    True \\
MLCM (10 OBIS CQs) &                  RSD &   -0.2193 &    0.0 & -0.2313 & -0.2072 &    True \\
MLCM (20 OBIS CQs) &  MLCM (20 na\"{i}ve CQs) &   -0.0743 &    0.0 & -0.0863 & -0.0623 &    True \\
MLCM (20 OBIS CQs) & MLCM (20 random CQs) &   -0.1108 &    0.0 & -0.1228 & -0.0987 &    True \\
MLCM (20 OBIS CQs) &                  RSD &   -0.2484 &    0.0 & -0.2604 & -0.2363 &    True \\
MLCM (20 na\"{i}ve CQs) & MLCM (20 random CQs) &   -0.0365 &    0.0 & -0.0485 & -0.0244 &    True \\
MLCM (20 na\"{i}ve CQs) &                  RSD &   -0.1741 &    0.0 & -0.1861 & -0.1620 &    True \\
MLCM (20 random CQs) &                  RSD &   -0.1376 &    0.0 & -0.1496 & -0.1256 &    True \\
\bottomrule
\end{tabular}

}
\end{sc}
\vskip 0.1cm
\caption{Post-hoc tukey test for minimum student utility. Supply ratio 1.25, 9 popular courses. Significance level for \textsc{Reject} column was set to $0.025$.}
\label{tab:tukey_4}
\vskip -0.5cm
\end{table}

\begin{table}[h!]
	\robustify\bfseries
	\centering
	\begin{sc}
	\resizebox{0.75\columnwidth}{!}{
	\setlength\tabcolsep{4pt}
\begin{tabular}{llrrrrr}
\toprule
       group1 &        group2 &  meandiff &  p-adj &   lower &   upper &  reject \\
\midrule
CM &   MLCM (10 OBIS CQs) &    0.0872 &    0.0 &  0.0830 &  0.0914 &    True \\
CM &   MLCM (20 OBIS CQs) &    0.1122 &    0.0 &  0.1080 &  0.1164 &    True \\
CM &  MLCM (20 na\"{i}ve CQs) &    0.1120 &    0.0 &  0.1078 &  0.1162 &    True \\
CM & MLCM (20 random CQs) &    0.0148 &    0.0 &  0.0106 &  0.0190 &    True \\
CM &                  RSD &   -0.0125 &    0.0 & -0.0167 & -0.0083 &    True \\
MLCM (10 OBIS CQs) &   MLCM (20 OBIS CQs) &    0.0250 &    0.0 &  0.0208 &  0.0292 &    True \\
MLCM (10 OBIS CQs) &  MLCM (20 na\"{i}ve CQs) &    0.0248 &    0.0 &  0.0206 &  0.0290 &    True \\
MLCM (10 OBIS CQs) & MLCM (20 random CQs) &   -0.0724 &    0.0 & -0.0766 & -0.0682 &    True \\
MLCM (10 OBIS CQs) &                  RSD &   -0.0998 &    0.0 & -0.1040 & -0.0956 &    True \\
MLCM (20 OBIS CQs) &  MLCM (20 na\"{i}ve CQs) &   -0.0003 &    1.0 & -0.0045 &  0.0040 &   False \\
MLCM (20 OBIS CQs) & MLCM (20 random CQs) &   -0.0974 &    0.0 & -0.1016 & -0.0932 &    True \\
MLCM (20 OBIS CQs) &                  RSD &   -0.1248 &    0.0 & -0.1290 & -0.1206 &    True \\
MLCM (20 na\"{i}ve CQs) & MLCM (20 random CQs) &   -0.0972 &    0.0 & -0.1014 & -0.0930 &    True \\
MLCM (20 na\"{i}ve CQs) &                  RSD &   -0.1245 &    0.0 & -0.1287 & -0.1203 &    True \\
MLCM (20 random CQs) &                  RSD &   -0.0274 &    0.0 & -0.0316 & -0.0232 &    True \\
\bottomrule
\end{tabular}
}
\end{sc}
\vskip 0.1cm
\caption{Post-hoc tukey test for average student utility. Supply ratio 1.5, 9 popular courses. Significance level for \textsc{Reject} column was set to $0.025$.}
\label{tab:tukey_5}
\vskip -0.5cm
\end{table}

\begin{table}[h!]
	\robustify\bfseries
	\centering
	\begin{sc}
	\resizebox{0.75\columnwidth}{!}{
	\setlength\tabcolsep{4pt}
\begin{tabular}{llrrrrr}
\toprule
       group1 &        group2 &  meandiff &  p-adj &   lower &   upper &  reject \\
\midrule
CM &   MLCM (10 OBIS CQs) &    0.1170 &    0.0 &  0.1055 &  0.1285 &    True \\
CM &   MLCM (20 OBIS CQs) &    0.1427 &    0.0 &  0.1312 &  0.1542 &    True \\
CM &  MLCM (20 na\"{i}ve CQs) &    0.0642 &    0.0 &  0.0527 &  0.0757 &    True \\
CM & MLCM (20 random CQs) &    0.0269 &    0.0 &  0.0154 &  0.0384 &    True \\
CM &                  RSD &   -0.0392 &    0.0 & -0.0507 & -0.0277 &    True \\
MLCM (10 OBIS CQs) &   MLCM (20 OBIS CQs) &    0.0257 &    0.0 &  0.0142 &  0.0372 &    True \\
MLCM (10 OBIS CQs) &  MLCM (20 na\"{i}ve CQs) &   -0.0528 &    0.0 & -0.0643 & -0.0413 &    True \\
MLCM (10 OBIS CQs) & MLCM (20 random CQs) &   -0.0901 &    0.0 & -0.1016 & -0.0786 &    True \\
MLCM (10 OBIS CQs) &                  RSD &   -0.1562 &    0.0 & -0.1677 & -0.1447 &    True \\
MLCM (20 OBIS CQs) &  MLCM (20 na\"{i}ve CQs) &   -0.0785 &    0.0 & -0.0900 & -0.0670 &    True \\
MLCM (20 OBIS CQs) & MLCM (20 random CQs) &   -0.1158 &    0.0 & -0.1273 & -0.1043 &    True \\
MLCM (20 OBIS CQs) &                  RSD &   -0.1819 &    0.0 & -0.1934 & -0.1704 &    True \\
MLCM (20 na\"{i}ve CQs) & MLCM (20 random CQs) &   -0.0373 &    0.0 & -0.0488 & -0.0259 &    True \\
MLCM (20 na\"{i}ve CQs) &                  RSD &   -0.1035 &    0.0 & -0.1149 & -0.0920 &    True \\
MLCM (20 random CQs) &                  RSD &   -0.0661 &    0.0 & -0.0776 & -0.0546 &    True \\
\bottomrule
\end{tabular}

}
\end{sc}
\vskip 0.1cm
\caption{Post-hoc tukey test for minimum student utility. Supply ratio 1.5, 9 popular courses. Significance level for \textsc{Reject} column was set to $0.025$.}
\label{tab:tukey_6}
\vskip -0.25cm
\end{table}

\begin{table}[h!]
	\robustify\bfseries
	\centering
	\begin{sc}
	\resizebox{0.75\columnwidth}{!}{
	\setlength\tabcolsep{4pt}
\begin{tabular}{llrrrrr}
\toprule
       group1 &        group2 &  meandiff &  p-adj &   lower &   upper &  reject \\
\midrule
CM &   MLCM (10 OBIS CQs) &    0.0553 &    0.0 &  0.0508 &  0.0598 &    True \\
CM &   MLCM (20 OBIS CQs) &    0.0737 &    0.0 &  0.0692 &  0.0782 &    True \\
CM &  MLCM (20 na\"{i}ve CQs) &    0.0230 &    0.0 &  0.0185 &  0.0275 &    True \\
CM & MLCM (20 random CQs) &   -0.0099 &    0.0 & -0.0144 & -0.0054 &    True \\
CM &                  RSD &   -0.0470 &    0.0 & -0.0515 & -0.0425 &    True \\
MLCM (10 OBIS CQs) &   MLCM (20 OBIS CQs) &    0.0184 &    0.0 &  0.0139 &  0.0229 &    True \\
MLCM (10 OBIS CQs) &  MLCM (20 na\"{i}ve CQs) &   -0.0323 &    0.0 & -0.0367 & -0.0278 &    True \\
MLCM (10 OBIS CQs) & MLCM (20 random CQs) &   -0.0652 &    0.0 & -0.0697 & -0.0607 &    True \\
MLCM (10 OBIS CQs) &                  RSD &   -0.1022 &    0.0 & -0.1067 & -0.0977 &    True \\
MLCM (20 OBIS CQs) &  MLCM (20 na\"{i}ve CQs) &   -0.0507 &    0.0 & -0.0552 & -0.0462 &    True \\
MLCM (20 OBIS CQs) & MLCM (20 random CQs) &   -0.0836 &    0.0 & -0.0881 & -0.0791 &    True \\
MLCM (20 OBIS CQs) &                  RSD &   -0.1207 &    0.0 & -0.1252 & -0.1162 &    True \\
MLCM (20 na\"{i}ve CQs) & MLCM (20 random CQs) &   -0.0329 &    0.0 & -0.0374 & -0.0284 &    True \\
MLCM (20 na\"{i}ve CQs) &                  RSD &   -0.0700 &    0.0 & -0.0745 & -0.0655 &    True \\
MLCM (20 random CQs) &                  RSD &   -0.0370 &    0.0 & -0.0415 & -0.0325 &    True \\
\bottomrule
\end{tabular}

}
\end{sc}
\vskip 0.1cm
\caption{Post-hoc tukey test for average student utility. Supply ratio 1.1, 6 popular courses. Significance level for \textsc{Reject} column was set to $0.025$.}
\label{tab:tukey_7}
\vskip -0.25cm
\end{table}

\begin{table}[h!]
	\robustify\bfseries
	\centering
	\begin{sc}
	\resizebox{0.75\columnwidth}{!}{
	\setlength\tabcolsep{4pt}
\begin{tabular}{llrrrrr}
\toprule
       group1 &        group2 &  meandiff &  p-adj &   lower &   upper &  reject \\
\midrule
CM &   MLCM (10 OBIS CQs) &    0.0869 & 0.0000 &  0.0742 &  0.0997 &    True \\
CM &   MLCM (20 OBIS CQs) &    0.1023 & 0.0000 &  0.0896 &  0.1150 &    True \\
CM &  MLCM (20 na\"{i}ve CQs) &    0.0101 & 0.1384 & -0.0026 &  0.0229 &   False \\
CM & MLCM (20 random CQs) &   -0.0012 & 0.9997 & -0.0139 &  0.0116 &   False \\
CM &                  RSD &   -0.2530 & 0.0000 & -0.2657 & -0.2402 &    True \\
MLCM (10 OBIS CQs) &   MLCM (20 OBIS CQs) &    0.0154 & 0.0028 &  0.0026 &  0.0281 &    True \\
MLCM (10 OBIS CQs) &  MLCM (20 na\"{i}ve CQs) &   -0.0768 & 0.0000 & -0.0895 & -0.0641 &    True \\
MLCM (10 OBIS CQs) & MLCM (20 random CQs) &   -0.0881 & 0.0000 & -0.1009 & -0.0754 &    True \\
MLCM (10 OBIS CQs) &                  RSD &   -0.3399 & 0.0000 & -0.3526 & -0.3272 &    True \\
MLCM (20 OBIS CQs) &  MLCM (20 na\"{i}ve CQs) &   -0.0922 & 0.0000 & -0.1049 & -0.0794 &    True \\
MLCM (20 OBIS CQs) & MLCM (20 random CQs) &   -0.1035 & 0.0000 & -0.1162 & -0.0907 &    True \\
MLCM (20 OBIS CQs) &                  RSD &   -0.3553 & 0.0000 & -0.3680 & -0.3425 &    True \\
MLCM (20 na\"{i}ve CQs) & MLCM (20 random CQs) &   -0.0113 & 0.0679 & -0.0240 &  0.0014 &   False \\
MLCM (20 na\"{i}ve CQs) &                  RSD &   -0.2631 & 0.0000 & -0.2758 & -0.2504 &    True \\
MLCM (20 random CQs) &                  RSD &   -0.2518 & 0.0000 & -0.2645 & -0.2391 &    True \\
\bottomrule
\end{tabular}

}
\end{sc}
\vskip 0.1cm
\caption{Post-hoc tukey test for minimum student utility. Supply ratio 1.1, 6 popular courses. Significance level for \textsc{Reject} column was set to $0.025$.}
\label{tab:tukey_8}
\vskip -0.25cm
\end{table}

\begin{table}[h!]
	\robustify\bfseries
	\centering
	\begin{sc}
	\resizebox{0.75\columnwidth}{!}{
	\setlength\tabcolsep{4pt}
\begin{tabular}{llrrrrr}
\toprule
       group1 &        group2 &  meandiff &  p-adj &   lower &   upper &  reject \\
\midrule
CM &   MLCM (10 OBIS CQs) &    0.0561 &    0.0 &  0.0514 &  0.0607 &    True \\
CM &   MLCM (20 OBIS CQs) &    0.0742 &    0.0 &  0.0695 &  0.0788 &    True \\
CM &  MLCM (20 na\"{i}ve CQs) &    0.0327 &    0.0 &  0.0280 &  0.0374 &    True \\
CM & MLCM (20 random CQs) &   -0.0079 &    0.0 & -0.0126 & -0.0032 &    True \\
CM &                  RSD &   -0.0359 &    0.0 & -0.0406 & -0.0312 &    True \\
MLCM (10 OBIS CQs) &   MLCM (20 OBIS CQs) &    0.0181 &    0.0 &  0.0134 &  0.0228 &    True \\
MLCM (10 OBIS CQs) &  MLCM (20 na\"{i}ve CQs) &   -0.0233 &    0.0 & -0.0280 & -0.0187 &    True \\
MLCM (10 OBIS CQs) & MLCM (20 random CQs) &   -0.0640 &    0.0 & -0.0686 & -0.0593 &    True \\
MLCM (10 OBIS CQs) &                  RSD &   -0.0920 &    0.0 & -0.0966 & -0.0873 &    True \\
MLCM (20 OBIS CQs) &  MLCM (20 na\"{i}ve CQs) &   -0.0414 &    0.0 & -0.0461 & -0.0368 &    True \\
MLCM (20 OBIS CQs) & MLCM (20 random CQs) &   -0.0821 &    0.0 & -0.0867 & -0.0774 &    True \\
MLCM (20 OBIS CQs) &                  RSD &   -0.1101 &    0.0 & -0.1147 & -0.1054 &    True \\
MLCM (20 na\"{i}ve CQs) & MLCM (20 random CQs) &   -0.0406 &    0.0 & -0.0453 & -0.0359 &    True \\
MLCM (20 na\"{i}ve CQs) &                  RSD &   -0.0686 &    0.0 & -0.0733 & -0.0639 &    True \\
MLCM (20 random CQs) &                  RSD &   -0.0280 &    0.0 & -0.0327 & -0.0233 &    True \\
\bottomrule
\end{tabular}

}
\end{sc}
\vskip 0.1cm
\caption{Post-hoc tukey test for average student utility. Supply ratio 1.25, 6 popular courses. Significance level for \textsc{Reject} column was set to $0.025$.}
\label{tab:tukey_9}
\vskip -0.25cm
\end{table}

\begin{table}[h!]
	\robustify\bfseries
	\centering
	\begin{sc}
	\resizebox{0.75\columnwidth}{!}{
	\setlength\tabcolsep{4pt}
\begin{tabular}{llrrrrr}
\toprule
       group1 &        group2 &  meandiff &  p-adj &   lower &   upper &  reject \\
\midrule
CM &   MLCM (10 OBIS CQs) &    0.0821 & 0.0000 &  0.0694 &  0.0949 &    True \\
CM &   MLCM (20 OBIS CQs) &    0.1019 & 0.0000 &  0.0891 &  0.1146 &    True \\
CM &  MLCM (20 na\"{i}ve CQs) &    0.0153 & 0.0030 &  0.0025 &  0.0281 &    True \\
CM & MLCM (20 random CQs) &    0.0021 & 0.9962 & -0.0107 &  0.0148 &   False \\
CM &                  RSD &   -0.1832 & 0.0000 & -0.1960 & -0.1705 &    True \\
MLCM (10 OBIS CQs) &   MLCM (20 OBIS CQs) &    0.0197 & 0.0000 &  0.0069 &  0.0325 &    True \\
MLCM (10 OBIS CQs) &  MLCM (20 na\"{i}ve CQs) &   -0.0668 & 0.0000 & -0.0796 & -0.0541 &    True \\
MLCM (10 OBIS CQs) & MLCM (20 random CQs) &   -0.0801 & 0.0000 & -0.0928 & -0.0673 &    True \\
MLCM (10 OBIS CQs) &                  RSD &   -0.2654 & 0.0000 & -0.2781 & -0.2526 &    True \\
MLCM (20 OBIS CQs) &  MLCM (20 na\"{i}ve CQs) &   -0.0865 & 0.0000 & -0.0993 & -0.0738 &    True \\
MLCM (20 OBIS CQs) & MLCM (20 random CQs) &   -0.0998 & 0.0000 & -0.1126 & -0.0870 &    True \\
MLCM (20 OBIS CQs) &                  RSD &   -0.2851 & 0.0000 & -0.2978 & -0.2723 &    True \\
MLCM (20 na\"{i}ve CQs) & MLCM (20 random CQs) &   -0.0133 & 0.0172 & -0.0260 & -0.0005 &    True \\
MLCM (20 na\"{i}ve CQs) &                  RSD &   -0.1985 & 0.0000 & -0.2113 & -0.1858 &    True \\
MLCM (20 random CQs) &                  RSD &   -0.1853 & 0.0000 & -0.1981 & -0.1725 &    True \\
\bottomrule
\end{tabular}

}
\end{sc}
\vskip 0.1cm
\caption{Post-hoc tukey test for minimum student utility. Supply ratio 1.25, 6 popular courses. Significance level for \textsc{Reject} column was set to $0.025$.}
\label{tab:tukey_10}
\vskip -0.25cm
\end{table}

\begin{table}[h!]
	\robustify\bfseries
	\centering
	\begin{sc}
	\resizebox{0.75\columnwidth}{!}{
	\setlength\tabcolsep{4pt}
\begin{tabular}{llrrrrr}
\toprule
       group1 &        group2 &  meandiff &  p-adj &   lower &   upper &  reject \\
\midrule
CM &   MLCM (10 OBIS CQs) &    0.0610 & 0.0000 &  0.0561 &  0.0658 &    True \\
CM &   MLCM (20 OBIS CQs) &    0.0782 & 0.0000 &  0.0734 &  0.0831 &    True \\
CM &  MLCM (20 na\"{i}ve CQs) &    0.0488 & 0.0000 &  0.0440 &  0.0537 &    True \\
CM & MLCM (20 random CQs) &   -0.0062 & 0.0011 & -0.0111 & -0.0014 &    True \\
CM &                  RSD &   -0.0300 & 0.0000 & -0.0348 & -0.0251 &    True \\
MLCM (10 OBIS CQs) &   MLCM (20 OBIS CQs) &    0.0172 & 0.0000 &  0.0124 &  0.0221 &    True \\
MLCM (10 OBIS CQs) &  MLCM (20 na\"{i}ve CQs) &   -0.0122 & 0.0000 & -0.0170 & -0.0073 &    True \\
MLCM (10 OBIS CQs) & MLCM (20 random CQs) &   -0.0672 & 0.0000 & -0.0720 & -0.0623 &    True \\
MLCM (10 OBIS CQs) &                  RSD &   -0.0910 & 0.0000 & -0.0958 & -0.0861 &    True \\
MLCM (20 OBIS CQs) &  MLCM (20 na\"{i}ve CQs) &   -0.0294 & 0.0000 & -0.0343 & -0.0246 &    True \\
MLCM (20 OBIS CQs) & MLCM (20 random CQs) &   -0.0844 & 0.0000 & -0.0893 & -0.0796 &    True \\
MLCM (20 OBIS CQs) &                  RSD &   -0.1082 & 0.0000 & -0.1131 & -0.1034 &    True \\
MLCM (20 na\"{i}ve CQs) & MLCM (20 random CQs) &   -0.0550 & 0.0000 & -0.0599 & -0.0502 &    True \\
MLCM (20 na\"{i}ve CQs) &                  RSD &   -0.0788 & 0.0000 & -0.0836 & -0.0739 &    True \\
MLCM (20 random CQs) &                  RSD &   -0.0238 & 0.0000 & -0.0286 & -0.0189 &    True \\
\bottomrule
\end{tabular}

}
\end{sc}
\vskip 0.1cm
\caption{Post-hoc tukey test for average student utility. Supply ratio 1.5, 6 popular courses. Significance level for \textsc{Reject} column was set to $0.025$.}
\label{tab:tukey_11}
\vskip -0.25cm
\end{table}

\begin{table}[h!]
	\robustify\bfseries
	\centering
	\begin{sc}
	\resizebox{0.75\columnwidth}{!}{
	\setlength\tabcolsep{4pt}
\begin{tabular}{llrrrrr}
\toprule
       group1 &        group2 &  meandiff &  p-adj &   lower &   upper &  reject \\
\midrule
CM &   MLCM (10 OBIS CQs) &    0.0925 &   0.00 &  0.0800 &  0.1050 &    True \\
CM &   MLCM (20 OBIS CQs) &    0.1148 &   0.00 &  0.1023 &  0.1273 &    True \\
CM &  MLCM (20 na\"{i}ve CQs) &    0.0299 &   0.00 &  0.0174 &  0.0423 &    True \\
CM & MLCM (20 random CQs) &    0.0048 &   0.84 & -0.0077 &  0.0173 &   False \\
CM &                  RSD &   -0.1454 &   0.00 & -0.1579 & -0.1329 &    True \\
MLCM (10 OBIS CQs) &   MLCM (20 OBIS CQs) &    0.0223 &   0.00 &  0.0098 &  0.0348 &    True \\
MLCM (10 OBIS CQs) &  MLCM (20 na\"{i}ve CQs) &   -0.0627 &   0.00 & -0.0751 & -0.0502 &    True \\
MLCM (10 OBIS CQs) & MLCM (20 random CQs) &   -0.0877 &   0.00 & -0.1002 & -0.0752 &    True \\
MLCM (10 OBIS CQs) &                  RSD &   -0.2379 &   0.00 & -0.2504 & -0.2255 &    True \\
MLCM (20 OBIS CQs) &  MLCM (20 na\"{i}ve CQs) &   -0.0849 &   0.00 & -0.0974 & -0.0724 &    True \\
MLCM (20 OBIS CQs) & MLCM (20 random CQs) &   -0.1100 &   0.00 & -0.1224 & -0.0975 &    True \\
MLCM (20 OBIS CQs) &                  RSD &   -0.2602 &   0.00 & -0.2727 & -0.2477 &    True \\
MLCM (20 na\"{i}ve CQs) & MLCM (20 random CQs) &   -0.0250 &   0.00 & -0.0375 & -0.0125 &    True \\
MLCM (20 na\"{i}ve CQs) &                  RSD &   -0.1753 &   0.00 & -0.1878 & -0.1628 &    True \\
MLCM (20 random CQs) &                  RSD &   -0.1503 &   0.00 & -0.1627 & -0.1378 &    True \\
\bottomrule
\end{tabular}

}
\end{sc}
\vskip 0.1cm
\caption{Post-hoc tukey test for minimum student utility. Supply ratio 1.5, 6 popular courses. Significance level for \textsc{Reject} column was set to $0.025$.}
\label{tab:tukey_12}
\vskip -0.25cm
\end{table}

\begin{table}[h!]
        \vskip 0.5cm
	\robustify\bfseries
	\centering
	\begin{sc}
	\resizebox{0.75\columnwidth}{!}{
	\setlength\tabcolsep{4pt}
\begin{tabular}{llrrrrr}
\toprule
       group1 &        group2 &  meandiff &  p-adj &   lower &   upper &  reject \\
\midrule
$\text{CM}^*$ &                 CM &   -0.1639 &    0.0 & -0.1654 & -0.1625 &    True \\
$\text{CM}^*$ & MLCM (10 OBIS CQs) &   -0.0756 &    0.0 & -0.0770 & -0.0742 &    True \\
$\text{CM}^*$ & MLCM (20 OBIS CQs) &   -0.0568 &    0.0 & -0.0582 & -0.0553 &    True \\
CM & MLCM (10 OBIS CQs) &    0.0883 &    0.0 &  0.0869 &  0.0898 &    True \\
CM & MLCM (20 OBIS CQs) &    0.1072 &    0.0 &  0.1057 &  0.1086 &    True \\
MLCM (10 OBIS CQs) & MLCM (20 OBIS CQs) &    0.0188 &    0.0 &  0.0174 &  0.0203 &    True \\
\bottomrule
\end{tabular}

}
\end{sc}
\vskip 0.1cm
\caption{Post-hoc tukey test for average student utility. Supply ratio 1.1, 9 popular courses, additive preferences. Significance level for \textsc{Reject} column was set to $0.025$.}
\label{tab:tukey_13}
\vskip -0.5cm
\end{table}

\begin{table}[h!]
        \vskip 0.5cm
	\robustify\bfseries
	\centering
	\begin{sc}
	\resizebox{0.75\columnwidth}{!}{
	\setlength\tabcolsep{4pt}
\begin{tabular}{llrrrrr}
\toprule
       group1 &        group2 &  meandiff &  p-adj &   lower &   upper &  reject \\
\midrule
$\text{CM}^*$ &                 CM &   -0.3174 &    0.0 & -0.3260 & -0.3087 &    True \\
$\text{CM}^*$ & MLCM (10 OBIS CQs) &   -0.1773 &    0.0 & -0.1859 & -0.1687 &    True \\
$\text{CM}^*$ & MLCM (20 OBIS CQs) &   -0.1523 &    0.0 & -0.1609 & -0.1437 &    True \\
CM & MLCM (10 OBIS CQs) &    0.1400 &    0.0 &  0.1314 &  0.1487 &    True \\
CM & MLCM (20 OBIS CQs) &    0.1651 &    0.0 &  0.1565 &  0.1737 &    True \\
MLCM (10 OBIS CQs) & MLCM (20 OBIS CQs) &    0.0250 &    0.0 &  0.0164 &  0.0337 &    True \\
\bottomrule
\end{tabular}

}
\end{sc}
\vskip 0.1cm
\caption{Post-hoc tukey test for minimum student utility. Supply ratio 1.1, 9 popular courses, additive preferences. Significance level for \textsc{Reject} column was set to $0.025$.}
\label{tab:tukey_14}
\vskip -0.5cm
\end{table}

\begin{table}[h!]
	\robustify\bfseries
	\centering
	\begin{sc}
	\resizebox{0.75\columnwidth}{!}{
	\setlength\tabcolsep{4pt}
\begin{tabular}{llrrrrr}
\toprule
       group1 &        group2 &  meandiff &  p-adj &   lower &   upper &  reject \\
\midrule
$\text{CM}^*$ &                 CM &   -0.1454 &    0.0 & -0.1468 & -0.1439 &    True \\
$\text{CM}^*$ & MLCM (10 OBIS CQs) &   -0.0685 &    0.0 & -0.0699 & -0.0671 &    True \\
$\text{CM}^*$ & MLCM (20 OBIS CQs) &   -0.0507 &    0.0 & -0.0521 & -0.0493 &    True \\
CM & MLCM (10 OBIS CQs) &    0.0769 &    0.0 &  0.0754 &  0.0783 &    True \\
CM & MLCM (20 OBIS CQs) &    0.0947 &    0.0 &  0.0932 &  0.0961 &    True \\
MLCM (10 OBIS CQs) & MLCM (20 OBIS CQs) &    0.0178 &    0.0 &  0.0164 &  0.0192 &    True \\
\bottomrule
\end{tabular}

}
\end{sc}
\vskip 0.1cm
\caption{Post-hoc tukey test for average student utility. Supply ratio 1.25, 9 popular courses, additive preferences. Significance level for \textsc{Reject} column was set to $0.025$.}
\label{tab:tukey_15}
\vskip -0.5cm
\end{table}

\begin{table}[h!]
	\robustify\bfseries
	\centering
	\begin{sc}
	\resizebox{0.75\columnwidth}{!}{
	\setlength\tabcolsep{4pt}
\begin{tabular}{llrrrrr}
\toprule
       group1 &        group2 &  meandiff &  p-adj &   lower &   upper &  reject \\
\midrule
$\text{CM}^*$ &                 CM &   -0.2994 &    0.0 & -0.3078 & -0.2909 &    True \\
$\text{CM}^*$ & MLCM (10 OBIS CQs) &   -0.1664 &    0.0 & -0.1749 & -0.1579 &    True \\
$\text{CM}^*$ & MLCM (20 OBIS CQs) &   -0.1389 &    0.0 & -0.1474 & -0.1305 &    True \\
CM & MLCM (10 OBIS CQs) &    0.1330 &    0.0 &  0.1245 &  0.1415 &    True \\
CM & MLCM (20 OBIS CQs) &    0.1604 &    0.0 &  0.1519 &  0.1689 &    True \\
MLCM (10 OBIS CQs) & MLCM (20 OBIS CQs) &    0.0274 &    0.0 &  0.0189 &  0.0359 &    True \\
\bottomrule
\end{tabular}

}
\end{sc}
\vskip 0.1cm
\caption{Post-hoc tukey test for minimum student utility. Supply ratio 1.25, 9 popular courses, additive preferences. Significance level for \textsc{Reject} column was set to $0.025$.}
\label{tab:tukey_16}
\vskip -0.5cm
\end{table}

\begin{table}[h!]
	\robustify\bfseries
	\centering
	\begin{sc}
	\resizebox{0.75\columnwidth}{!}{
	\setlength\tabcolsep{4pt}
\begin{tabular}{llrrrrr}
\toprule
       group1 &        group2 &  meandiff &  p-adj &   lower &   upper &  reject \\
\midrule
$\text{CM}^*$ &                 CM &   -0.1191 &    0.0 & -0.1205 & -0.1176 &    True \\
$\text{CM}^*$ & MLCM (10 OBIS CQs) &   -0.0588 &    0.0 & -0.0602 & -0.0573 &    True \\
$\text{CM}^*$ & MLCM (20 OBIS CQs) &   -0.0430 &    0.0 & -0.0445 & -0.0415 &    True \\
CM & MLCM (10 OBIS CQs) &    0.0603 &    0.0 &  0.0588 &  0.0618 &    True \\
CM & MLCM (20 OBIS CQs) &    0.0761 &    0.0 &  0.0746 &  0.0775 &    True \\
MLCM (10 OBIS CQs) & MLCM (20 OBIS CQs) &    0.0157 &    0.0 &  0.0143 &  0.0172 &    True \\
\bottomrule
\end{tabular}
}
\end{sc}
\vskip 0.1cm
\caption{Post-hoc tukey test for average student utility. Supply ratio 1.5, 9 popular courses, additive preferences. Significance level for \textsc{Reject} column was set to $0.025$.}
\label{tab:tukey_17}
\vskip -0.5cm
\end{table}

\begin{table}[h!]
	\robustify\bfseries
	\centering
	\begin{sc}
	\resizebox{0.75\columnwidth}{!}{
	\setlength\tabcolsep{4pt}
\begin{tabular}{llrrrrr}
\toprule
       group1 &        group2 &  meandiff &  p-adj &   lower &   upper &  reject \\
\midrule
$\text{CM}^*$ &                 CM &   -0.2693 &    0.0 & -0.2781 & -0.2605 &    True \\
$\text{CM}^*$ & MLCM (10 OBIS CQs) &   -0.1538 &    0.0 & -0.1626 & -0.1450 &    True \\
$\text{CM}^*$ & MLCM (20 OBIS CQs) &   -0.1288 &    0.0 & -0.1375 & -0.1200 &    True \\
CM & MLCM (10 OBIS CQs) &    0.1155 &    0.0 &  0.1068 &  0.1243 &    True \\
CM & MLCM (20 OBIS CQs) &    0.1406 &    0.0 &  0.1318 &  0.1493 &    True \\
MLCM (10 OBIS CQs) & MLCM (20 OBIS CQs) &    0.0250 &    0.0 &  0.0163 &  0.0338 &    True \\
\bottomrule
\end{tabular}

}
\end{sc}
\vskip 0.1cm
\caption{Post-hoc tukey test for minimum student utility. Supply ratio 1.5, 9 popular courses, additive preferences. Significance level for \textsc{Reject} column was set to $0.025$.}
\label{tab:tukey_18}
\vskip -0.25cm
\end{table}


\newpage

\end{document}